\documentclass[aps,pra,reprint,superscriptaddress,nobibnotes]{revtex4-1}
\usepackage[english]{babel}
\usepackage[utf8x]{inputenc}
\usepackage[T1]{fontenc}
\usepackage{url}
\usepackage{hyperref}
\usepackage{graphicx}
\graphicspath{{figures/}}
\usepackage{amsmath} 
\usepackage{amssymb} 
\usepackage[detect-all]{siunitx}
\usepackage{bm}
\usepackage{mathtools}
\usepackage{color}
\usepackage{layouts}
\renewcommand{\d}{\mathrm{d}}
\usepackage{amsthm}

\usepackage{fancyhdr}
\begin{document}

\title{Twisted domain walls and skyrmions in perpendicularly magnetized multilayers}
\author{Ivan Lemesh}
\email{ivan.g.lemesh@gmail.com}
\affiliation{Department of Materials Science and Engineering, Massachusetts Institute of Technology, Cambridge, Massachusetts 02139, USA}
\author{Geoffrey S. D. Beach}
\affiliation{Department of Materials Science and Engineering, Massachusetts Institute of Technology, Cambridge, Massachusetts 02139, USA}
\date{\today}
\begin{abstract}
We present an analytical theory to describe three-dimensional magnetic textures in perpendicularly magnetized magnetic multilayers that arise in the presence of magnetostatic interactions and the Dzyaloshinskii-Moriya interaction (DMI).  We demonstrate that domain walls in multilayers develop a complex twisted structure, which persists even for films with strong DMI. The origin of this twist is surface-volume stray field interactions that manifest as a depth-dependent effective field whose form mimics the DMI effective field. We find that the wall twist has a minor impact on the equilibrium skyrmion or domain size, but can significantly affect current-driven dynamics. Our conclusions are based on the derived analytical expressions for the magnetostatic energy and confirmed by micromagnetic simulations.
\end{abstract}

\maketitle 
\thispagestyle{fancy}
\section{Introduction}
Magnetic thin films with chiral exchange interactions can host a variety of topological spin textures such as homochiral domain walls (DWs)~\cite{Heide2008a,Thiaville2012,Emori2013f} and magnetic skyrmions~\cite{Rossler2006,Muhlbauer2009,Fert2013} with rich fundamental behaviors.  Although usually considered as two-dimensional (2D) systems, thin films with competing surface and volume interactions can exhibit more complex three-dimensional (3D) textures, as recently realized in the case of cubic helimagnets~\cite{Rybakov2013,Meynell2014,Leonov2016,Rybakov2016,McGrouther2016,Zhang2018} with bulk Dzyaloshinskii-Moriya interaction (DMI).  In the case of heavy-metal/ferromagnet bilayers with perpendicular magnetic anisotropy (PMA) and interfacial DMI, the ferromagnet thickness is typically much less than the exchange length so the spin textures are truly 2D~\cite{Yoshida2012,Romming2015,Wiesendanger2016}.  However, recent efforts to stabilize such textures at room temperature have employed multilayers in which the 2D textures are coupled from layer to layer by dipolar fields~\cite{Buttner2015b,Woo2016,Moreau-Luchaire2016a,Litzius2016,Wiesendanger2016,Buttner2017e}.  Such composite spin textures  are usually treated two-dimensionally with magnetic properties scaled using an effective medium approach~\cite{Suna1986a,Woo2016,Lemesh2017} and with the assumption of a layer-independent magnetization profile (the 2D model). However, recently~\cite{Dovzhenko2016,Montoya2017}, it has been argued that the actual magnetic configuration of multilayers is rather different, and that the equilibrium DW width $\Delta$ and angle $\psi$ vary from one layer to another. Previously, such an idea of twisted DWs has already been explored theoretically by Schl\"omann~\cite{Schlomann1973a,Schlomann1973b}, who found a similar magnetization distribution in thick magnetic single layer films ($\mathcal{T}>l_{ex}$). 

In this paper, we show DW twists (see Fig.~\ref{fig:1}a) emerge as a general feature in magnetic thin film multilayers due to chiral stray field interactions. We solve the multilayer stray field integrals analytically and find that the twist is caused by the previously ignored mutual surface-volume stray field interactions, which mathematically resemble a layer-dependent interfacial Dzyaloshinskii-Moriya interaction (DMI). We develop an analytical 3D model to accurately predict the equilibrium structure of domains and skyrmions, as well as to describe current-driven skyrmion dynamics.

\section{Twisted straight domain wall}
\begin{figure}[ht]
  	\includegraphics[scale = 0.53]{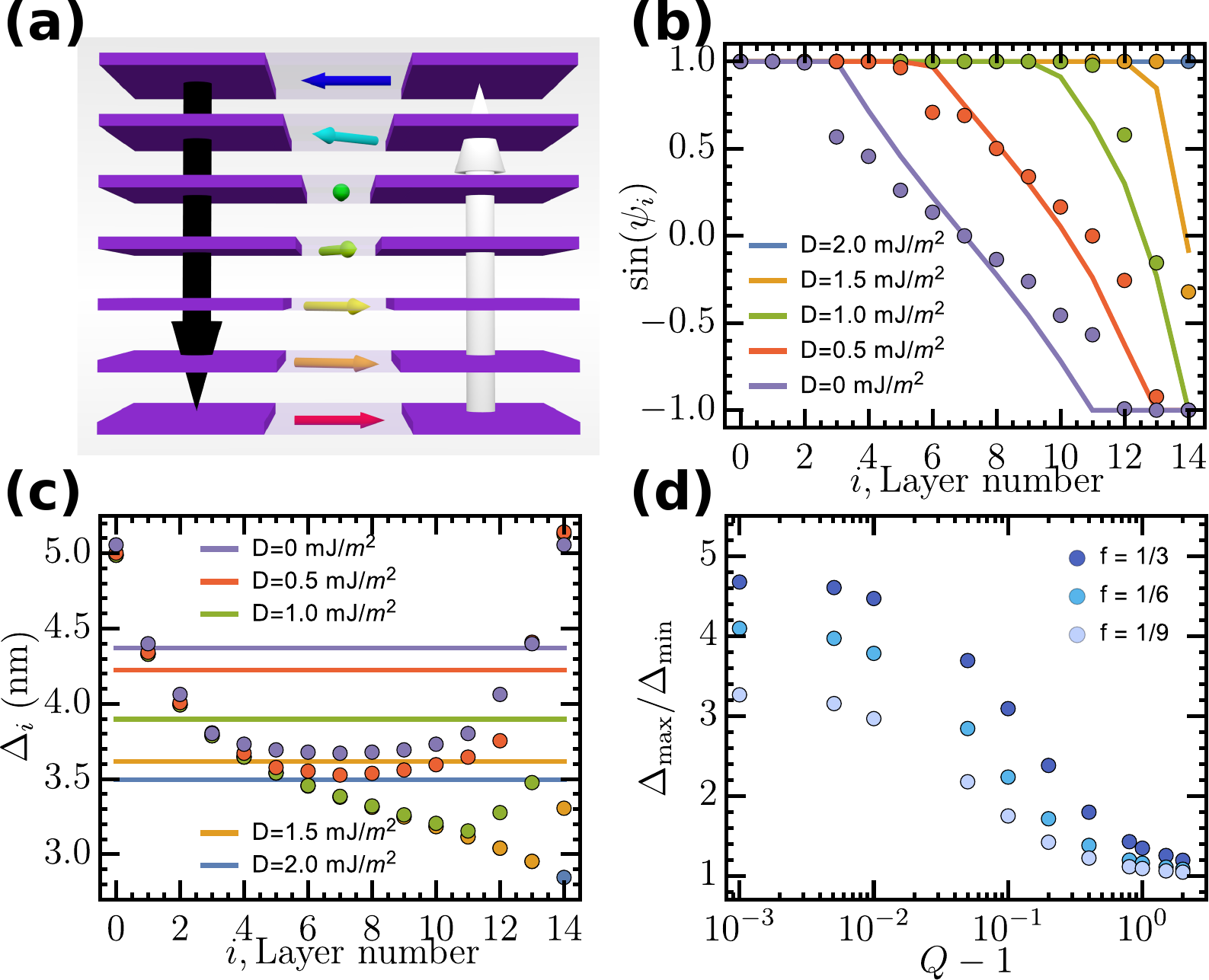}
  	\caption{{\bf DW twist.} {\bf(a)} Schematic plot of the ($\downarrow|\uparrow$) twisted  DW. {\bf(b)} $\psi_i$ and {\bf(c)}  $\Delta_i$ as a function of the layer number and interfacial DMI for a film with $Q=1.4$. Points represent the simulated results, continuous lines show the numerical solution of the proposed twisted wall theory. {\bf(d)} $\Delta_{\rm{max}}/\Delta_{\rm{min}}$ ratio as a function of $Q $ and a scaling factor $f=\mathcal{T}/\mathcal{P}$. }\label{fig:1}
  \end{figure}

First, consider an isolated straight DW in a multilayer film comprised of magnetic and nonmagnetic layers, where $\mathcal{T}$ is the magnetic layer thickness, $\mathcal{P}$ is the multilayer period, and $\mathcal{N}$ is the number of multilayer repeats.  Micromagnetic simulations for a representative Co-based  multilayer~\cite{Buttner2015b,Woo2016,Moreau-Luchaire2016a,Litzius2016,Wiesendanger2016,Buttner2017e}, (saturation magnetization $M_s = \SI{1.4e6}{A/m}$, exchange stiffness $A = \SI{1.0e-11}{ J/m}$, quality factor $Q = 2K_u/(\mu_0M_s^2)=1.4$, and with $\mathcal{N}=15$, $\mathcal{T}=1~\rm{nm}$, and $\mathcal{P} = 6~\rm{nm}$), summarized in Figs.~\ref{fig:1}a-d, reveal that both $\Delta_i$ and the DW angle $\psi_i$ varies from layer to layer $(i)$.  When the DMI constant $D=0$, the DWs in the top and the bottom layers have N\'eel profile with opposite chiralities and larger $\Delta$. In contrast, the middle layers exhibit Bloch DWs with smaller $\Delta$. Increasing the DMI shifts the position of the Bloch layer towards one surface, and at very high DMI all the layers saturate to a homochiral N\'eel state.  

\begin{figure*}[ht]
  	\includegraphics[scale = 0.69]{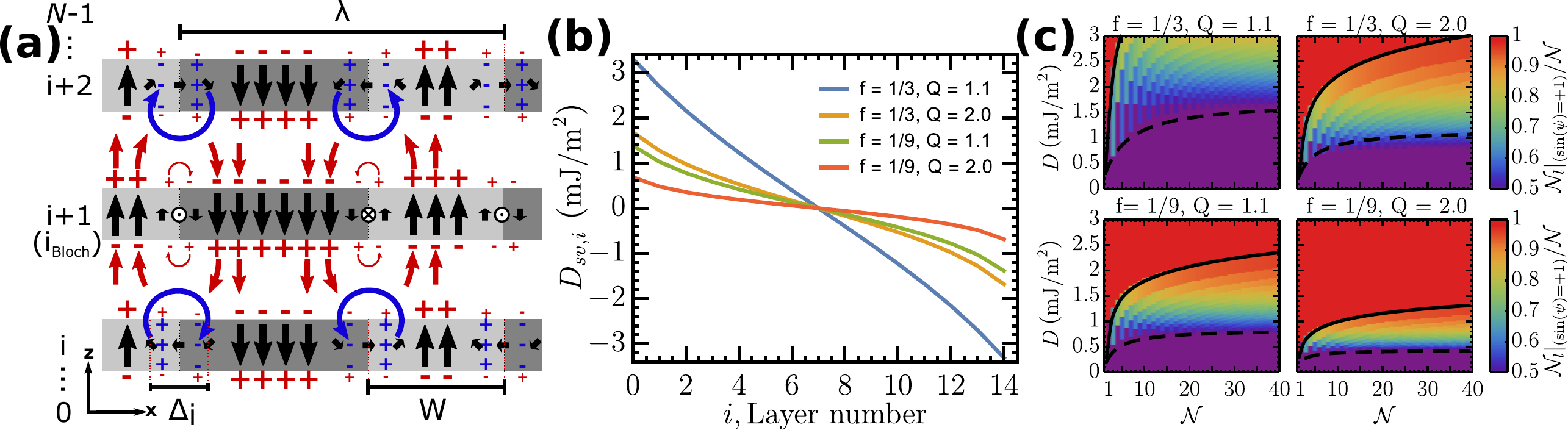}
  	\caption{{\bf DW twist.} {\bf(a)} The schematic distribution of the surface (volume) charges depicted with red (blue) signs in the layers surrounding the Bloch layer.  {\bf(b)} The surface-volume stray field interaction term $D_{sv}$ as a function of the layer number $(i)$, for various values of $f$ and $Q$ for ($\downarrow|\uparrow$) twisted  DW. {\bf(c)} The fraction of layers with $\sin(\psi_i)=+1$  as a function of DMI, $\mathcal{N}$,  $f$, and $Q$. The dashed curves in {\bf(c)} correspond to the 2D model prediction of the threshold for purely N\'eel DWs. The solid curves give the 3D analytical model prediction derived here.}\label{fig:2}
  \end{figure*}

Figure~\ref{fig:2}a shows schematically the stray fields around the Bloch layer ($i_{\rm{Bloch}}$), explaining the origin of the wall twist. In the adjacent top and bottom nonmagnetic layers, the surface stray fields of the neighboring domains are antiparallel. The energy of the system is minimized if these fields are co-aligned with the stray fields from the neighboring layers, giving rise to domain coupling.  This tendency also favors the creation of corresponding volume charges $\rho_v=-\nabla \cdot \mathbf{M} $ (shown in blue), which results in the observed DW twist. The stray fields tend to increase (decrease) $\Delta$ when they are parallel (antiparallel) to the DW magnetization, hence leading to the observed thickness-dependent $\Delta$.  

To quantify these effects, one must calculate the corresponding surface-volume stray field integral~\footnote{In our notations, $\sigma^{\mathcal{M}, \mathcal{N}}$ indicates the  energy (per multilayer cross-section) of $\mathcal{M}$ DWs in the film with $\mathcal{N}$ multilayer repeats}
\begin{equation}
\sigma_{d, sv}^{ 1, \mathcal{N}} = \frac{\mu_0}{4\pi \mathcal{N}\mathcal{P}L_y} \iint\d^3 r \d^3 r^\prime \rho_s(\mathbf{r})\frac{1}{{|\mathbf{r}-\mathbf{r}^\prime|}}\rho_v(\mathbf{r}^\prime)\label{eq:svgeneral}
\end{equation}
We assume that the DW in each layer $i$ can be described by its wall angle $\psi_i$ and polar angle $(\theta)$ through $\theta_i(x) = \arctan\{\exp[\mp(x-q)/\Delta_i]\}$\footnote{Our approximation assumes that $\mathcal{T}<l_{\rm{ex}}$.}, where upper (lower) sign stands for $\downarrow|\uparrow$  $(\uparrow|\downarrow)$ DW state.  Micromagnetic simulations indicate that the wall angle $\psi_i$ also varies as a function of coordinate~\cite{Legrand2017a}, $\psi_i=\psi_i(x)$ (see Supplemental Material~\cite{Supp}). However, this effect occurs dominantly in the tails of the DW, and we therefore neglect it in our analytical model.  For the purpose of comparison between micromagnetics simulations and our analytical model, we fitted all the simulation data with this simplified DW profile, in which case the fitted $\psi_i$ are dominated by the region near the DW center.

 As shown in the Supplemental Material~\cite{Supp}, Eq.~\eqref{eq:svgeneral} for an infinitely extended film ($L_x$, $L_y\rightarrow \infty$) reduces to
\begin{equation}
\sigma_{d, sv}^{ 1, \mathcal{N}}=\mp  \frac{\pi f}{\mathcal{N}}\sum_{i = 0}^{\mathcal{N}-1} \sin(\psi_i) D_{sv, i} (\Delta_0, ...,\Delta_{\mathcal{N}-1})\label{eq:effectiveDMI}
\end{equation}
with $f=\mathcal{T}/\mathcal{P}$ being a scaling factor. A key result is that this expression follows the exact functional form of a (layer-dependent) interfacial DMI. That is, surface-volume stray fields manifest as a chiral magnetostatic interaction that promotes homochiral textures within each individual layer, even in the absence of DMI.  The twist develops as a consequence of the fact that $D_{sv, i}$ is an asymmetric function with respect to $i$, ranging from zero at the middle layer to its maximum magnitude at the top and the bottom layers (with the opposite signs, as depicted in Fig.~\ref{fig:2}b). Adding interfacial DMI simply offsets $D_{sv, i}$ by $D$ in every layer, which leads to a net shift of the Bloch layer away from the center. 

Since the $D_{sv, i}$ each depend on every $\Delta_j$, the coupled magnetostatic integrals in Eq.~\eqref{eq:effectiveDMI} involve $2\mathcal{N}$ independent variables, $\Delta_i$, $\psi_i$, leading to analytically intractable magnetostatic integrals (See Eq.~\eqref{eq:A1}).  However, micromagnetic simulations (Fig.~\ref{fig:1}~d) reveal that $\Delta_{\rm{max}}/\Delta_{\rm{min}}$ differs significantly from 1 only for relatively low $Q$.  We henceforth treat $\Delta$ as constant across the layers, which allows for analytical solutions for the $\psi_i$ to be obtained.  The total magnetostatic energy of the isolated DW (including volume-volume, surface-surface, and surface-volume components) can then be reduced to
 \begin{eqnarray}
 \sigma_{d}^{1, \mathcal{N}}  (\Delta, \psi_i)= \sum_{i = 0}^{\mathcal{N}-1} \sum_{j = 0}^{\mathcal{N}-1} \left\{ \sin(\psi_i)\sin(\psi_{j})  F_{v, ij} (\Delta)\notag \right.\\\left.+F_{s, ij} (\Delta)\pm \sin(\psi_i)\text{sgn}(i-j)F_{sv, ij} (\Delta)\right\}  \label{eq:exactstray}
 \end{eqnarray}
 with functions $F_{\alpha, ij}$ derived in the Supplemental Material~\cite{Supp} and defined analytically in Eq.~\eqref{eq:A2}. Here, we treat the layered structure explicitly rather than through the effective medium approximation~\cite{Suna1986a,Woo2016}, as we find that the intrinsic error of that approach affects the prediction accuracy of $\Delta,\psi_i$ (and more importantly, the sizes of domains and skyrmions~\cite{Lemesh2017}). The total micromagnetic energy $\sigma_{tot}^{ 1, \mathcal{N}}(\Delta, \psi_i)$ then reads
\begin{equation} \sigma_{tot}^{ 1, \mathcal{N}} =\sigma_{d}^{1, \mathcal{N}} + \frac{2A}{\Delta} f +2K_{u} \Delta f \mp  \frac{\pi D f}{\mathcal{N}}\sum_{i = 0}^{\mathcal{N}-1}  \sin(\psi_i).\label{eq:energy}
\end{equation}
The equilibrium profile  is obtained by setting $\frac{\partial \sigma_{tot}^{1, \mathcal{N}}}{\partial \Delta}=0$, $\frac{\partial \sigma_{tot}^{1, \mathcal{N}}}{\partial \psi_i}=0$ for $i=0,...\ \mathcal{N}-1$, which after introducing the matrix formalism (shown in the  Supplemental Material~\cite{Supp}) reduces to
\begin{eqnarray}
\frac{2A}{\Delta^2}f-2K_{u} f = \mp   \frac{\pi f}{\mathcal{N}}\sum_{i = 0}^{\mathcal{N}-1} \sin(\psi_i) \frac{\partial{D_{sv, i}} }{\partial \Delta}\notag \\  +\sum_{i=0}^{\mathcal{N}-1}\sum_{j=0}^{\mathcal{N}-1} \left\{\frac{\partial F_{s, ij}}{\partial \Delta} + \sin(\psi_i)\sin(\psi_j)\frac{\partial F_{v, ij}}{\partial \Delta}\right\} \label{eq:delta},\\
{\sin(\psi_{i})} =   \pm \tilde{f}\left( \frac{\pi  f}{\mathcal{N}}  \hat{\varkappa}_{v}^{-1} \cdot [ \vec{D}_{sv} +\vec{1} D ]\right)_i 
  \label{eq:psi},
\end{eqnarray}    
where we introduced a helper function $\tilde{f}(x)$ that becomes $x$, when $|x|\leq 1$ and  $\text{sgn}(x)$ otherwise, and defined the matrix $\hat{\varkappa}_v$ and vector  $\vec{D}_{sv}$ as:
\begin{eqnarray}
\varkappa_{v, ij}=\left(1+\delta_{ij}\right)   F_{v, ij}(\Delta)\label{eq:kappa}\\ D_{sv, i}=-\frac{\mathcal{N}}{\pi  f}\sum_{j=0}^{\mathcal{N}-1}  F_{sv, ij}(\Delta)\text{sgn}(i-j). \label{eq:Dsv}
\end{eqnarray} 

Equations~\eqref{eq:delta} and \eqref{eq:psi} constitute an implicit  relation for the equilibrium $\Delta$, which can be disentangled from $\sin(\psi_i)$ through separation of variables. The equilibrium $\psi_i$ can then be found by plugging the obtained $\Delta$ directly into Eq.~\eqref{eq:psi}.
The resulting analytical solutions of $\Delta$, $\psi_i$ for films with various $D$ are plotted in Figs.~\ref{fig:1}b,~c. We find that $\Delta$ in our model correctly predicts the average DW width $\sum_{0}^{\mathcal{N}-1}\Delta_i/\mathcal{N}$, and our constant $\Delta$ approximation permits quite accurate prediction of the layer-dependent $\psi_i$ (even when  $\Delta_{\rm{max}}/\Delta_{\rm{min}}\sim 4$  as shown for a film with $Q=1.01$, $f=1/6$ in Supplemental Fig.~1~\cite{Supp}).

We find that the values of $D_{sv}$ in typical multilayers are comparable to values of interfacial DMI found experimentally, as shown in Fig.~\ref{fig:2}b, where energies on the order of $\SI{1}{mJ/m^2}$ are seen. Its magnitude increases with increasing $f$ and decreasing $Q$. Figure~\ref{fig:2}c shows that as a result, much larger values of DMI are required to saturate domain walls in a purely N\'eel state than would be expected from a 2D treatment. There, we analyze multilayers with various $Q$ and $f$ and plot the fraction of layers with right-handed N\'eel walls (here, $\sin(\psi_i)=+1$) as a function of DMI and $\mathcal{N}$. We find that films with the smallest $f$ and Q are easier to saturate to the complete N\'eel state.  We also find that the threshold $D_{\rm{thr.}}^{2D}$, at which the wall in every layer becomes completely N\'eel in the 2D model, \cite{Lemesh2017,Rohart2013a} applied to multilayers using an effective medium approach (Eq.~\eqref{eq:A8}), (dashed curve in Fig.~\ref{fig:2}c) significantly underestimates the actual threshold. In the Supplemental Material~\cite{Supp} we derive a more precise numerical relation for $D_{\rm{thr.}}^{3D}$  (Eqs. ~\eqref{eq:A6}, ~\eqref{eq:A7}), plotted as continuous curves in Fig.~\ref{fig:2}c. Notably, we find that the critical DMI strength required to ensure uniform N\'eel character is more than a factor of 2 greater than would be estimated from a 2D treatment. We note that an analytical treatment to determine the threshold for the onset of a twist was also presented recently in Ref.~\citenum{Legrand2017a}. 

One can see from the form of Eq.~\eqref{eq:psi} that volume-volume stray fields, accounted for by $\hat{\varkappa}_{v}^{-1}$,  also influence the layer dependent $\psi_i$. However, if $D_{sv}$ is neglected, the volume-volume interactions alone would predict a twist only in the case of nonzero DMI and that twist would be symmetric, since the matrix  $\hat{\varkappa}_{v}^{-1}$  is centrosymmetric (see Supplemental Information~\cite{Supp}). It is, in fact, the surface-volume stray fields that lead to the experimentally observed asymmetric twist~~\cite{Dovzhenko2016,Montoya2017,Legrand2017a}, since the vector $D_{sv}$ in Eq.~\eqref{eq:psi} is antisymmetric. 

\section{Domain size}
We now consider a multidomain state with twisted DWs, with domain period $\lambda$ and minority domain width $W$. One can anticipate that shifting  from the 2D model to the 3D model should result in first-order corrections to the intra- and interwall energetics of the system, which would lead to more accurate predictions of $W$~\cite{Lemesh2017, Woo2016}.  To evaluate the impact of this effect, we first identified the ground state for multilayer films with low DMI using micromagnetic simulations with various densities of stripes. After performing a relaxation procedure~\cite{Lemesh2017}, we find that the state with minimum total energy is the one in which the intralayer DW chirality is conserved. This effect is also induced by the surface-volume stray field interactions as depicted in Fig.~\ref{fig:2}a. 
\begin{figure*}
	\includegraphics[scale = 0.73]{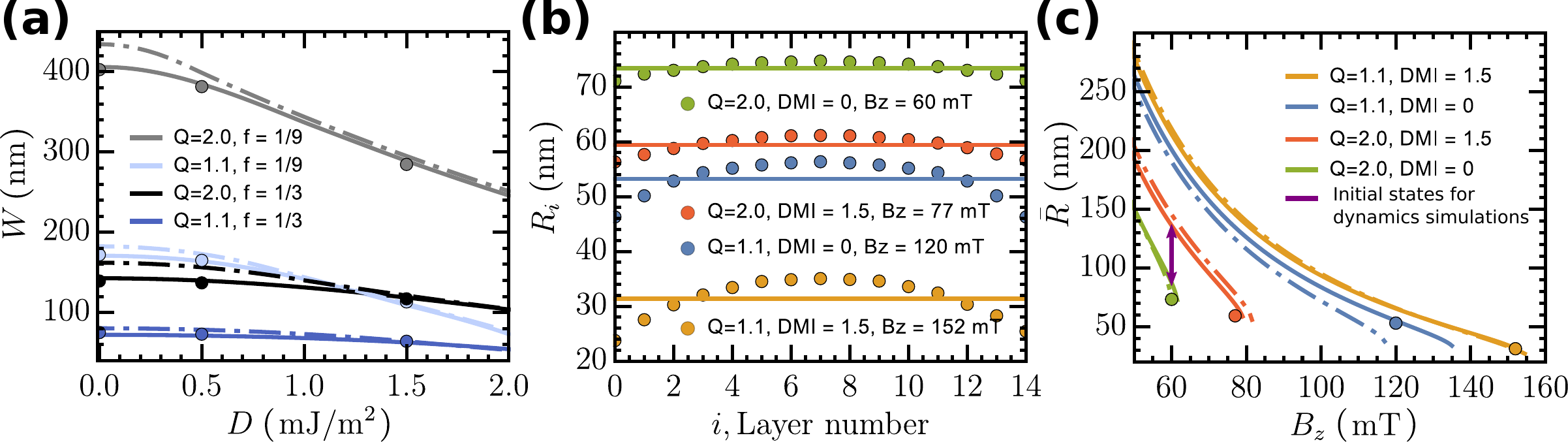}
	\caption{{\bf Magnetic domains and skyrmions with twisted walls.}  {\bf(a)}   Equilibrium domain width as a function of interfacial DMI, $Q$ and $f$.  {\bf(b)} $R_i$ as a function of the layer number. {\bf(c)} Average skyrmion radius $\sum_{0}^{\mathcal{N}-1}R_i/\mathcal{N}$ as a function of applied field, $Q$ and $D$ for films with $f = 1/6$. DMI constants $D$ are in units of $\SI{}{mJ/m^2}$. Solid (dashed) lines represent the numerical solution for 3D (2D~\cite{Lemesh2017,Buttner2017}) theory, dots represent multilayer simulations, with explicit spacer layers.}\label{fig:3}
\end{figure*}

Based on this ground state, we derive in the Supplemental Material~\cite{Supp} the exact magnetostatic energy of the magnetized multidomain phase with a wall twist, $\sigma_{d}^{\infty, \mathcal{N}}(\lambda, W, \Delta, \psi_i)$  (Eq.~\eqref{eq:B1}). We then derive expressions for the equilibrium domain parameters by minimizing the total energy $\mathcal{E}_{tot}^{ \infty, \mathcal{N}}$ with respect to $\lambda, W, \Delta, \psi_i$ (see Eqs.~\eqref{eq:B6}-Eq.~\eqref{eq:B10}). In Fig.~\ref{fig:3}a we plot $W$ as a function of $D$ for the demagnetized state ($\lambda = 2W$). We find that the full 3D treatment closely matches the 2D theory~\cite{Lemesh2017}. The largest deviation occurs for films with high $Q$ and weak DMI and is caused by two effects: (i) surface-volume interactions, which are inherently ignored in the effective medium approach, and (ii) the intrinsic error of the effective medium approach~\cite{Lemesh2017}, both of which have a comparable first-order effect on $W$. Note that the slope of the $W=W(D)$ curve approaches zero in the region of small DMI, which means that using domain width measurements for the extraction of small values of DMI is impractical.

\section{Twisted skyrmions}
We next treat isolated skyrmions analytically using the wall-energy model~\cite{Cape1971}, incorporating the twisted DW energy density derived above.  Micromagnetic simulations reveal a layer-dependent radius $R_{i}$, which we plot in Fig.~\ref{fig:3}b, for the case $f = 1/6$ with several values of $Q$ and $D$ (with fields $B_z$ applied to yield similar radii). The skyrmion radius reaches a minimum at the top and the bottom layers, and a maximum closer to the middle layer. This effect, similarly to the DW twist, is also caused by stray field interactions. 

Since the interlayer variation of $R_{i}$ and $\Delta_i$ is difficult to evaluate analytically, we approximate them as constant through the thickness (equal to $R$ and $\Delta$, respectively).
Assuming that the DW energy is independent of $R$ (valid for skyrmions with $R> \mathcal{O}(\Delta)$), we can express its total energy $E_{\rm{tot}}^{\text{sk}, \mathcal{N}}  (R, \Delta, \psi_i, B_z) $ analogously to the 2D expression derived in Ref~\citenum{Buttner2017}, where the 3D twist is incorporated in the DW energy term:
  \begin{equation}
  E_{\rm{tot}}^{\text{sk}, \mathcal{N}}  =2\pi d R \sigma_{tot}^{ 1, \mathcal{N}}+aR-bR\ln\left(R/d\right)+cB_z R^2,\label{eq:skyrmionenergy}
  \end{equation}
where $\sigma_{tot}^{ 1, \mathcal{N}}(\Delta, \psi_i)$ is taken from Eq.~\eqref{eq:energy}, and constants are defined in Eqs.~\eqref{eq:C1}-\eqref{eq:C4}. The equilibrium $R$ can be determined by simply plugging the equilibrium parameters $\Delta$, $\psi_i$ found from the straight DW theory (Eqs.~\eqref{eq:delta},~\eqref{eq:psi}) into Eq.~\eqref{eq:skyrmionenergy} and minimizing the resulting expression with respect to $R$. Note that skyrmions with topological charge $N=1$ ($N=-1$) correspond to the lower (upper) sign in Eqs.~\eqref{eq:psi}.  We find that $R$ predicted by our 
analytical theory is very close to the average $R$ obtained from the explicit multilayer simulations (Figs.~\ref{fig:3}~b,~c). For comparison, the prediction of the 2D model derived in Ref.~\citenum{Buttner2017} (Fig.~\ref{fig:3}c applied by treating the multilayer using effective medium scaling) is seen to be quantitatively inaccurate due to the intrinsic error of the effective medium approach~\cite{Lemesh2017,Woo2016}) and the ignored surface-volume stray field interactions.

Finally, we examine current-induced dynamics of twisted skyrmions analytically and through micromagnetic simulations. For simplicity we consider only damping-like spin-orbit torque (SOT).  Treating the skyrmion as a rigid texture whose static configuration is preserved while moving, we use the Thiele equation~\cite{Thiele1973} to derive analytical expressions for the steady state skyrmion velocity $v$ and Hall angle $\xi'$, similarly to the approach in Ref.~\cite{Buttner2017}.  By summing up the forces acting on each individual skyrmion in the multilayer, we arrive at (see Supplemental Material~\cite{Supp} and Eqs.~\eqref{eq:C5}-\eqref{eq:C8}):
\begin{eqnarray}
|v|  = j \frac{\pi \hbar \gamma \Delta  \theta_{\text{SH}} I_{D}(\rho)}{2 e M_s \mathcal{T}\sqrt{\tilde{G}^2+\tilde{D}^2\alpha^2}} \tilde{f}\label{eq:v}\\
\xi'  = \rm{atan}2(\tilde{G},\tilde{D}\alpha) -(\tilde{\psi}-\pi/2)+\pi\Theta(\theta_{\rm{SH}}N).\label{eq:xi}
\end{eqnarray}
 The constants $\tilde{f}$, $\tilde{\psi} $ capture the influence of the DW twist:
\begin{eqnarray}
\tilde{f}  = \frac{1}{\mathcal{N}}\sqrt{\left(\sum_{i=0}^{\mathcal{N}-1} \cos(\psi_i)\right)^2+\left(\sum_{i=0}^{\mathcal{N}-1} \sin(\psi_i)\right)^2} \\
\tilde{\psi}  = \rm{atan2} \left(\sum_{i=0}^{\mathcal{N}-1} \sin(\psi_i), \sum_{i=0}^{\mathcal{N}-1} \cos(\psi_i)\right)
\end{eqnarray}
For the 2D model, these constants become $\tilde{f} = 1$ and $\tilde{\psi}=\psi_{\rm{2D}}$~\cite{Buttner2017}. Hence, even if the 2D model could predict the equilibrium $R$ exactly, its predictions of skyrmion dynamics would still deviate from our multilayer treatment as $v_{\rm{3D}}/v_{\rm{2D}} = \tilde{f}$ and $\xi'_{\rm{3D}}-\xi'_{\rm{2D}} \equiv  \tilde{\psi}-\psi_{\rm{2D}}$. 
   \begin{figure}
   	\includegraphics[scale = 0.55]{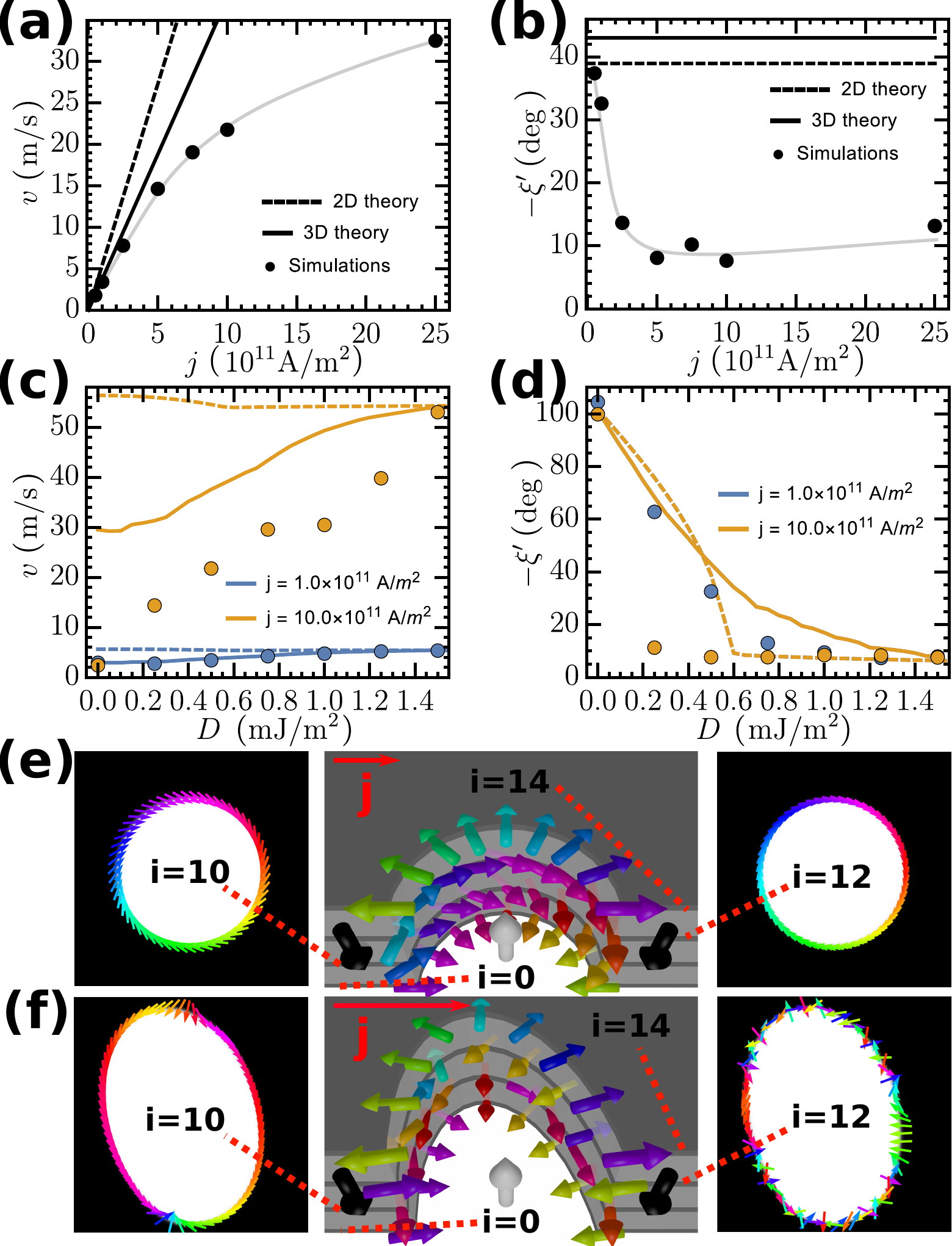}
   	\caption{{\bf Dynamics of skyrmions with twisted walls }  {\bf(a)} Skyrmion velocity $v$ and {\bf(b)} skyrmion hall angle $\xi'$ as a function of current density $j$ for films with  $D=0.5~\rm{mJ/m}^2$ with gray continuous lines representing guides to the eye. {\bf(c)} $v$ and {\bf(d)} $\xi'$ as a function of DMI. Continuous (dotted) lines depict 3D (2D~\cite{Buttner2017}) model. {\bf(e)}, {\bf(f)} 3D cuts of multilayer skyrmions at $j=\SI{1.0e11}{A/m^2}$ and $j=\SI{1.0e12}{A/m^2}$ for films with $D=0.5~\rm{mJ/m}^2$ (at $t=9~\rm{ns}$). Simulation parameters are $Q=2.0$, $f = 1/6$ (i.e. $D_{\rm{thr}}^{3D}=1.47~\rm{mJ/m}^2$), $\theta_{\rm{SH}}=0.1$, $\alpha=0.3$. }\label{fig:4}
   \end{figure}

Figures~\ref{fig:4}a,~b compare the values of $v$, $\xi'$ predicted by these two theories, with the ones extracted from the explicit multilayer simulations for films with $f = 1/6$, $Q = 2.0$, $B_z=59~\rm{mT}$, $D=0.5~\rm{mJ/m}^2$. Both theories provide a reasonable estimate of the skyrmion Hall angle, however the velocity predictions in our 3D model are in much better agreement with the explicit multilayer simulations than are those of the 2D model~\cite{Buttner2017}, especially in the low current regime.  The low-$j$ deviations of $\xi'$ in the 3D model are attributed to the slight underestimation of the $\psi_i$  predicted by our model. 

Micromagnetic simulations show that for small $j$, the skyrmion profile preserves its static configuration (Fig.~\ref{fig:4}e). By contrast, at higher $j$, for some layers, $\psi_i$ becomes non-uniform across the perimeter of the skyrmion, which leads to a reduced net force acting on the skyrmion tube.  We generally find that the closer the static configuration in a layer is to being Bloch, the higher the likelihood that at high $j$ the skyrmion in that layer accumulates pairs of Bloch lines (as depicted in Fig.~\ref{fig:4}f) and exhibits nonuniform precession and oscillations during current injection (as demonstrated for $D=0.5~\rm{mJ/m}^2$ in the Supplemental Videos~\cite{Supp}). Both velocity and skyrmion hall angle, particularly at high currents, are many times smaller than they would be in the absence of these factors, i.e for the 2D model, or even for our (rigid) twisted wall-energy model (Figs.~\ref{fig:4}a,~b). These high-$j$ phenomena affect the resulting dynamics of multilayer skyrmions, especially at low DMI. There, only a fraction of skyrmions contribute to the net force, since skyrmions in the upper and lower layers have opposite chiralities so that the forces tend to cancel. What is left are the transient and Bloch skyrmions that contribute only weakly due to the development of Bloch lines or wall angle oscillations~\cite{Malozemoff1979}, leading to significantly lower velocities. Such defects or oscillations are absent in layers with N\'eel walls, which is why our high-DMI predictions of $v$ are always accurate (Figs.~\ref{fig:4}c,~d). Finally, we find that high currents also lead to distorted skyrmions shapes, as well as to their slight magnetostatic decoupling along the film. Such high-SOT effects may also contribute to the observed deviations of our 3D dynamics model. 

\section{Summary}
We have explicitly demonstrated that DWs and skyrmions in magnetic multilayers generally form a twisted structure with varying  $\psi_i$ $\Delta_i$ and $R_i$ due to the mutual surface-volume stray field interactions. We have calculated the wall twist analytically, assuming a varying $\psi_i$, but a fixed $\Delta_i=\Delta$ across the layers. We have found that 2D treatments, in addition to completely ignoring the wall twist, yield quantitative errors in domain spacing and isolated skyrmion sizes, though in most cases the error is relatively modest.  However, these twisted states, and the variation strength of stabilization of DW angle through the thickness, leads to markedly different dynamics from what 2D treatments would predict.  We derived analytical expressions for skyrmion velocity and Hall angle accounting for the twisted states, which works well at low current but fails at higher currents due to complex dynamical changes in the spin textures that cannot be captured by rigid models. Our work provides key insights into the novel static and dynamic layer-dependent phenomena in PMA multilayers. 
 
\begin{acknowledgments}
We thank Dr. Felix B\"uttner for optimizing the efficiency of the numerical script, providing the code templates to plot the figures and to extract the skyrmion parameters. This work was supported by the U.S. Department of Energy (DOE), Office of Science, Basic Energy Sciences (BES) under Award \#DE-SC0012371 (development of domain wall twist model) and by the DARPA TEE program (application to magnetic skyrmion statics and dynamics).
\end{acknowledgments}

\onecolumngrid
\appendix

\section{Twisted straight domain wall}\label{sec:isolated}
Films with layer dependent $\Delta_i, \psi_i$ develop an effective DMI, which stems from the surface-volume stray fields, and looks as follows (see Eq.~S67 in the Supplemental Information~\cite{Supp})
\begin{align}
D_{sv, i}&  = -\frac{2\mu_{0}M_s^2 \Delta_i}{\mathcal{T}}  \sum_{j=0}^{\mathcal{N}-1} \Delta_j \text{sgn}(i-j)\int_{0}^{\infty}dk\frac{1}{k}\frac{  e^{-k|(i-j)P+\mathcal{T}|}+e^{-k|(i-j)P-\mathcal{T}|}-2 e^{-k\mathcal{P}|i-j|} }{4\sinh\left(\frac{\pi \Delta_j k}{2}\right)\cosh\left(\frac{\pi \Delta_i k}{2}\right)},\label{eq:A1}
\end{align}
which after a constant $\Delta$ assumption reduces to Eq.~\eqref{eq:Dsv}. The generic function $F_{\alpha, ij}$ used in the expression for the total magnetostatic energy of the isolated domain wall $\sigma_{d}^{1, \mathcal{N}}$ (Eq.~\eqref{eq:exactstray}) is derived in Supplemental Information~\cite{Supp} (Eqs.~S35,~S52,~S69). It can be summarized as 
\begin{align}
F_{\alpha, ij} (\mathcal{T},\mathcal{P}, \Delta)=\frac{\pi \mu_{0}M_{s}^2 \Delta^2}{\mathcal{N}\mathcal{P}}\left[ G_{\alpha}\left(\frac{\left|(i-j)\mathcal{P}+\mathcal{T}\right|}{2\pi \Delta}\right)+ G_{\alpha}\left(\frac{\left|(i-j)\mathcal{P}-\mathcal{T}\right|}{2\pi \Delta}\right)  -2G_{\alpha}\left(\frac{\left|(i-j)\mathcal{P}\right|}{2\pi \Delta}\right)\right]\label{eq:A2}
\end{align}
 with functions $G_{\alpha}(x)$ defined analytically as follows (for $G_v(x)$):
\begin{align}
G_v(x)&= -2\left\{\Psi^{-2}(x+1)-\Psi^{-2}\left(x+\frac{1}{2}\right)- x \ \ln(\Gamma(x+1)) 
+x \ln\left[\Gamma\left(x+\frac{1}{2}\right)\right] \right. \notag\\ &\left.- \Psi^{-2}(1)+\Psi^{-2}\left(\frac{1}{2}\right) 
\right\}\label{eq:A3}\\
G_s(x) &=-\left\{\Psi ^{(-2)}(2 x)+x^2 (2  \log (x)+\log (4)-1)-x(1+2  \ln[\Gamma (2 x)])\right\}\label{eq:A4}\\
G_{sv}(x) & = 2 \ln\left[\Gamma\left(x+\frac{1}{2}\right)\right]\label{eq:A5},
\end{align}
where the volume-volume stray field component $G_v(x)$ has been originally derived for homochiral multilayers in Ref.~\citenum{buttner2015}. The value of DMI at which all the layers are saturated to the homochiral N\'eel state ($ D_{\text{thr}}$~\cite{Lemesh2017}) can be derived from the following equations (with  $\Delta_{\text{thr}}$ and $ D_{\text{thr}}^{3D}$ being the unknown variables).
\begin{align}
2K_{u} f-\frac{2A}{\Delta_{\text{thr}}^2}f+ \sum_{i=0}^{\mathcal{N}-1}\sum_{j=0}^{\mathcal{N}-1} \left[\frac{\partial F_{s, ij}}{\partial \Delta_{\text{thr}}} +\frac{\partial F_{v, ij}}{\partial \Delta_{\text{thr}}}  \pm   \text{sgn}(i-j)\frac{\partial F_{sv, ij}}{\partial \Delta_{\text{thr}}}\right] =0,\label{eq:A6}
\end{align}    
\begin{align}
\left(\sum_{j=0}^{\mathcal{N}-1} [\left(1+\delta_{ij}\right)   F_{v, ij}(\Delta_{\text{thr}})]^{-1}\cdot\left[  \frac{\pi D_{\text{thr}}^{3D}  f}{\mathcal{N}}   1_j  - \sum_{k=0}^{\mathcal{N}-1} F_{sv, jk}(\Delta_{\text{thr}})\text{sgn}(j-k) \right]\right)_{\mathcal{N}-1}-1=0 \label{eq:A7},
\end{align}
where the sign ``$^{-1}$'' represents the matrix inversion operation. This value can be compared with the value given by the 2D-model~\cite{Lemesh2017}, extending it to multilayers via the effective medium approach~\cite{Suna1986a,Woo2016,Lemesh2017}:
\begin{equation}
D_{\text{thr}}^{2D} = \frac{2\mu_{0}M^2_s f}{\frac{\pi^2}{\mathcal{P}\mathcal{N} \ln(2)}+\pi \sqrt{\frac{K_u-\frac{\mu_0M_s^2}{2}+\mu_0M_s^2f}{A}}}. \label{eq:A8}
\end{equation}

\section{Domain size}
The total magnetostatic energy of magnetized multidomain multilayers is derived in Supplemental Information~\cite{Supp} (Eqs. S101, S118, S130) and can be expressed as 
\begin{align}
\sigma_{d}^{\infty, \mathcal{N}}&=  \frac{\lambda}{4}\mu_0M_s^2 \left(\frac{2W}{\lambda}-1\right)^2\frac{\mathcal{T}}{\mathcal{P}}+ \sum_{i=0}^{\mathcal{N}-1}\sum_{j=0}^{\mathcal{N}-1} \tilde{F}_{s, ij} (\mathcal{T},\mathcal{P}, \Delta, \lambda, W)\notag\\&+ \sum_{i=0}^{\mathcal{N}-1}\sum_{j=0}^{\mathcal{N}-1} \left\{ \sin(\psi_i)\sin(\psi_j)\tilde{F}_{v, ij} (\mathcal{T},\mathcal{P}, \Delta, \lambda, W)+ \sin(\psi_i)\text{sgn}(i-j)\tilde{F}_{sv, ij} (\mathcal{T},\mathcal{P}, \Delta, \lambda, W)\right\}\label{eq:B1}
\end{align}
with a generic  function $\tilde{F}_{\alpha, ij} $ and its dependencies  defined as follows
\begin{align}
\tilde{F}_{\alpha, ij} & = \frac{\pi \mu_{0}M_s^2\Delta^2}{\mathcal{N}\mathcal{P} }\sum_{n=1}^{\infty}\frac{  \sin^2\left(\frac{\pi n W}{\lambda}\right)}{n}\tilde{G}_{\alpha, ijn}(\mathcal{T},\mathcal{P}, \Delta, \lambda),\label{eq:B2}\\
\tilde{G}_{v, ijn} & = \frac{2\sinh^2(\frac{\pi n \mathcal{T}}{\lambda})e^{-\frac{2\pi n \mathcal{P}|i-j|}{\lambda}}(1-\delta_{i-j, 0})+(e^{-\frac{2\pi n \mathcal{T}}{\lambda}}+\frac{2\pi n \mathcal{T}}{\lambda}-1)\delta_{ij}}{\cosh^2\left(\frac{\pi^2 n \Delta }{\lambda}\right)}\label{eq:B3}\\
\tilde{G}_{s, ijn} & = \frac{2e^{-\frac{2\pi |(i-j)\mathcal{P}|n}{\lambda}}-e^{-\frac{2\pi|\mathcal{T}-(i-j)\mathcal{P}|n}{\lambda}}-e^{-\frac{2\pi |\mathcal{T}+(i-j)\mathcal{P}|n}{\lambda}}}{ 2\sinh^2\left(\frac{\pi^2 n \Delta}{\lambda}\right)}\label{eq:B4}\\
\tilde{G}_{sv, ijn} & = \frac{8\sinh^2(\frac{\pi n \mathcal{T}}{\lambda})e^{-\frac{2\pi n\mathcal{P}|i-j|}{\lambda}} }{\sinh\left(\frac{2\pi^2 n \Delta}{\lambda}\right)}\label{eq:B5}
\end{align}

Assuming the magnetic field applied in z direction (in the absence of currents), the total volumetric energy per single domain wall per layer therefore can be expressed as:
\begin{align}\mathcal{E}_{tot}^{ \infty, \mathcal{N}}(\lambda, W, \Delta, \psi_i)=\frac{2}{\lambda}\left[     \frac{2A}{\Delta}f+2K_{u}\Delta f -M_s\left(1-\frac{2W}{\lambda}\right)B_z\frac{f\lambda}{2}+\sigma_{d}^{\infty, \mathcal{N}}(\lambda, W, \Delta, \psi_i)-\frac{\pi D  f}{\mathcal{N}}  \sum_{i=0}^{\mathcal{N}-1}\sin(\psi_i)\label{eq:B6} \right]
\end{align}

By performing the energy minimization (as shown in Supplemental Information~\cite{Supp}), we will have the system of four equations that define the equilibrium $\lambda, W, \Delta, \psi$:
\begin{align}
\sum_{i=0}^{\mathcal{N}-1}\sum_{j=0}^{\mathcal{N}-1} \left\{\left[\tilde{F}_{s, ij}-\lambda\frac{\partial \tilde{F}_{s, ij}}{\partial \lambda} \right]+\sin(\psi_{i})  \sin(\psi_{j}) \left[\tilde{F}_{v, ij}-\lambda\frac{\partial \tilde{F}_{v, ij}}{\partial \lambda} \right]  + \sin(\psi_{i})  \text{sgn}(i-j)\left[\tilde{F}_{sv, ij} -\lambda\frac{\partial \tilde{F}_{sv, ij}}{\partial \lambda} \right]\right\}\notag\\
+\left[\frac{2A}{\Delta}f+2K_{u}\Delta f- \frac{\pi D  f}{\mathcal{N}}  \sum_{i=0}^{\mathcal{N}-1}\sin(\psi_i)+WM_sB_zf+\mu_0M_s^2Wf\left(\frac{2W}{\lambda}-1\right) \right]=0\label{eq:B7}
\end{align}     
\begin{align}
M_s f \left[B_z+\mu_0M_s\left(\frac{2W}{\lambda}-1\right)\right]+\sum_{i=0}^{\mathcal{N}-1}\sum_{j=0}^{\mathcal{N}-1} \left\{\frac{\partial \tilde{F}_{s, ij}}{\partial W} +\sin(\psi_{i})  \sin(\psi_{j})  \frac{\partial \tilde{F}_{v, ij}}{\partial W}   + \sin(\psi_{i})  \text{sgn}(i-j)\frac{\partial \tilde{F}_{sv, ij}}{\partial W} \right\}=0 \label{eq:B8}
\end{align}  
\begin{align}
-\frac{2A}{\Delta^2}f+2K_{u} f+ \sum_{i=0}^{\mathcal{N}-1}\sum_{j=0}^{\mathcal{N}-1} \frac{\partial \tilde{F}_{s, ij}}{\partial \Delta} +\sum_{i=0}^{\mathcal{N}-1}\sum_{j=0}^{\mathcal{N}-1}\sin(\psi_{i})  \sin(\psi_{j})  \frac{\partial \tilde{F}_{v, ij}}{\partial \Delta}   + \sum_{i=0}^{\mathcal{N}-1}\sum_{j=0}^{\mathcal{N}-1}  \sin(\psi_{i})  \text{sgn}(i-j)\frac{\partial \tilde{F}_{sv, ij}}{\partial \Delta} =0\label{eq:B9}
\end{align}     
\begin{align}
\sin(\psi_{i}) =   \tilde{f}\left(\sum_{j=0}^{\mathcal{N}-1} [\left(1+\delta_{ij}\right)   \tilde{F}_{v, ij}(\Delta, W)]^{-1}\left[   \frac{\pi D  f}{\mathcal{N}}   1_j -\sum_{k=0}^{\mathcal{N}-1} \tilde{F}_{sv, jk}(\Delta, W)\text{sgn}(j-k) \right]\right)\label{eq:B10},
\end{align} 	 
where the sign ``$^{-1}$'' represents the matrix inversion operation.

\section{Twisted skyrmions}
For the skyrmion statics expressions (Eq.~\eqref{eq:skyrmionenergy} and Eq.~(S150)), we have used the constants defined in Supplemental Information~\cite{Supp} and Ref.~\citenum{Buttner2017} as:
\begin{align}
a & = -\mu_0 M_s^2 (\mathcal{P}\mathcal{N})^2[6\ln(2)-1]\label{eq:C1}\\
b& = 2\mu_0 M_s^2(\mathcal{P}\mathcal{N})^2\label{eq:C2}\\
c& = -2\pi \mathcal{P}\mathcal{N} M_s\label{eq:C3}\\
d& = \mathcal{P}\mathcal{N}\label{eq:C4}
\end{align}
Similarly, the constants for the skyrmion dynamics (Eqs.~\eqref{eq:v},~\eqref{eq:xi}) are~\cite{Buttner2017} 
\begin{align}
\tilde{G} &= -4\pi N\label{eq:C5}\\
\tilde{D} &= \pi I_{A}(R/\Delta)\label{eq:C6}\\
I_{\rm{A}} (\rho) &= 2\rho+\frac{2}{\rho}+1.93(\rho-0.65)\exp[-1.48(\rho-0.65)]\label{eq:C7}\\
I_{\rm{D}} (\rho) &= \pi \rho+\frac{1}{2}\exp(-\rho)\label{eq:C8}
\end{align}

\section{Methods}
For simulating magnetic textures (isolated domain walls, perpendicular stripes, and skyrmions), the micromagnetic MuMax3 solver~\cite{Vansteenkiste2014} was used with the magnetic parameters given in the manuscript. The cell size is $1~\rm{nm}\times 1~\rm{nm}\times1~\rm{nm}$  and the simulation size is $1~\mu m\times 1~\mu m\times \mathcal{N}\mathcal{P}$. For skyrmion dynamics simulations, Zhang-Li torque has been disabled, and the modified Slonczewski-like torque module has been used (with the enabled damping-like torque and disabled field-like torque). Spin hall angle is $\Theta_{\rm{SH}} = 0.1$, damping constant $\alpha=0.3$, fixed layer polarization is along -y direction.

\twocolumngrid


\begin{thebibliography}{34}%
	\makeatletter
	\providecommand \@ifxundefined [1]{%
		\@ifx{#1\undefined}
	}%
	\providecommand \@ifnum [1]{%
		\ifnum #1\expandafter \@firstoftwo
		\else \expandafter \@secondoftwo
		\fi
	}%
	\providecommand \@ifx [1]{%
		\ifx #1\expandafter \@firstoftwo
		\else \expandafter \@secondoftwo
		\fi
	}%
	\providecommand \natexlab [1]{#1}%
	\providecommand \enquote  [1]{``#1''}%
	\providecommand \bibnamefont  [1]{#1}%
	\providecommand \bibfnamefont [1]{#1}%
	\providecommand \citenamefont [1]{#1}%
	\providecommand \href@noop [0]{\@secondoftwo}%
	\providecommand \href [0]{\begingroup \@sanitize@url \@href}%
	\providecommand \@href[1]{\@@startlink{#1}\@@href}%
	\providecommand \@@href[1]{\endgroup#1\@@endlink}%
	\providecommand \@sanitize@url [0]{\catcode `\\12\catcode `\$12\catcode
		`\&12\catcode `\#12\catcode `\^12\catcode `\_12\catcode `\%12\relax}%
	\providecommand \@@startlink[1]{}%
	\providecommand \@@endlink[0]{}%
	\providecommand \url  [0]{\begingroup\@sanitize@url \@url }%
	\providecommand \@url [1]{\endgroup\@href {#1}{\urlprefix }}%
	\providecommand \urlprefix  [0]{URL }%
	\providecommand \Eprint [0]{\href }%
	\providecommand \doibase [0]{http://dx.doi.org/}%
	\providecommand \selectlanguage [0]{\@gobble}%
	\providecommand \bibinfo  [0]{\@secondoftwo}%
	\providecommand \bibfield  [0]{\@secondoftwo}%
	\providecommand \translation [1]{[#1]}%
	\providecommand \BibitemOpen [0]{}%
	\providecommand \bibitemStop [0]{}%
	\providecommand \bibitemNoStop [0]{.\EOS\space}%
	\providecommand \EOS [0]{\spacefactor3000\relax}%
	\providecommand \BibitemShut  [1]{\csname bibitem#1\endcsname}%
	\let\auto@bib@innerbib\@empty
	\bibitem [{\citenamefont {Heide}\ \emph {et~al.}(2008)\citenamefont {Heide},
		\citenamefont {Bihlmayer},\ and\ \citenamefont {Bl{\"{u}}gel}}]{Heide2008a}%
	\BibitemOpen
	\bibfield  {author} {\bibinfo {author} {\bibfnamefont {M.}~\bibnamefont
			{Heide}}, \bibinfo {author} {\bibfnamefont {G.}~\bibnamefont {Bihlmayer}}, \
		and\ \bibinfo {author} {\bibfnamefont {S.}~\bibnamefont {Bl{\"{u}}gel}},\
	}\href {\doibase 10.1103/PhysRevB.78.140403} {\bibfield  {journal} {\bibinfo
		{journal} {Physical Review B}\ }\textbf {\bibinfo {volume} {78}},\ \bibinfo
	{pages} {140403} (\bibinfo {year} {2008})}\BibitemShut {NoStop}%
\bibitem [{\citenamefont {Thiaville}\ \emph {et~al.}(2012)\citenamefont
	{Thiaville}, \citenamefont {Rohart}, \citenamefont {Ju{\'{e}}}, \citenamefont
	{Cros},\ and\ \citenamefont {Fert}}]{Thiaville2012}%
\BibitemOpen
\bibfield  {author} {\bibinfo {author} {\bibfnamefont {A.}~\bibnamefont
		{Thiaville}}, \bibinfo {author} {\bibfnamefont {S.}~\bibnamefont {Rohart}},
	\bibinfo {author} {\bibfnamefont {{\'{E}}.}~\bibnamefont {Ju{\'{e}}}},
	\bibinfo {author} {\bibfnamefont {V.}~\bibnamefont {Cros}}, \ and\ \bibinfo
	{author} {\bibfnamefont {A.}~\bibnamefont {Fert}},\ }\href {\doibase
	10.1209/0295-5075/100/57002} {\bibfield  {journal} {\bibinfo  {journal}
		{EPL}\ }\textbf {\bibinfo {volume} {100}},\ \bibinfo {pages} {57002}
	(\bibinfo {year} {2012})}\BibitemShut {NoStop}%
\bibitem [{\citenamefont {Emori}\ \emph {et~al.}(2013)\citenamefont {Emori},
	\citenamefont {Bauer}, \citenamefont {Ahn}, \citenamefont {Martinez},\ and\
	\citenamefont {Beach}}]{Emori2013f}%
\BibitemOpen
\bibfield  {author} {\bibinfo {author} {\bibfnamefont {S.}~\bibnamefont
		{Emori}}, \bibinfo {author} {\bibfnamefont {U.}~\bibnamefont {Bauer}},
	\bibinfo {author} {\bibfnamefont {S.~M.}\ \bibnamefont {Ahn}}, \bibinfo
	{author} {\bibfnamefont {E.}~\bibnamefont {Martinez}}, \ and\ \bibinfo
	{author} {\bibfnamefont {G.~S.~D.}\ \bibnamefont {Beach}},\ }\href {\doibase
	10.1038/nmat3675} {\bibfield  {journal} {\bibinfo  {journal} {Nature
			Materials}\ }\textbf {\bibinfo {volume} {12}},\ \bibinfo {pages} {611}
	(\bibinfo {year} {2013})}\BibitemShut {NoStop}%
\bibitem [{\citenamefont {R{\"{o}}{\ss}ler}\ \emph {et~al.}(2006)\citenamefont
	{R{\"{o}}{\ss}ler}, \citenamefont {Bogdanov},\ and\ \citenamefont
	{Pfleiderer}}]{Rossler2006}%
\BibitemOpen
\bibfield  {author} {\bibinfo {author} {\bibfnamefont {U.~K.}\ \bibnamefont
		{R{\"{o}}{\ss}ler}}, \bibinfo {author} {\bibfnamefont {A.~N.}\ \bibnamefont
		{Bogdanov}}, \ and\ \bibinfo {author} {\bibfnamefont {C.}~\bibnamefont
		{Pfleiderer}},\ }\href {\doibase 10.1038/nature05056} {\bibfield  {journal}
	{\bibinfo  {journal} {Nature}\ }\textbf {\bibinfo {volume} {442}},\ \bibinfo
	{pages} {797} (\bibinfo {year} {2006})}\BibitemShut {NoStop}%
\bibitem [{\citenamefont {M{\"{u}}hlbauer}\ \emph {et~al.}(2009)\citenamefont
	{M{\"{u}}hlbauer}, \citenamefont {Binz}, \citenamefont {Jonietz},
	\citenamefont {Pfleiderer}, \citenamefont {Rosch}, \citenamefont {Neubauer},
	\citenamefont {Georgii},\ and\ \citenamefont {B{\"{o}}ni}}]{Muhlbauer2009}%
\BibitemOpen
\bibfield  {author} {\bibinfo {author} {\bibfnamefont {S.}~\bibnamefont
		{M{\"{u}}hlbauer}}, \bibinfo {author} {\bibfnamefont {B.}~\bibnamefont
		{Binz}}, \bibinfo {author} {\bibfnamefont {F.}~\bibnamefont {Jonietz}},
	\bibinfo {author} {\bibfnamefont {C.}~\bibnamefont {Pfleiderer}}, \bibinfo
	{author} {\bibfnamefont {A.}~\bibnamefont {Rosch}}, \bibinfo {author}
	{\bibfnamefont {A.}~\bibnamefont {Neubauer}}, \bibinfo {author}
	{\bibfnamefont {R.}~\bibnamefont {Georgii}}, \ and\ \bibinfo {author}
	{\bibfnamefont {P.}~\bibnamefont {B{\"{o}}ni}},\ }\href {\doibase
	10.1126/science.1166767} {\bibfield  {journal} {\bibinfo  {journal}
		{Science}\ }\textbf {\bibinfo {volume} {323}},\ \bibinfo {pages} {915}
	(\bibinfo {year} {2009})}\BibitemShut {NoStop}%
\bibitem [{\citenamefont {Fert}\ \emph {et~al.}(2013)\citenamefont {Fert},
	\citenamefont {Cros},\ and\ \citenamefont {Sampaio}}]{Fert2013}%
\BibitemOpen
\bibfield  {author} {\bibinfo {author} {\bibfnamefont {A.}~\bibnamefont
		{Fert}}, \bibinfo {author} {\bibfnamefont {V.}~\bibnamefont {Cros}}, \ and\
	\bibinfo {author} {\bibfnamefont {J.}~\bibnamefont {Sampaio}},\ }\href
{\doibase 10.1038/nnano.2013.29} {\bibfield  {journal} {\bibinfo  {journal}
		{Nature Nanotechnology}\ }\textbf {\bibinfo {volume} {8}},\ \bibinfo {pages}
	{152} (\bibinfo {year} {2013})}\BibitemShut {NoStop}%
\bibitem [{\citenamefont {Rybakov}\ \emph {et~al.}(2013)\citenamefont
	{Rybakov}, \citenamefont {Borisov},\ and\ \citenamefont
	{Bogdanov}}]{Rybakov2013}%
\BibitemOpen
\bibfield  {author} {\bibinfo {author} {\bibfnamefont {F.~N.}\ \bibnamefont
		{Rybakov}}, \bibinfo {author} {\bibfnamefont {A.~B.}\ \bibnamefont
		{Borisov}}, \ and\ \bibinfo {author} {\bibfnamefont {A.~N.}\ \bibnamefont
		{Bogdanov}},\ }\href {\doibase 10.1103/PhysRevB.87.094424} {\bibfield
	{journal} {\bibinfo  {journal} {Physical Review B}\ }\textbf {\bibinfo
		{volume} {87}},\ \bibinfo {pages} {094424} (\bibinfo {year}
	{2013})}\BibitemShut {NoStop}%
\bibitem [{\citenamefont {Meynell}\ \emph {et~al.}(2014)\citenamefont
	{Meynell}, \citenamefont {Wilson}, \citenamefont {Fritzsche}, \citenamefont
	{Bogdanov},\ and\ \citenamefont {Monchesky}}]{Meynell2014}%
\BibitemOpen
\bibfield  {author} {\bibinfo {author} {\bibfnamefont {S.~A.}\ \bibnamefont
		{Meynell}}, \bibinfo {author} {\bibfnamefont {M.~N.}\ \bibnamefont {Wilson}},
	\bibinfo {author} {\bibfnamefont {H.}~\bibnamefont {Fritzsche}}, \bibinfo
	{author} {\bibfnamefont {A.~N.}\ \bibnamefont {Bogdanov}}, \ and\ \bibinfo
	{author} {\bibfnamefont {T.~L.}\ \bibnamefont {Monchesky}},\ }\href {\doibase
	10.1103/PhysRevB.90.014406} {\bibfield  {journal} {\bibinfo  {journal}
		{Physical Review B}\ }\textbf {\bibinfo {volume} {90}},\ \bibinfo {pages}
	{014406} (\bibinfo {year} {2014})}\BibitemShut {NoStop}%
\bibitem [{\citenamefont {Leonov}\ \emph {et~al.}(2016)\citenamefont {Leonov},
	\citenamefont {Togawa}, \citenamefont {Monchesky}, \citenamefont {Bogdanov},
	\citenamefont {Kishine}, \citenamefont {Kousaka}, \citenamefont {Miyagawa},
	\citenamefont {Koyama}, \citenamefont {Akimitsu}, \citenamefont {Koyama},
	\citenamefont {Harada}, \citenamefont {Mori}, \citenamefont {McGrouther},
	\citenamefont {Lamb}, \citenamefont {Krajnak}, \citenamefont {McVitie},
	\citenamefont {Stamps},\ and\ \citenamefont {Inoue}}]{Leonov2016}%
\BibitemOpen
\bibfield  {author} {\bibinfo {author} {\bibfnamefont {A.~O.}\ \bibnamefont
		{Leonov}}, \bibinfo {author} {\bibfnamefont {Y.}~\bibnamefont {Togawa}},
	\bibinfo {author} {\bibfnamefont {T.~L.}\ \bibnamefont {Monchesky}}, \bibinfo
	{author} {\bibfnamefont {A.~N.}\ \bibnamefont {Bogdanov}}, \bibinfo {author}
	{\bibfnamefont {J.}~\bibnamefont {Kishine}}, \bibinfo {author} {\bibfnamefont
		{Y.}~\bibnamefont {Kousaka}}, \bibinfo {author} {\bibfnamefont
		{M.}~\bibnamefont {Miyagawa}}, \bibinfo {author} {\bibfnamefont
		{T.}~\bibnamefont {Koyama}}, \bibinfo {author} {\bibfnamefont
		{J.}~\bibnamefont {Akimitsu}}, \bibinfo {author} {\bibfnamefont
		{T.}~\bibnamefont {Koyama}}, \bibinfo {author} {\bibfnamefont
		{K.}~\bibnamefont {Harada}}, \bibinfo {author} {\bibfnamefont
		{S.}~\bibnamefont {Mori}}, \bibinfo {author} {\bibfnamefont {D.}~\bibnamefont
		{McGrouther}}, \bibinfo {author} {\bibfnamefont {R.}~\bibnamefont {Lamb}},
	\bibinfo {author} {\bibfnamefont {M.}~\bibnamefont {Krajnak}}, \bibinfo
	{author} {\bibfnamefont {S.}~\bibnamefont {McVitie}}, \bibinfo {author}
	{\bibfnamefont {R.~L.}\ \bibnamefont {Stamps}}, \ and\ \bibinfo {author}
	{\bibfnamefont {K.}~\bibnamefont {Inoue}},\ }\href {\doibase
	10.1103/PhysRevLett.117.087202} {\bibfield  {journal} {\bibinfo  {journal}
		{Physical Review Letters}\ }\textbf {\bibinfo {volume} {117}},\ \bibinfo
	{pages} {087202} (\bibinfo {year} {2016})}\BibitemShut {NoStop}%
\bibitem [{\citenamefont {Rybakov}\ \emph {et~al.}(2016)\citenamefont
	{Rybakov}, \citenamefont {Borisov}, \citenamefont {Bl{\"{u}}gel},\ and\
	\citenamefont {Kiselev}}]{Rybakov2016}%
\BibitemOpen
\bibfield  {author} {\bibinfo {author} {\bibfnamefont {F.~N.}\ \bibnamefont
		{Rybakov}}, \bibinfo {author} {\bibfnamefont {A.~B.}\ \bibnamefont
		{Borisov}}, \bibinfo {author} {\bibfnamefont {S.}~\bibnamefont
		{Bl{\"{u}}gel}}, \ and\ \bibinfo {author} {\bibfnamefont {N.~S.}\
		\bibnamefont {Kiselev}},\ }\href {\doibase 10.1088/1367-2630/18/4/045002}
{\bibfield  {journal} {\bibinfo  {journal} {New Journal of Physics}\ }\textbf
	{\bibinfo {volume} {18}},\ \bibinfo {pages} {045002} (\bibinfo {year}
	{2016})}\BibitemShut {NoStop}%
\bibitem [{\citenamefont {McGrouther}\ \emph {et~al.}(2016)\citenamefont
	{McGrouther}, \citenamefont {Lamb}, \citenamefont {Krajnak}, \citenamefont
	{McFadzean}, \citenamefont {McVitie}, \citenamefont {Stamps}, \citenamefont
	{Leonov}, \citenamefont {Bogdanov},\ and\ \citenamefont
	{Togawa}}]{McGrouther2016}%
\BibitemOpen
\bibfield  {author} {\bibinfo {author} {\bibfnamefont {D.}~\bibnamefont
		{McGrouther}}, \bibinfo {author} {\bibfnamefont {R.~J.}\ \bibnamefont
		{Lamb}}, \bibinfo {author} {\bibfnamefont {M.}~\bibnamefont {Krajnak}},
	\bibinfo {author} {\bibfnamefont {S.}~\bibnamefont {McFadzean}}, \bibinfo
	{author} {\bibfnamefont {S.}~\bibnamefont {McVitie}}, \bibinfo {author}
	{\bibfnamefont {R.~L.}\ \bibnamefont {Stamps}}, \bibinfo {author}
	{\bibfnamefont {A.~O.}\ \bibnamefont {Leonov}}, \bibinfo {author}
	{\bibfnamefont {A.~N.}\ \bibnamefont {Bogdanov}}, \ and\ \bibinfo {author}
	{\bibfnamefont {Y.}~\bibnamefont {Togawa}},\ }\href {\doibase
	10.1088/1367-2630/18/9/095004} {\bibfield  {journal} {\bibinfo  {journal}
		{New Journal of Physics}\ }\textbf {\bibinfo {volume} {18}},\ \bibinfo
	{pages} {095004} (\bibinfo {year} {2016})}\BibitemShut {NoStop}%
\bibitem [{\citenamefont {Zhang}\ \emph {et~al.}(2018)\citenamefont {Zhang},
	\citenamefont {van~der Laan}, \citenamefont {Wang}, \citenamefont
	{Haghighirad},\ and\ \citenamefont {Hesjedal}}]{Zhang2018}%
\BibitemOpen
\bibfield  {author} {\bibinfo {author} {\bibfnamefont {S.~L.}\ \bibnamefont
		{Zhang}}, \bibinfo {author} {\bibfnamefont {G.}~\bibnamefont {van~der Laan}},
	\bibinfo {author} {\bibfnamefont {W.~W.}\ \bibnamefont {Wang}}, \bibinfo
	{author} {\bibfnamefont {A.~A.}\ \bibnamefont {Haghighirad}}, \ and\ \bibinfo
	{author} {\bibfnamefont {T.}~\bibnamefont {Hesjedal}},\ }\href {\doibase
	10.1103/PhysRevLett.120.227202} {\bibfield  {journal} {\bibinfo  {journal}
		{Physical Review Letters}\ }\textbf {\bibinfo {volume} {120}},\ \bibinfo
	{pages} {227202} (\bibinfo {year} {2018})}\BibitemShut {NoStop}%
\bibitem [{\citenamefont {Yoshida}\ \emph {et~al.}(2012)\citenamefont
	{Yoshida}, \citenamefont {Schroder}, \citenamefont {Ferriani}, \citenamefont
	{Serrate}, \citenamefont {Kubetzka}, \citenamefont {{von Bergmann}},
	\citenamefont {Heinze},\ and\ \citenamefont {Wiesendanger}}]{Yoshida2012}%
\BibitemOpen
\bibfield  {author} {\bibinfo {author} {\bibfnamefont {Y.}~\bibnamefont
		{Yoshida}}, \bibinfo {author} {\bibfnamefont {S.}~\bibnamefont {Schroder}},
	\bibinfo {author} {\bibfnamefont {P.}~\bibnamefont {Ferriani}}, \bibinfo
	{author} {\bibfnamefont {D.}~\bibnamefont {Serrate}}, \bibinfo {author}
	{\bibfnamefont {A.}~\bibnamefont {Kubetzka}}, \bibinfo {author}
	{\bibfnamefont {K.}~\bibnamefont {{von Bergmann}}}, \bibinfo {author}
	{\bibfnamefont {S.}~\bibnamefont {Heinze}}, \ and\ \bibinfo {author}
	{\bibfnamefont {R.}~\bibnamefont {Wiesendanger}},\ }\href {\doibase
	10.1103/PhysRevLett.108.087205} {\bibfield  {journal} {\bibinfo  {journal}
		{Physical Review Letters}\ }\textbf {\bibinfo {volume} {108}},\ \bibinfo
	{pages} {087205} (\bibinfo {year} {2012})}\BibitemShut {NoStop}%
\bibitem [{\citenamefont {Romming}\ \emph {et~al.}(2015)\citenamefont
	{Romming}, \citenamefont {Kubetzka}, \citenamefont {Hanneken}, \citenamefont
	{{von Bergmann}},\ and\ \citenamefont {Wiesendanger}}]{Romming2015}%
\BibitemOpen
\bibfield  {author} {\bibinfo {author} {\bibfnamefont {N.}~\bibnamefont
		{Romming}}, \bibinfo {author} {\bibfnamefont {A.}~\bibnamefont {Kubetzka}},
	\bibinfo {author} {\bibfnamefont {C.}~\bibnamefont {Hanneken}}, \bibinfo
	{author} {\bibfnamefont {K.}~\bibnamefont {{von Bergmann}}}, \ and\ \bibinfo
	{author} {\bibfnamefont {R.}~\bibnamefont {Wiesendanger}},\ }\href {\doibase
	10.1103/PhysRevLett.114.177203} {\bibfield  {journal} {\bibinfo  {journal}
		{Physical Review Letters}\ }\textbf {\bibinfo {volume} {114}},\ \bibinfo
	{pages} {177203} (\bibinfo {year} {2015})}\BibitemShut {NoStop}%
\bibitem [{\citenamefont {Wiesendanger}(2016)}]{Wiesendanger2016}%
\BibitemOpen
\bibfield  {author} {\bibinfo {author} {\bibfnamefont {R.}~\bibnamefont
		{Wiesendanger}},\ }\href {\doibase 10.1038/natrevmats.2016.44} {\enquote
	{\bibinfo {title} {{Nanoscale magnetic skyrmions in metallic films and
				multilayers: A new twist for spintronics}},}\ } (\bibinfo {year}
{2016})\BibitemShut {NoStop}%
\bibitem [{\citenamefont {B{\"{u}}ttner}\ \emph {et~al.}(2015)\citenamefont
	{B{\"{u}}ttner}, \citenamefont {Moutafis}, \citenamefont {Schneider},
	\citenamefont {Kr{\"{u}}ger}, \citenamefont {G{\"{u}}nther}, \citenamefont
	{Geilhufe}, \citenamefont {Schmising}, \citenamefont {Mohanty}, \citenamefont
	{Pfau}, \citenamefont {Schaffert}, \citenamefont {Bisig}, \citenamefont
	{Foerster}, \citenamefont {Schulz}, \citenamefont {Vaz}, \citenamefont
	{Franken}, \citenamefont {Swagten}, \citenamefont {Kl{\"{a}}ui},\ and\
	\citenamefont {Eisebitt}}]{Buttner2015b}%
\BibitemOpen
\bibfield  {author} {\bibinfo {author} {\bibfnamefont {F.}~\bibnamefont
		{B{\"{u}}ttner}}, \bibinfo {author} {\bibfnamefont {C.}~\bibnamefont
		{Moutafis}}, \bibinfo {author} {\bibfnamefont {M.}~\bibnamefont {Schneider}},
	\bibinfo {author} {\bibfnamefont {B.}~\bibnamefont {Kr{\"{u}}ger}}, \bibinfo
	{author} {\bibfnamefont {C.~M.}\ \bibnamefont {G{\"{u}}nther}}, \bibinfo
	{author} {\bibfnamefont {J.}~\bibnamefont {Geilhufe}}, \bibinfo {author}
	{\bibfnamefont {C.~V.~K.}\ \bibnamefont {Schmising}}, \bibinfo {author}
	{\bibfnamefont {J.}~\bibnamefont {Mohanty}}, \bibinfo {author} {\bibfnamefont
		{B.}~\bibnamefont {Pfau}}, \bibinfo {author} {\bibfnamefont {S.}~\bibnamefont
		{Schaffert}}, \bibinfo {author} {\bibfnamefont {A.}~\bibnamefont {Bisig}},
	\bibinfo {author} {\bibfnamefont {M.}~\bibnamefont {Foerster}}, \bibinfo
	{author} {\bibfnamefont {T.}~\bibnamefont {Schulz}}, \bibinfo {author}
	{\bibfnamefont {C.~a.~F.}\ \bibnamefont {Vaz}}, \bibinfo {author}
	{\bibfnamefont {J.~H.}\ \bibnamefont {Franken}}, \bibinfo {author}
	{\bibfnamefont {H.~J.~M.}\ \bibnamefont {Swagten}}, \bibinfo {author}
	{\bibfnamefont {M.}~\bibnamefont {Kl{\"{a}}ui}}, \ and\ \bibinfo {author}
	{\bibfnamefont {S.}~\bibnamefont {Eisebitt}},\ }\href {\doibase
	10.1038/nphys3234} {\bibfield  {journal} {\bibinfo  {journal} {Nature
			Physics}\ }\textbf {\bibinfo {volume} {11}},\ \bibinfo {pages} {225}
	(\bibinfo {year} {2015})}\BibitemShut {NoStop}%
\bibitem [{\citenamefont {Woo}\ \emph {et~al.}(2016)\citenamefont {Woo},
	\citenamefont {Litzius}, \citenamefont {Kr{\"{u}}ger}, \citenamefont {Im},
	\citenamefont {Caretta}, \citenamefont {Richter}, \citenamefont {Mann},
	\citenamefont {Krone}, \citenamefont {Reeve}, \citenamefont {Weigand},
	\citenamefont {Agrawal}, \citenamefont {Lemesh}, \citenamefont {Mawass},
	\citenamefont {Fischer}, \citenamefont {Kl{\"{a}}ui},\ and\ \citenamefont
	{Beach}}]{Woo2016}%
\BibitemOpen
\bibfield  {author} {\bibinfo {author} {\bibfnamefont {S.}~\bibnamefont
		{Woo}}, \bibinfo {author} {\bibfnamefont {K.}~\bibnamefont {Litzius}},
	\bibinfo {author} {\bibfnamefont {B.}~\bibnamefont {Kr{\"{u}}ger}}, \bibinfo
	{author} {\bibfnamefont {M.-Y.}\ \bibnamefont {Im}}, \bibinfo {author}
	{\bibfnamefont {L.}~\bibnamefont {Caretta}}, \bibinfo {author} {\bibfnamefont
		{K.}~\bibnamefont {Richter}}, \bibinfo {author} {\bibfnamefont
		{M.}~\bibnamefont {Mann}}, \bibinfo {author} {\bibfnamefont {A.}~\bibnamefont
		{Krone}}, \bibinfo {author} {\bibfnamefont {R.~M.}\ \bibnamefont {Reeve}},
	\bibinfo {author} {\bibfnamefont {M.}~\bibnamefont {Weigand}}, \bibinfo
	{author} {\bibfnamefont {P.}~\bibnamefont {Agrawal}}, \bibinfo {author}
	{\bibfnamefont {I.}~\bibnamefont {Lemesh}}, \bibinfo {author} {\bibfnamefont
		{M.-A.}\ \bibnamefont {Mawass}}, \bibinfo {author} {\bibfnamefont
		{P.}~\bibnamefont {Fischer}}, \bibinfo {author} {\bibfnamefont
		{M.}~\bibnamefont {Kl{\"{a}}ui}}, \ and\ \bibinfo {author} {\bibfnamefont
		{G.~S.~D.}\ \bibnamefont {Beach}},\ }\href {\doibase 10.1038/nmat4593}
{\bibfield  {journal} {\bibinfo  {journal} {Nature Materials}\ }\textbf
	{\bibinfo {volume} {15}},\ \bibinfo {pages} {501} (\bibinfo {year}
	{2016})}\BibitemShut {NoStop}%
\bibitem [{\citenamefont {Moreau-Luchaire}\ \emph {et~al.}(2016)\citenamefont
	{Moreau-Luchaire}, \citenamefont {Moutafis}, \citenamefont {Reyren},
	\citenamefont {Sampaio}, \citenamefont {Vaz}, \citenamefont {{Van Horne}},
	\citenamefont {Bouzehouane}, \citenamefont {Garcia}, \citenamefont
	{Deranlot}, \citenamefont {Warnicke}, \citenamefont {Wohlh{\"{u}}ter},
	\citenamefont {George}, \citenamefont {Weigand}, \citenamefont {Raabe},
	\citenamefont {Cros},\ and\ \citenamefont {Fert}}]{Moreau-Luchaire2016a}%
\BibitemOpen
\bibfield  {author} {\bibinfo {author} {\bibfnamefont {C.}~\bibnamefont
		{Moreau-Luchaire}}, \bibinfo {author} {\bibfnamefont {C.}~\bibnamefont
		{Moutafis}}, \bibinfo {author} {\bibfnamefont {N.}~\bibnamefont {Reyren}},
	\bibinfo {author} {\bibfnamefont {J.}~\bibnamefont {Sampaio}}, \bibinfo
	{author} {\bibfnamefont {C.~A.}\ \bibnamefont {Vaz}}, \bibinfo {author}
	{\bibfnamefont {N.}~\bibnamefont {{Van Horne}}}, \bibinfo {author}
	{\bibfnamefont {K.}~\bibnamefont {Bouzehouane}}, \bibinfo {author}
	{\bibfnamefont {K.}~\bibnamefont {Garcia}}, \bibinfo {author} {\bibfnamefont
		{C.}~\bibnamefont {Deranlot}}, \bibinfo {author} {\bibfnamefont
		{P.}~\bibnamefont {Warnicke}}, \bibinfo {author} {\bibfnamefont
		{P.}~\bibnamefont {Wohlh{\"{u}}ter}}, \bibinfo {author} {\bibfnamefont
		{J.~M.}\ \bibnamefont {George}}, \bibinfo {author} {\bibfnamefont
		{M.}~\bibnamefont {Weigand}}, \bibinfo {author} {\bibfnamefont
		{J.}~\bibnamefont {Raabe}}, \bibinfo {author} {\bibfnamefont
		{V.}~\bibnamefont {Cros}}, \ and\ \bibinfo {author} {\bibfnamefont
		{A.}~\bibnamefont {Fert}},\ }\href {\doibase 10.1038/nnano.2015.313}
{\bibfield  {journal} {\bibinfo  {journal} {Nature Nanotechnology}\ }\textbf
	{\bibinfo {volume} {11}},\ \bibinfo {pages} {444} (\bibinfo {year}
	{2016})}\BibitemShut {NoStop}%
\bibitem [{\citenamefont {Litzius}\ \emph {et~al.}(2016)\citenamefont
	{Litzius}, \citenamefont {Lemesh}, \citenamefont {Kr{\"{u}}ger},
	\citenamefont {Caretta}, \citenamefont {Richter}, \citenamefont
	{B{\"{u}}ttner}, \citenamefont {Bassirian}, \citenamefont {F{\"{o}}rster},
	\citenamefont {Reeve}, \citenamefont {Weigand}, \citenamefont {Bykova},
	\citenamefont {Stoll}, \citenamefont {Sch{\"{u}}tz}, \citenamefont {Beach},\
	and\ \citenamefont {Kl{\"{a}}ui}}]{Litzius2016}%
\BibitemOpen
\bibfield  {author} {\bibinfo {author} {\bibfnamefont {K.}~\bibnamefont
		{Litzius}}, \bibinfo {author} {\bibfnamefont {I.}~\bibnamefont {Lemesh}},
	\bibinfo {author} {\bibfnamefont {B.}~\bibnamefont {Kr{\"{u}}ger}}, \bibinfo
	{author} {\bibfnamefont {L.}~\bibnamefont {Caretta}}, \bibinfo {author}
	{\bibfnamefont {K.}~\bibnamefont {Richter}}, \bibinfo {author} {\bibfnamefont
		{F.}~\bibnamefont {B{\"{u}}ttner}}, \bibinfo {author} {\bibfnamefont
		{P.}~\bibnamefont {Bassirian}}, \bibinfo {author} {\bibfnamefont
		{J.}~\bibnamefont {F{\"{o}}rster}}, \bibinfo {author} {\bibfnamefont {R.~M.}\
		\bibnamefont {Reeve}}, \bibinfo {author} {\bibfnamefont {M.}~\bibnamefont
		{Weigand}}, \bibinfo {author} {\bibfnamefont {I.}~\bibnamefont {Bykova}},
	\bibinfo {author} {\bibfnamefont {H.}~\bibnamefont {Stoll}}, \bibinfo
	{author} {\bibfnamefont {G.}~\bibnamefont {Sch{\"{u}}tz}}, \bibinfo {author}
	{\bibfnamefont {G.~S.~D.}\ \bibnamefont {Beach}}, \ and\ \bibinfo {author}
	{\bibfnamefont {M.}~\bibnamefont {Kl{\"{a}}ui}},\ }\href {\doibase
	10.1038/nphys4000} {\bibfield  {journal} {\bibinfo  {journal} {Nature
			Physics}\ }\textbf {\bibinfo {volume} {13}},\ \bibinfo {pages} {170}
	(\bibinfo {year} {2016})}\BibitemShut {NoStop}%
\bibitem [{\citenamefont {B{\"{u}}ttner}\ \emph {et~al.}(2017)\citenamefont
	{B{\"{u}}ttner}, \citenamefont {Lemesh}, \citenamefont {Schneider},
	\citenamefont {Pfau}, \citenamefont {G{\"{u}}nther}, \citenamefont {Hessing},
	\citenamefont {Geilhufe}, \citenamefont {Caretta}, \citenamefont {Engel},
	\citenamefont {Kr{\"{u}}ger}, \citenamefont {Viefhaus}, \citenamefont
	{Eisebitt},\ and\ \citenamefont {Beach}}]{Buttner2017e}%
\BibitemOpen
\bibfield  {author} {\bibinfo {author} {\bibfnamefont {F.}~\bibnamefont
		{B{\"{u}}ttner}}, \bibinfo {author} {\bibfnamefont {I.}~\bibnamefont
		{Lemesh}}, \bibinfo {author} {\bibfnamefont {M.}~\bibnamefont {Schneider}},
	\bibinfo {author} {\bibfnamefont {B.}~\bibnamefont {Pfau}}, \bibinfo {author}
	{\bibfnamefont {C.~M.}\ \bibnamefont {G{\"{u}}nther}}, \bibinfo {author}
	{\bibfnamefont {P.}~\bibnamefont {Hessing}}, \bibinfo {author} {\bibfnamefont
		{J.}~\bibnamefont {Geilhufe}}, \bibinfo {author} {\bibfnamefont
		{L.}~\bibnamefont {Caretta}}, \bibinfo {author} {\bibfnamefont
		{D.}~\bibnamefont {Engel}}, \bibinfo {author} {\bibfnamefont
		{B.}~\bibnamefont {Kr{\"{u}}ger}}, \bibinfo {author} {\bibfnamefont
		{J.}~\bibnamefont {Viefhaus}}, \bibinfo {author} {\bibfnamefont
		{S.}~\bibnamefont {Eisebitt}}, \ and\ \bibinfo {author} {\bibfnamefont
		{G.~S.~D.}\ \bibnamefont {Beach}},\ }\href {\doibase 10.1038/nnano.2017.178}
{\bibfield  {journal} {\bibinfo  {journal} {Nature Nanotechnology}\ }\textbf
	{\bibinfo {volume} {12}},\ \bibinfo {pages} {1040} (\bibinfo {year}
	{2017})}\BibitemShut {NoStop}%
\bibitem [{\citenamefont {Suna}(1986)}]{Suna1986a}%
\BibitemOpen
\bibfield  {author} {\bibinfo {author} {\bibfnamefont {A.}~\bibnamefont
		{Suna}},\ }\href {\doibase 10.1063/1.336684} {\bibfield  {journal} {\bibinfo
		{journal} {Journal of Applied Physics}\ }\textbf {\bibinfo {volume} {59}},\
	\bibinfo {pages} {313} (\bibinfo {year} {1986})}\BibitemShut {NoStop}%
\bibitem [{\citenamefont {Lemesh}\ \emph {et~al.}(2017)\citenamefont {Lemesh},
	\citenamefont {B{\"{u}}ttner},\ and\ \citenamefont {Beach}}]{Lemesh2017}%
\BibitemOpen
\bibfield  {author} {\bibinfo {author} {\bibfnamefont {I.}~\bibnamefont
		{Lemesh}}, \bibinfo {author} {\bibfnamefont {F.}~\bibnamefont
		{B{\"{u}}ttner}}, \ and\ \bibinfo {author} {\bibfnamefont {G.~S.~D.}\
		\bibnamefont {Beach}},\ }\href {\doibase 10.1103/PhysRevB.95.174423}
{\bibfield  {journal} {\bibinfo  {journal} {Physical Review B}\ }\textbf
	{\bibinfo {volume} {95}},\ \bibinfo {pages} {174423} (\bibinfo {year}
	{2017})}\BibitemShut {NoStop}%
\bibitem [{\citenamefont {Dovzhenko}\ \emph {et~al.}(2016)\citenamefont
	{Dovzhenko}, \citenamefont {Casola}, \citenamefont {Schlotter}, \citenamefont
	{Zhou}, \citenamefont {B{\"{u}}ttner}, \citenamefont {Walsworth},
	\citenamefont {Beach},\ and\ \citenamefont {Yacoby}}]{Dovzhenko2016}%
\BibitemOpen
\bibfield  {author} {\bibinfo {author} {\bibfnamefont {Y.}~\bibnamefont
		{Dovzhenko}}, \bibinfo {author} {\bibfnamefont {F.}~\bibnamefont {Casola}},
	\bibinfo {author} {\bibfnamefont {S.}~\bibnamefont {Schlotter}}, \bibinfo
	{author} {\bibfnamefont {T.~X.}\ \bibnamefont {Zhou}}, \bibinfo {author}
	{\bibfnamefont {F.}~\bibnamefont {B{\"{u}}ttner}}, \bibinfo {author}
	{\bibfnamefont {R.~L.}\ \bibnamefont {Walsworth}}, \bibinfo {author}
	{\bibfnamefont {G.~S.~D.}\ \bibnamefont {Beach}}, \ and\ \bibinfo {author}
	{\bibfnamefont {A.}~\bibnamefont {Yacoby}},\ }\href {\doibase 10.1038/s41467-018-05158-9}
{\bibfield  {journal} {\bibinfo  {journal} {Nature communications}\ }\textbf
	{\bibinfo {volume} {9.1}},\ \bibinfo {pages} {2712} (\bibinfo {year}
	{2018})}\BibitemShut {NoStop}%
\bibitem [{\citenamefont {Montoya}\ \emph {et~al.}(2017)\citenamefont
	{Montoya}, \citenamefont {Couture}, \citenamefont {Chess}, \citenamefont
	{Lee}, \citenamefont {Kent}, \citenamefont {Henze}, \citenamefont {Sinha},
	\citenamefont {Im}, \citenamefont {Kevan}, \citenamefont {Fischer},
	\citenamefont {McMorran}, \citenamefont {Lomakin}, \citenamefont {Roy},\ and\
	\citenamefont {Fullerton}}]{Montoya2017}%
\BibitemOpen
\bibfield  {author} {\bibinfo {author} {\bibfnamefont {S.~A.}\ \bibnamefont
		{Montoya}}, \bibinfo {author} {\bibfnamefont {S.}~\bibnamefont {Couture}},
	\bibinfo {author} {\bibfnamefont {J.~J.}\ \bibnamefont {Chess}}, \bibinfo
	{author} {\bibfnamefont {J.~C.~.T}\ \bibnamefont {Lee}}, \bibinfo {author}
	{\bibfnamefont {N.}~\bibnamefont {Kent}}, \bibinfo {author} {\bibfnamefont
		{D.}~\bibnamefont {Henze}}, \bibinfo {author} {\bibfnamefont {S.~K.}\
		\bibnamefont {Sinha}}, \bibinfo {author} {\bibfnamefont {M.~Y.}\ \bibnamefont
		{Im}}, \bibinfo {author} {\bibfnamefont {S.~D.}\ \bibnamefont {Kevan}},
	\bibinfo {author} {\bibfnamefont {P.}~\bibnamefont {Fischer}}, \bibinfo
	{author} {\bibfnamefont {B.~J.}\ \bibnamefont {McMorran}}, \bibinfo {author}
	{\bibfnamefont {V.}~\bibnamefont {Lomakin}}, \bibinfo {author} {\bibfnamefont
		{S.}~\bibnamefont {Roy}}, \ and\ \bibinfo {author} {\bibfnamefont {E.~E.}\
		\bibnamefont {Fullerton}},\ }\href {\doibase 10.1103/PhysRevB.95.024415}
{\bibfield  {journal} {\bibinfo  {journal} {Physical Review B}\ }\textbf
	{\bibinfo {volume} {95}},\ \bibinfo {pages} {024415} (\bibinfo {year}
	{2017})}\BibitemShut {NoStop}%
\bibitem [{\citenamefont
	{Schl{\"{o}}mann}(1973{\natexlab{a}})}]{Schlomann1973a}%
\BibitemOpen
\bibfield  {author} {\bibinfo {author} {\bibfnamefont {E.}~\bibnamefont
		{Schl{\"{o}}mann}},\ }\href {\doibase 10.1063/1.1662460} {\bibfield
	{journal} {\bibinfo  {journal} {Journal of Applied Physics}\ }\textbf
	{\bibinfo {volume} {44}},\ \bibinfo {pages} {1837} (\bibinfo {year}
	{1973}{\natexlab{a}})}\BibitemShut {NoStop}%
\bibitem [{\citenamefont
	{Schl{\"{o}}mann}(1973{\natexlab{b}})}]{Schlomann1973b}%
\BibitemOpen
\bibfield  {author} {\bibinfo {author} {\bibfnamefont {E.}~\bibnamefont
		{Schl{\"{o}}mann}},\ }\href {\doibase 10.1063/1.1662461} {\bibfield
	{journal} {\bibinfo  {journal} {Journal of Applied Physics}\ }\textbf
	{\bibinfo {volume} {44}},\ \bibinfo {pages} {1850} (\bibinfo {year}
	{1973}{\natexlab{b}})}\BibitemShut {NoStop}%
\bibitem [{Note1()}]{Note1}%
\BibitemOpen
\bibinfo {note} {In our notations, $\sigma ^{\protect \mathcal {M}, \protect
		\mathcal {N}}$ indicates the energy (per multilayer cross-section) of
	$\protect \mathcal {M}$ DWs in the film with $\protect \mathcal {N}$
	multilayer repeats}\BibitemShut {NoStop}%
\bibitem [{Note2()}]{Note2}%
\BibitemOpen
\bibinfo {note} { Our approximation assumes that $\mathcal{T}<l_{\rm{ex}}$.}\BibitemShut {NoStop}%
\bibitem [{\citenamefont {Legrand}\ \emph {et~al.}(2017)\citenamefont
	{Legrand}, \citenamefont {Chauleau}, \citenamefont {Maccariello},
	\citenamefont {Reyren}, \citenamefont {Collin}, \citenamefont {Bouzehouane},
	\citenamefont {Jaouen}, \citenamefont {Cros},\ and\ \citenamefont
	{Fert}}]{Legrand2017a}%
\BibitemOpen
\bibfield  {author} {\bibinfo {author} {\bibfnamefont {W.}~\bibnamefont
		{Legrand}}, \bibinfo {author} {\bibfnamefont {J.-Y.}\ \bibnamefont
		{Chauleau}}, \bibinfo {author} {\bibfnamefont {D.}~\bibnamefont
		{Maccariello}}, \bibinfo {author} {\bibfnamefont {N.}~\bibnamefont {Reyren}},
	\bibinfo {author} {\bibfnamefont {S.}~\bibnamefont {Collin}}, \bibinfo
	{author} {\bibfnamefont {K.}~\bibnamefont {Bouzehouane}}, \bibinfo {author}
	{\bibfnamefont {N.}~\bibnamefont {Jaouen}}, \bibinfo {author} {\bibfnamefont
		{V.}~\bibnamefont {Cros}}, \ and\ \bibinfo {author} {\bibfnamefont
		{A.}~\bibnamefont {Fert}},\ }\href {http://arxiv.org/abs/1712.05978}
{\bibfield
	{journal} {\bibinfo  {journal} {Science advances}\ }\textbf {\bibinfo
		{volume} {4.7}}, (\bibinfo {year} {2018})}\BibitemShut
{NoStop}%
\bibitem{Supp}%
\BibitemOpen
\bibfield  {author} {\bibinfo {author} {See Supplemental Material at \url{http://link.aps.org/supplemental/10.1103/PhysRevB.98.104402} for the detailed derivation of energetics of twisted domain walls, domains, and skyrmions in PMA multilayers and videos which demonstrate the dynamic accumulation of Bloch line pairs in twisted skyrmions in high-j regime (with $j=\SI{1.0e12}{j/m^2}$, Spin Hall angle 0.1, $D=\SI{0.5}{mJ/m^2}$, $\alpha=0.3$, and $B_z=\SI{59}{mT}$)}}\BibitemShut
{NoStop}%
\bibitem [{\citenamefont {Rohart}\ and\ \citenamefont
	{Thiaville}(2013)}]{Rohart2013a}%
\BibitemOpen
\bibfield  {author} {\bibinfo {author} {\bibfnamefont {S.}~\bibnamefont
		{Rohart}}\ and\ \bibinfo {author} {\bibfnamefont {A.}~\bibnamefont
		{Thiaville}},\ }\href@noop {} {\bibfield  {journal} {\bibinfo  {journal}
		{Physical Review B}\ }\textbf {\bibinfo {volume} {88}},\ \bibinfo {pages}
	{184422} (\bibinfo {year} {2013})}\BibitemShut {NoStop}%
\bibitem [{\citenamefont {B{\"{u}}ttner}\ \emph {et~al.}(2018)\citenamefont
	{B{\"{u}}ttner}, \citenamefont {Lemesh},\ and\ \citenamefont
	{Beach}}]{Buttner2017}%
\BibitemOpen
\bibfield  {author} {\bibinfo {author} {\bibfnamefont {F.}~\bibnamefont
		{B{\"{u}}ttner}}, \bibinfo {author} {\bibfnamefont {I.}~\bibnamefont
		{Lemesh}}, \ and\ \bibinfo {author} {\bibfnamefont {G.~S.~D.}\ \bibnamefont
		{Beach}},\ }\href {\doibase 10.1038/s41598-018-22242-8} {\bibfield  {journal}
	{\bibinfo  {journal} {Scientific Reports}\ }\textbf {\bibinfo {volume} {8}},\
	\bibinfo {pages} {1} (\bibinfo {year} {2018})}\BibitemShut {NoStop}%
\bibitem [{\citenamefont {Cape}\ and\ \citenamefont {Lehman}(1971)}]{Cape1971}%
\BibitemOpen
\bibfield  {author} {\bibinfo {author} {\bibfnamefont {J.~A.}\ \bibnamefont
		{Cape}}\ and\ \bibinfo {author} {\bibfnamefont {G.~W.}\ \bibnamefont
		{Lehman}},\ }\href {\doibase 10.1063/1.1660007} {\bibfield  {journal}
	{\bibinfo  {journal} {Journal of Applied Physics}\ }\textbf {\bibinfo
		{volume} {42}},\ \bibinfo {pages} {5732} (\bibinfo {year}
	{1971})}\BibitemShut {NoStop}%
\bibitem [{\citenamefont {Thiele}(1973)}]{Thiele1973}%
\BibitemOpen
\bibfield  {author} {\bibinfo {author} {\bibfnamefont {A.~A.}\ \bibnamefont
		{Thiele}},\ }\href {\doibase 10.1103/PhysRevLett.30.230} {\bibfield
	{journal} {\bibinfo  {journal} {Physical Review Letters}\ }\textbf {\bibinfo
		{volume} {30}},\ \bibinfo {pages} {230} (\bibinfo {year} {1973})}\BibitemShut
{NoStop}%
\bibitem [{\citenamefont {Malozemoff}\ and\ \citenamefont
	{Slonczewski}(1979)}]{Malozemoff1979}%
\BibitemOpen
\bibfield  {author} {\bibinfo {author} {\bibfnamefont {A.}~\bibnamefont
		{Malozemoff}}\ and\ \bibinfo {author} {\bibfnamefont {J.}~\bibnamefont
		{Slonczewski}},\ }\href {\doibase 10.1016/B978-0-12-002951-8.50014-5} {\emph
	{\bibinfo {title} {Magnetic Domain Walls in Bubble Materials}}}\ (\bibinfo
{publisher} {Academic Press},\ \bibinfo {year} {1979})\ pp.\ \bibinfo {pages}
{149--152}\BibitemShut {NoStop}%
	\bibitem [{\citenamefont {B{\"{u}}ttner}\ \emph {et~al.}(2015)\citenamefont
		{B{\"{u}}ttner}, \citenamefont {Kr{\"{u}}ger}, \citenamefont {Eisebitt},\
		and\ \citenamefont {Kl{\"{a}}ui}}]{buttner2015}%
	\BibitemOpen
	\bibfield  {author} {\bibinfo {author} {\bibfnamefont {F.}~\bibnamefont
			{B{\"{u}}ttner}}, \bibinfo {author} {\bibfnamefont {B.}~\bibnamefont
			{Kr{\"{u}}ger}}, \bibinfo {author} {\bibfnamefont {S.}~\bibnamefont
			{Eisebitt}}, \ and\ \bibinfo {author} {\bibfnamefont {M.}~\bibnamefont
			{Kl{\"{a}}ui}},\ }\href {\doibase 10.1103/PhysRevB.92.054408} {\bibfield
		{journal} {\bibinfo  {journal} {Physical Review B}\ }\textbf {\bibinfo
			{volume} {92}},\ \bibinfo {pages} {054408} (\bibinfo {year}
		{2015})}\BibitemShut {NoStop}%
\bibitem [{\citenamefont {Vansteenkiste}\ \emph {et~al.}(2014)\citenamefont
	{Vansteenkiste}, \citenamefont {Leliaert}, \citenamefont {Dvornik},
	\citenamefont {Helsen}, \citenamefont {Garcia-Sanchez},\ and\ \citenamefont
	{{Van Waeyenberge}}}]{Vansteenkiste2014}%
\BibitemOpen
\bibfield  {author} {\bibinfo {author} {\bibfnamefont {A.}~\bibnamefont
		{Vansteenkiste}}, \bibinfo {author} {\bibfnamefont {J.}~\bibnamefont
		{Leliaert}}, \bibinfo {author} {\bibfnamefont {M.}~\bibnamefont {Dvornik}},
	\bibinfo {author} {\bibfnamefont {M.}~\bibnamefont {Helsen}}, \bibinfo
	{author} {\bibfnamefont {F.}~\bibnamefont {Garcia-Sanchez}}, \ and\ \bibinfo
	{author} {\bibfnamefont {B.}~\bibnamefont {{Van Waeyenberge}}},\ }\href
{\doibase 10.1063/1.4899186} {\bibfield  {journal} {\bibinfo  {journal} {AIP
			Advances}\ }\textbf {\bibinfo {volume} {4}},\ \bibinfo {pages} {107133}
	(\bibinfo {year} {2014})}\BibitemShut {NoStop}%
\end{thebibliography}
\end{document}


\title{Supplementary material: \\ Twisted domain walls and skyrmions in perpendicularly magnetized multilayers} 
	
	\author{Ivan Lemesh}
	\email{ivan.g.lemesh@gmail.com}
	\affiliation{Department of Materials Science and Engineering, Massachusetts Institute of Technology, Cambridge, Massachusetts 02139, USA}
	\author{Geoffrey S. D. Beach}
	\affiliation{Department of Materials Science and Engineering, Massachusetts Institute of Technology, Cambridge, Massachusetts 02139, USA}
	\date{\today}	

	
	\pacs{75.60.Ch,75.70.-i}
	
	\maketitle 
	\thispagestyle{fancy}
\tableofcontents
\section{Theory of isolated twisted domain walls in multilayers}	

	\subsection{Energetics of isolated twisted wall}
The total micromagnetic energy density of the thin magnetic film with the magnetization distribution $\mathbf{M}(x) = M_s \mathbf{m}(x)$ can be expressed as follows:
\begin{align}
E_{tot}^{ 1, \mathcal{N}} (\mathbf{m}, \partial \mathbf{m}/\partial x)& = E_{\text{exch}}+E_{\text{DMI}}+E_{\text{anis}}+E_{\text{Zeeman}}+E_{d}^{1, \mathcal{N}} \notag \\&= A  \ \left[\left(\frac{\partial m_{x}}{\partial x}\right)^2+\left(\frac{\partial m_{y}}{\partial x}\right)^2+\left(\frac{\partial m_{z}}{\partial x}\right)^2\right] -D  \left[m_{x}\frac{\partial m_{z}}{\partial x}-m_{z}\frac{\partial m_{x}}{\partial x}\right] + K_{u} \left[(m_{x})^2+(m_{y})^2\right]  \notag \\
&-\mu_0M_s (\mathbf{m \cdot B})-\frac{1}{2}\mu_0M_s(\mathbf{m\cdot B_d}),\label{eq:zero}
\end{align}
where A is the exchange stiffness, $D$ is the interfacial DMI constant, $K_u$ is the uniaxial magnetic anisotropy, $\mathbf{B}$ is an external magnetic field and $\mathbf{B_{d}}=\mathbf{B_{d}}(x)$ is the demagnetized field. 	

Consider a multilayer film with a straight isolated domain wall, in which $\mathcal{N}$ magnetic layers of thickness $\mathcal{T}$ are alternated with the spacer layers of thickness $\mathcal{P}-\mathcal{T}$.
Assuming that the ferromagnetic coupling is strong enough to couple domains in all layers, and that the domain wall width is identical in all layers (i.e. $\Delta_i=\Delta$ for each layer i), we can use the well-known profile of the domain wall located at position $x=q$:
\begin{align}
\theta_i(x, q) & = 2\arctan\{\exp[\mp(x-q)/\Delta]\} \\\phi_i(t) & = \psi_i(t)-\pi/2,
\end{align}
which corresponds to the following magnetization components~\footnote{We can treat domain wall angle in a range $-\pi/2, \ \pi/2$, ignoring $\pi, \ 3\pi/2$ as it is energetically equivalent}:
\begin{align}
m_{i, x} &= \sin(\psi_i) \cosh^{-1}\left(\frac{x-q}{\Delta}\right)\label{eq:magprofilex}\\
m_{i, y} &= \cos(\psi_i) \cosh^{-1}\left(\frac{x-q}{\Delta}\right)\label{eq:magprofiley}\\
m_{i, z} &= \pm \tanh\left(\frac{x-q}{\Delta}\right)\label{eq:magprofilez}
\end{align}
The  cross-sectional domain wall energy can then be found by integrating Eq.~\eqref{eq:zero} with this profile, i.e.
\begin{align}
\sigma_{tot}^{ 1, \mathcal{N}} = \int_{-\infty}^{+\infty}E_{tot}^{ 1, \mathcal{N}}(x)\d x
\end{align}
The integration is trivial except for the magnetostatic term, which we derive in Section~\ref{sec:isolated} of the supplement. The final result reads as
\begin{align}
\sigma_{tot}^{1, \mathcal{N}}(\Delta, \psi_i) &=\frac{2A}{\Delta}f+2K_{u}\Delta f\pm 2B_zM_s f q \notag \\& +\frac{1}{\mathcal{N}}  \sum_{i=0}^{\mathcal{N}-1}\left\{\mp \pi D  f\sin(\psi_i)  -\pi\Delta M_s f(B_x\sin(\psi_i)+B_y \cos(\psi_i))\right\} \notag  \\& + \sum_{i=0}^{\mathcal{N}-1}\sum_{j=0}^{\mathcal{N}-1}\left\{F_{s, ij}(\mathcal{T},\mathcal{P}, \Delta) + \sin(\psi_i)\sin(\psi_{j})  F_{v, ij} (\mathcal{T},\mathcal{P}, \Delta)\pm \sin(\psi_i)\text{sgn}(i-j)F_{sv, ij} (\mathcal{T},\mathcal{P}, \Delta) \right\}\label{eq:E}
\end{align}

\subsection{Equilibrium structure of isolated twisted domain wall}\label{sec:dw}
Consider the stationary domain wall without applied magnetic fields or currents. We can take the derivative of Eq.~\eqref{eq:E}, with respect to $\Delta$ and $\psi_i$, resulting in the following system of two equations:

\begin{align}
2K_{u} f-\frac{2A}{\Delta^2}f+ \sum_{i=0}^{\mathcal{N}-1}\sum_{j=0}^{\mathcal{N}-1} \frac{\partial F_{s, ij}}{\partial \Delta} +\sum_{i=0}^{\mathcal{N}-1}\sum_{j=0}^{\mathcal{N}-1}\sin(\psi_{i})  \sin(\psi_{j})  \frac{\partial F_{v, ij}}{\partial \Delta}  \pm \sum_{i=0}^{\mathcal{N}-1}\sum_{j=0}^{\mathcal{N}-1}  \sin(\psi_{i})  \text{sgn}(i-j)\frac{\partial F_{sv, ij}}{\partial \Delta} =0\label{eq:delta}
\end{align}     

\begin{align}
\left[\mp \frac{1}{\mathcal{N}}  \pi D  f  + \sum_{j=0}^{\mathcal{N}-1}\left(1+\delta_{ij}\right)   F_{v, ij} (\Delta)\sin(\psi_{j})\pm \sum_{j=0}^{\mathcal{N}-1} F_{sv, ij}(\Delta)\text{sgn}(i-j) \right]\cos(\psi_i) =0\label{eq:psi0}
\end{align}
Solution of Eq.~\eqref{eq:delta} combined with Eq.~\eqref{eq:psi0} gives the equilibrium values of $\Delta$, $\psi_i$. However, the direct numerical solution of these equations can lead to significant numerical errors. It would be more desirable to separate the variables $\sin(\psi_i)$ from $\Delta$ as  has been done in the single layer case~\cite{Lemesh2017}. This becomes possible only after introducing the matrix operations, and treating $\sin(\psi_i)$ as a vector and dipolar components $F_{ij}$ as matrices.
For that, we can introduce the following vectorial and matrix notations (for $ i=0, ... ,\mathcal{N}-1$):
\begin{align}
\zeta_{i} & = \sin(\psi_i)\\
 D_{sv, i}&=-\frac{\mathcal{N}}{\pi  f}\sum_{j=0}^{\mathcal{N}-1}  F_{sv, ij}(\Delta)\text{sgn}(i-j)  \\
 \varkappa_{v, ij}&=\left(1+\delta_{ij}\right)   F_{v, ij}(\Delta)\\
\end{align} 
Discarding the trivial solutions of Eq.~\eqref{eq:psi0}, we can now express it in the matrix form as:
\begin{align}
\mp \vec{1}D+\frac{\mathcal{N}}{\pi f}\hat{\varkappa}_{v}\cdot \vec{\zeta}\mp\vec{D}_{sv}=0\label{eq:mtrx},
\end{align}
where $1_i$ is the vector of ones with the length $\mathcal{N}$. After multiplying Eq.~\eqref{eq:mtrx} by $\hat{\varkappa}_v^{-1}$ we will obtain
\begin{align}
\mp D \hat{\varkappa}_v^{-1}+\frac{\mathcal{N}}{\pi f} \vec{\zeta}\mp\hat{\varkappa}_v^{-1}\cdot \vec{D}_{sv}=0
\end{align}
After rearranging the terms
\begin{align}
 \vec{\zeta}=\pm \frac{\pi f}{\mathcal{N}} \hat{\varkappa}_v^{-1}\cdot (\vec{D}_{sv}+\vec{1}D)
\end{align}
The absolute value of each component of vector $\zeta_{i}  = \sin(\psi_i)$ should never exceed one, so by introducing a helper function $\tilde{f}(x)$
\begin{align}
\tilde{f}(x)&=
\begin{cases}
x, & x\leq |1|\\
\text{sgn}(x), & \text{else} \\
\end{cases}
\end{align} 
we finally obtain
\begin{align}
{\sin(\psi_{i})} &=   \pm \tilde{f}\left( \frac{\pi  f}{\mathcal{N}}  \hat{\varkappa}_{v}^{-1} \cdot [ \vec{D}_{sv} +\vec{1} D ]\right)_i \label{eq:sinpsi}
\end{align}  
Or going back to our original notations, we will have the following final equation for $\psi_i$
\begin{align}
\sin(\psi_{i}) =   \pm \tilde{f}\left(\sum_{j=0}^{\mathcal{N}-1} [\left(1+\delta_{ij}\right)   F_{v, ij}(\Delta)]^{-1}\cdot\left[  \frac{\pi D  f}{\mathcal{N}}    1_j - \sum_{k=0}^{\mathcal{N}-1} F_{sv, jk}(\Delta)\text{sgn}(j-k) \right]\right)\label{eq:psi}
\end{align} 
Now, plugging in Eq.~\eqref{eq:psi} into Eq.~\eqref{eq:delta} gives an implicit equation for $\Delta$, which can be solved in any available numerical software package.

Finally, we can derive $ D_{\text{thr}} $, which is the value of DMI at which all the layers (including the last one) saturate to the N\'eel state~\cite{Lemesh2017}. By setting $\sin(\psi_{\mathcal{N}-1})=1$ in Eqs~\eqref{eq:delta},~\eqref{eq:psi}, we obtain
\begin{align}
2K_{u} f-\frac{2A}{\Delta_{\text{thr}}^2}f+ \sum_{i=0}^{\mathcal{N}-1}\sum_{j=0}^{\mathcal{N}-1} \left[\frac{\partial F_{s, ij}}{\partial \Delta_{\text{thr}}} +\frac{\partial F_{v, ij}}{\partial \Delta_{\text{thr}}}  \pm   \text{sgn}(i-j)\frac{\partial F_{sv, ij}}{\partial \Delta_{\text{thr}}}\right] =0\label{eq:delta2}
\end{align}    

\begin{align}
\left(\sum_{j=0}^{\mathcal{N}-1} [\left(1+\delta_{ij}\right)   F_{v, ij}(\Delta_{\text{thr}})]^{-1}\cdot\left[  \frac{\pi D_{\text{thr}}  f}{\mathcal{N}}   1_j  - \sum_{k=0}^{\mathcal{N}-1} F_{sv, jk}(\Delta_{\text{thr}})\text{sgn}(j-k) \right]\right)_{\mathcal{N}-1}-1=0\label{eq:dthr},
\end{align}
where $\Delta_{\text{thr}}$ and $ D_{\text{thr}} $ are the unknown variables.

\begin{figure}[h!]
	\includegraphics[scale = 0.65]{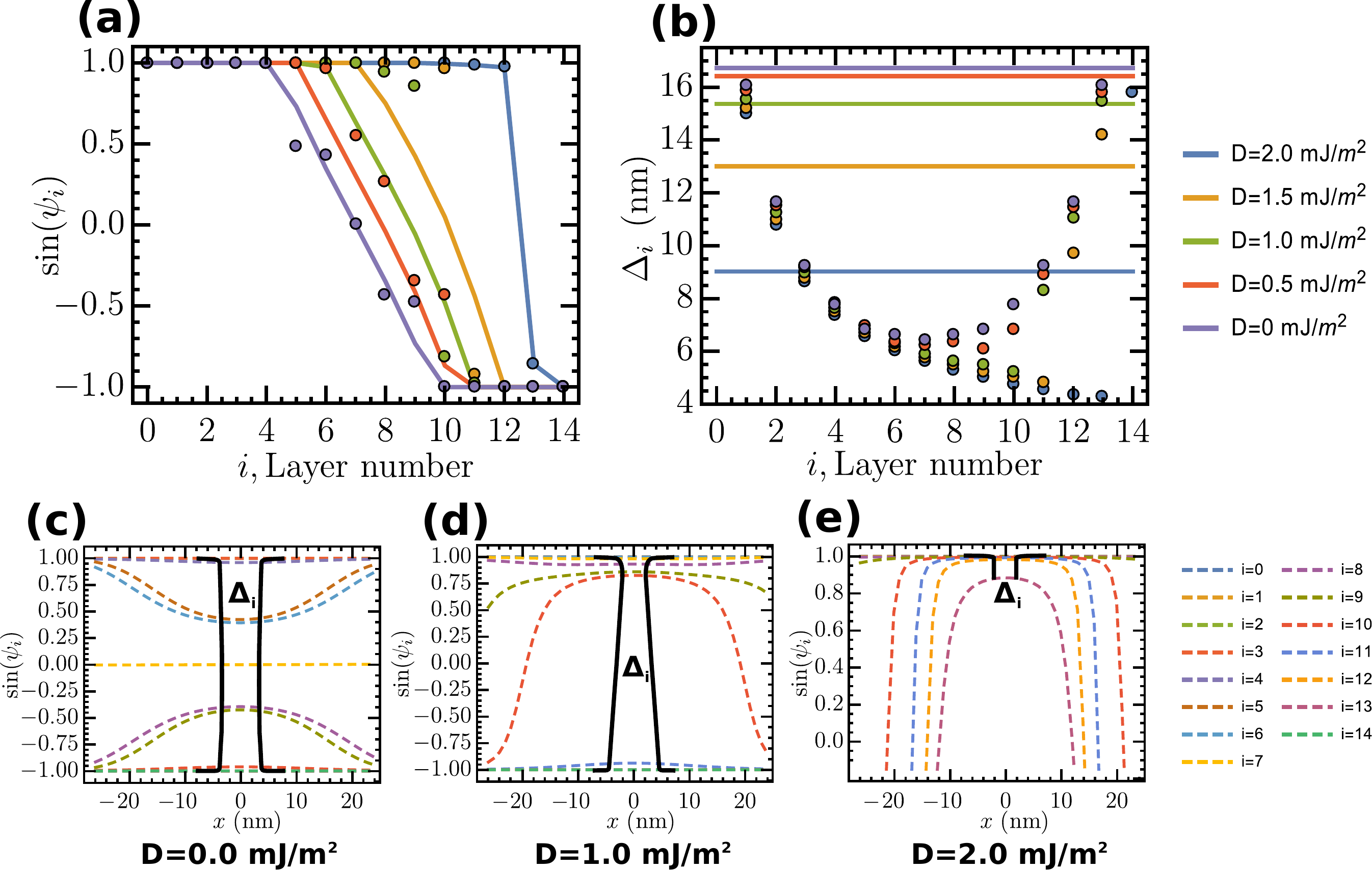}
	\caption{{\bf The DW twist}  predicted by theory (continuous lines) and micromagnetic simulations (points and dashed lines) for a film with $Q=1.01$, $f=1/6$, $M_s = 1.4\times10^6 \rm{A/m}$, $A = \SI{1.0e-11}{ J/m}$, $\mathcal{N}=15$, $\mathcal{T} = 1\rm{nm}$, with ($\downarrow|\uparrow$)  DW. {\bf(a)} DW angle $\psi_i$ and {\bf(b)} DW width $\Delta_i$ as a function of layer number. {\bf(c)} - {\bf(e)}  Variation of $\psi_i$ as a function of coordinate x, with the spread of $\Delta_i$ denoted with continuous black lines.}\label{fig:twist}
\end{figure}

We find that even though our analytical model relies on the constant $\Delta_i$ assumption, it can still reliably predict the equilibrium DW angles $\psi_i$. Figure~\ref{fig:twist}, depicts the comparison of our model with micromagnetic simulations for a case of very low quality factor ($Q=1.01$), when substantial variations in $\Delta_i$ are expected. We can clearly see that our predictions of $\psi_i$ are still accurate, even though $\Delta_{\rm{max}}/\Delta_{\rm{min}}\sim 4$ for this case. We also find that within each layer, $\psi_i$ varies as a function of distance from the DW center (Fig.~\ref{fig:twist}c-e). However, this variation is weak and manifests only in the tails. We note that part of this effect could be due to the boundary conditions, considering the finite size of the simulation volume, so we cannot exclude the possibility that this small variation observed here and also reported in Ref.~\citenum{Legrand2017a} is a simulation artefact. Nonetheless, since it is significant only in the distant tails in the DW, it has negligible effect on a total DW energy and hence, our assumption of a uniform $\psi_i$ is quite reasonable.

\subsection{Origins of DW wall twist}
\begin{figure}[h!]
		\includegraphics[scale = 0.85]{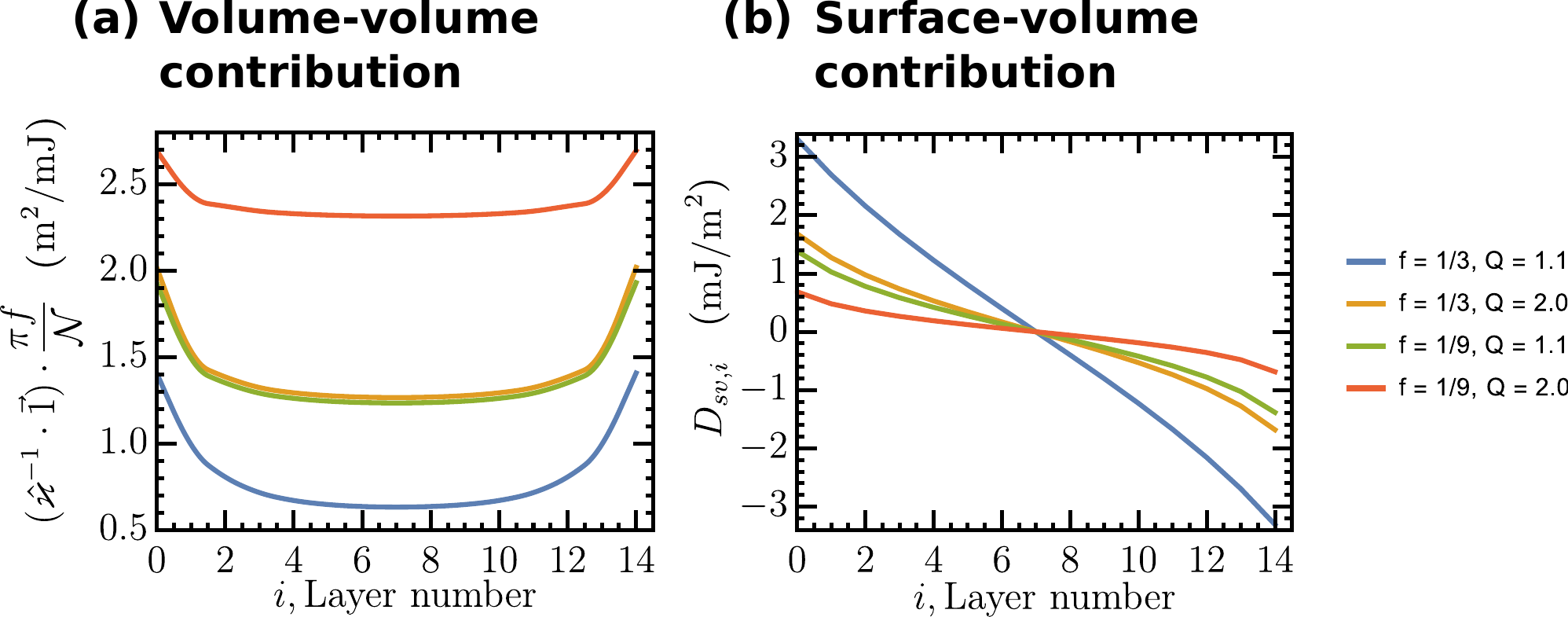}
	\caption{{\bf Contributors to the DW twist.} {\bf(a)} The components of the symmetric vector $ \frac{\pi  f}{\mathcal{N}}  \hat{\varkappa}_{v}^{-1} \vec{1}$. {\bf(b)} The components of the antisymmetric vector $D_{sv}$.}\label{fig:contributors}
\end{figure}

Let us now assess the contributions of surface-volume and volume-volume stray field interactions to  Eq.~\eqref{eq:sinpsi}. In the main text, we assert that in the absence of DMI, there is a Bloch layer ($i_{\rm{Bloch}}$) at the center of a film and adding DMI simply displaces the position of $i_{\rm{Bloch}}$. This would be an exact statement in the case that $\hat{\varkappa}_{v}^{-1}$ were a unity matrix, since then $D$ simply shifts all of the $\psi_i$  by the same offset. In actuality, $\hat{\varkappa}_{v}^{-1}$ is not a unity matrix. However, we find that for a broad range of material properties, the offset that DMI imparts to $\psi_i$ remains approximately constant for all of the layers, except those very near the top and bottom of the film. This can be seen in Fig.~\ref{fig:contributors}, where we plot the components of a vector  $\hat{\varkappa}_{v}^{-1}\cdot \vec{1}$ for a broad range of material parameters. As is evident, these curves are flat, except very near the film surfaces, thus validating our claim that the DMI simply offsets the position of the Bloch layer. This holds true, except when the Bloch layer approaches the surface of the film, i.e. near the transition to the purely N\'eel state.

Since the volume stray field contribution to $\hat{\varkappa}_{v}^{-1}$ is largely uniform throughout the film, the main contributor to the thickness-dependent variation of $\psi_i$ is $D_{sv}$ as claimed in the main text. (see Fig.~\ref{fig:contributors}b that reproduces Fig.~2b from the main text).

	\subsection{Derivation of the stray field energy of the straight isolated domain wall}\label{sec:isolated}
Our task is to calculate the total magnetostatic integral
	\begin{align}
	E_{d} &= \sum_{\alpha \beta = vv, ss, sv, vs}\frac{\mu_0}{8\pi}\iint\d^3 r \d^3 r^\prime \rho_\alpha(\mathbf{r})\frac{1}{{|\mathbf{r}-\mathbf{r}^\prime|}}\rho_\beta(\mathbf{r}^\prime)
	\end{align}
	expressed in units of energy per unit area of the domain wall (per layer):
	\begin{align}
	\sigma_{d}=\frac{E_{d}}{N\mathcal{P}L},
	\end{align}
	where $L$ is the length of the domain, measured in y direction and $\mathcal{P}$ is a multilayer periodicity. As we will show below, the total dipolar energy of the isolated domain wall can be expressed as:
	\begin{align}
	\sigma_{d}^{1, \mathcal{N}} &= \sum_{i = 0}^{\mathcal{N}-1} \sum_{j = 0}^{\mathcal{N}-1} \left\{ \sin(\psi_i)\sin(\psi_{j})  F_{v, ij} (\mathcal{T},\mathcal{P}, \Delta)+\sin(\psi_i)\text{sgn}(i-j)F_{sv, ij} (\mathcal{T},\mathcal{P}, \Delta)+F_{s, ij} (\mathcal{T},\mathcal{P}, \Delta)\right\}  \label{eq:exact}
	\end{align}
	with a generic function $F_{\alpha, ij}$ defined as
	\begin{align}
	F_{\alpha, ij} (\mathcal{T},\mathcal{P}, \Delta)=\frac{\pi \mu_{0}M_{s}^2 \Delta^2}{\mathcal{N}\mathcal{P}}\left[ G_{\alpha}\left(\frac{\left|(i-j)\mathcal{P}+\mathcal{T}\right|}{2\pi \Delta}\right)+ G_{\alpha}\left(\frac{\left|(i-j)\mathcal{P}-\mathcal{T}\right|}{2\pi \Delta}\right)  -2G_{\alpha}\left(\frac{\left|(i-j)\mathcal{P}\right|}{2\pi \Delta}\right)\right]
	\end{align}
	where  $\mathcal{T}$ is the single magnetic layer thickness, $\mathcal{P}$ is the reduced multilayer periodicity, and functions $G_{\alpha}(x)$ are defined analytically as follows:
	\begin{align}
	G_v(x)&= -2\left\{\Psi^{-2}(x+1)-\Psi^{-2}\left(x+\frac{1}{2}\right)- x \ \ln(\Gamma(x+1)) 
	+x \ln\left[\Gamma\left(x+\frac{1}{2}\right)\right] \right. \notag\\ &\left.- \Psi^{-2}(1)+\Psi^{-2}\left(\frac{1}{2}\right) 
	\right\}\\
	G_s(x) &=-\left\{\Psi ^{(-2)}(2 x)+x^2 (2  \log (x)+\log (4)-1)-x(1+2  \ln[\Gamma (2 x)])\right\}\\
	G_{sv}(x) & = 2 \ln\left[\Gamma\left(x+\frac{1}{2}\right)\right] 
	\end{align}
Below, we separately derive the volume-volume, surface-surface, and surface-volume components of Eq.~\eqref{eq:exact}.

		\subsubsection{Volume-Volume stray field energy}
We start from calculating the volume-volume term of the magnetostatic energy of the system, infinite in x and y directions, but finite in z direction:
\begin{align}
\sigma_{d, v}^{N_{dw}=1, N_l=\mathcal{N}}= \lim\limits_{L\rightarrow \infty}\frac{\mu_{0}}{8\pi  L  \mathcal{N}\mathcal{P}} \iint \d ^3 \mathbf{r} \d ^3 \mathbf{r'} \rho_v(\mathbf{r} ) \frac{1}{|\mathbf{r}-\mathbf{r'}|} \rho_v(\mathbf{r'} ), \label{eq:volvolsingle}
\end{align}		
where  $\rho_v=-\nabla \cdot \mathbf{M} $ is the volume charge distribution of the isolated domain wall in multilayers that can be expressed as:
\begin{align}		
\rho_v(\mathbf{r}) = &\sum_{j=0}^{\mathcal{N}-1}\sin(\psi_j) \frac{M_{s}}{\Delta_j}\frac{\tanh(x/\Delta_j)}{\cosh(x/\Delta_j)} \theta(\mathcal{T}/2-|z-j\mathcal{P}-\mathcal{T}/2|)\theta(L/2-|y-L/2|)\label{eq:volumechargesingle}
\end{align}		
The multilayer volume-volume stray field energy for an isolated domain wall $\sigma_{d, v}^{1, \mathcal{N}}$ has already been calculated by B\"uttner~\cite{buttner2015}. Slightly modifying in it to account for layer-dependent domain wall width and angle, we obtain:
\begin{align}
\sigma&_{d, v}^{N_{dw}=1, N_{l}=\mathcal{N}} = \frac{\pi \mu_{0}M_{s}^2 }{\mathcal{N}\mathcal{P}}\sum_{i = 0}^{\mathcal{N}-1} \sum_{i' = 0}^{\mathcal{N}-1}  \sin(\psi_i)\sin(\psi_{i'}) \notag\\& \times \left[G_{ii'}\left(\left|(i-{i'})\mathcal{P}+\mathcal{T}\right|\right)+ G_{ii'}\left(\left|(i-{i'})\mathcal{P}-\mathcal{T}\right|\right)  -2G_{ii'}\left(\left|(i-{i'})\mathcal{P}\right|\right)\right], \label{eq:exactvolumebuttner}
\end{align}
where  $\mathcal{T}$ is the single magnetic layer thickness, $\mathcal{P}$ is the reduced  multilayer periodicity, and $G(\alpha)$:
\begin{align}
G_{ij}&(\alpha, \Delta_i, \Delta_j)=\frac{\Delta_i \Delta_{i'}}{4} \int_{0}^{+\infty} dk \frac{e^{-k\alpha}+k\alpha-1}{k\cosh\left(\frac{\pi \Delta_i k}{2}\right)\cosh\left(\frac{\pi \Delta_j k}{2}\right)}\label{eq:g}
\end{align}
Let us assume that $\Delta_i$ is fixed. Then we can substitute the variables and simplify the equation as
\begin{align}
\sigma&_{d, v}^{1, \mathcal{N}} = \frac{\pi \mu_{0}M_{s}^2 \Delta^2}{\mathcal{N}\mathcal{P}}\sum_{i = 0}^{\mathcal{N}-1} \sum_{i' = 0}^{\mathcal{N}-1}  \sin(\psi_i)\sin(\psi_{i'})  \left[G_v\left(\frac{\left|(i-{i'})\mathcal{P}+\mathcal{T}\right|}{2\pi \Delta}\right)+ G_v\left(\frac{\left|(i-{i'})\mathcal{P}-\mathcal{T}\right|}{2\pi \Delta}\right)  -2G_v\left(\frac{\left|(i-{i'})\mathcal{P}\right|}{2\pi \Delta}\right)\right], \label{eq:exactvolumebuttner}
\end{align}
\begin{align}
G_v&(\alpha)= \frac{1}{4}\int_{0}^{+\infty} dq \frac{e^{-q\alpha}+q\alpha-1}{q\cosh^2(q/4)}=\notag \\&-2\left\{\Psi^{-2}(\alpha+1)-\Psi^{-2}\left(\alpha+\frac{1}{2}\right)- \alpha \ \ln(\Gamma(\alpha+1)) \right.\notag\\&\left.
+\alpha \ln\left[\Gamma\left(\alpha+\frac{1}{2}\right)\right] - \Psi^{-2}(1)+\Psi^{-2}\left(\frac{1}{2}\right) 
\right\}, \label{eq:g}
\end{align}
where $\Psi^{-2}(z)=\int_0^z \d t \ln\Gamma(t)$ is the second anti-derivative of the digamma function. 

		\subsubsection{Surface-Surface stray field energy}
Consider now the surface-surface dipolar term of the isolated domain wall, which can be generally expressed as:
\begin{align}
\sigma_{d, s}^{N_{dw}=1, N_l=\mathcal{N}}= \frac{\mu_{0}}{8\pi  L  \mathcal{N}\mathcal{P}} \iint \d ^3 \mathbf{r} \d ^3 \mathbf{r'} \rho_s(\mathbf{r} ) \frac{1}{|\mathbf{r}-\mathbf{r'}|} \rho_s(\mathbf{r'} ) \label{eq:surfsurfsingle}
\end{align}
The surface charge density $\rho_s$ is defined by the the out-of-plane component of the magnetization :
\begin{align}
\rho_s(\mathbf{r}) = & M_s\sum_{j=0}^{\mathcal{N}-1}\pm  \tanh(x/\Delta_j) \left[\delta(z-\mathcal{P}j-\mathcal{T})- \delta(z-\mathcal{P}j)\right]\theta(L/2-|y-L/2|)\label{eq:surfacechargesingle}
\end{align}
Subtracting the energy of a state with a sharp domain wall, Eq. \eqref{eq:surfsurfsingle} then becomes:
\begin{multline}
\sigma_{d, s}^{1, \mathcal{N}}= \frac{\mu_{0}M_s^2}{8\pi  L \mathcal{N}\mathcal{P} } \sum_{j=0}^{\mathcal{N}-1} \sum_{j'=0}^{\mathcal{N}-1}\iint \d ^3 \mathbf{r} \d ^3 \mathbf{r'} \left[\tanh(x/\Delta_j)\tanh(x'/\Delta_{j'})-\rm{sgn}(x)\rm{sgn}(x') \right]\frac{1}{|\mathbf{r}-\mathbf{r'}|}\times \\\times \left[\delta(z-\mathcal{P}j-\mathcal{T})- \delta(z-\mathcal{P}j)\right]\left[\delta(z'-\mathcal{P}j'-\mathcal{T})- \delta(z'-\mathcal{P}j')\right]
\end{multline}
after the substitution $z-\mathcal{P}j\rightarrow z$, $z'-\mathcal{P}j'\rightarrow z'$, we get
\begin{multline}
\sigma_{d, s}^{1, \mathcal{N}}= \frac{\mu_{0}M_s^2}{8\pi  L \mathcal{N}\mathcal{P}} \sum_{j=0}^{\mathcal{N}-1} \sum_{j'=0}^{\mathcal{N}-1} \iint \d ^3 \mathbf{r} \d ^3 \mathbf{r'} \frac{\left[\tanh(x/\Delta_j)\tanh(x'/\Delta_{j'})-\rm{sgn}(x)\rm{sgn}(x') \right]}{\sqrt{(x-x')^2+(y-y')^2+(z-z'+(j-j')\mathcal{P})^2}}\times \\\times \left[\delta(z-\mathcal{T})- \delta(z)\right]\left[\delta(z'-\mathcal{T})- \delta(z')\right]
\end{multline}
With the tools provided in Ref.~\cite{buttner2015}, the integration along $y$ and $z$ can be performed analytically. In the limit $L\rightarrow\infty$, the integration kernel reads
\begin{align}
h_s\left(x,\mathcal{T}, (j-j')\mathcal{P}\right)&=\lim_{L\rightarrow\infty}\frac{1}{2L}\int_{0}^{L}\int_{0}^{L}\d y \d y^\prime\int_{0}^\mathcal{T}\int_{0}^\mathcal{T}\d z \d z^\prime \frac{(\delta(z-\mathcal{T})-\delta(z))(\delta(z^\prime-\mathcal{T})-\delta(z^\prime))}{\sqrt{x^2+(y-y^\prime)^2+(z-z'+(j-j')\mathcal{P})^2}}\notag\\
&=\frac{1}{2}\left[2h\left(x, (j-j')\mathcal{P}\right)-h\left(x, \mathcal{T}-(j-j')\mathcal{P}\right)-h\left(x, \mathcal{T}+(j-j')\mathcal{P}\right)\right],\label{eq:hs} 
\end{align}
where $h(x, z)=-\ln(x^2+z^2)$.
We thus obtain
\begin{align}
\sigma_{d, s}^{1, \mathcal{N}}= \frac{\mu_{0}M_s^2}{4\pi  \mathcal{N}\mathcal{P} } \sum_{j=0}^{\mathcal{N}-1} \sum_{j'=0}^{\mathcal{N}-1}  \iint_{-\infty}^{\infty}\d x \d x'  \left[\tanh(x/\Delta_j)\tanh(x'/\Delta_{j'})-\rm{sgn}(x)\rm{sgn}(x') \right] h_s\left(x-x', \mathcal{T}\right),
\end{align}
Introducing $m_j(x) = \tanh(x/\Delta_j)$, $f(x) = \rm{sgn}(x)$, we can use the property of convolution to reduce the double integral into a single integral in Fourier space:
\begin{align}
\sigma_{d, s}^{1, \mathcal{N}}= \frac{\mu_{0}M_s^2}{4\pi  \mathcal{N}\mathcal{P} } \sqrt{2\pi}\sum_{j=0}^{\mathcal{N}-1} \sum_{j'=0}^{\mathcal{N}-1}  \int_{-\infty}^{\infty}\d k \left[\hat{m_j}(k)\hat{m_{j'}}^*(k) -\hat{f}^2(k)\right] h_s\left(k, \mathcal{T}, (j-j')\mathcal{P}\right)
\end{align}
The Fourier space functions are
\begin{align}
\hat{m_j}(k)\hat{m_{j'}}^*(k) &= \frac{\pi \Delta_j \Delta_{j'}  }{2}\frac{1}{\sinh\left(\frac{\pi  \Delta_j k}{2}\right)\sinh\left(\frac{\pi  \Delta_{j'}  k}{2}\right)}\\
\hat{f}^2(k) &= \frac{2}{\pi k^2}\\
\hat{h}_s\left(k,\mathcal{T}, j\mathcal{P}\right)&= \frac{\sqrt{2\pi}}{2|k|}\left(2e^{-|j\mathcal{P}||k|}-e^{-|\mathcal{T}-j\mathcal{P}||k|}-e^{-|\mathcal{T}+j\mathcal{P}||k|}\right)
\end{align}
Now, collecting all the terms, we obtain the final result for the stray field energy of a multilayer film with an isolated domain wall with variable domain wall width and angle
\begin{align}
\sigma_{d, s}^{1, \mathcal{N}}& = \frac{\mu_{0}M_s^2}{2\pi  \mathcal{N}\mathcal{P} } \sum_{j=0}^{\mathcal{N}-1} \sum_{j'=0}^{\mathcal{N}-1}  \int_{-\infty}^{\infty}\d k \left[\frac{\pi^2 }{4}\Delta_j\Delta_{j'}  k^2 \frac{1}{\sinh\left(\frac{\pi  \Delta _j k}{2}\right)\sinh\left(\frac{\pi  \Delta _{j'} k}{2}\right)} -1\right] \frac{1}{|k|^3}\times \notag\\ &\times \left(2e^{-|(j-j')\mathcal{P}||k|}-e^{-|\mathcal{T}-(j-j')\mathcal{P}||k|}-e^{-|\mathcal{T}+(j-j')\mathcal{P}||k|}\right)
\end{align}
Since the function is even with respect to k,
\begin{align}
\sigma_{d, s}^{1, \mathcal{N}}& = \frac{\mu_{0}M_s^2}{\pi  \mathcal{N}\mathcal{P} } \sum_{j=0}^{\mathcal{N}-1} \sum_{j'=0}^{\mathcal{N}-1}  \int_{0}^{\infty}\d k \left[\frac{\pi^2 }{4}\frac{\Delta_j \Delta_{j'}  k^2 }{\sinh\left(\frac{\pi  \Delta _j k}{2}\right)\sinh\left(\frac{\pi  \Delta _{j'} k}{2}\right)} -1\right] \times\frac{ 2e^{-|(j-j')\mathcal{P}|k}-e^{-|\mathcal{T}-(j-j')\mathcal{P}|k}-e^{-|\mathcal{T}+(j-j)\mathcal{P}|k}}{k^3}
\end{align}

If we assume that $\Delta$ is fixed, then we can use the analytical integration to reduce it to:
\begin{align}
\sigma_{d, s}^{1, \mathcal{N}}& = \frac{\mu_{0}M_s^2}{\pi  \mathcal{N}\mathcal{P} } \sum_{j=0}^{\mathcal{N}-1} \sum_{j'=0}^{\mathcal{N}-1}  \int_{0}^{\infty}\d k \left[\frac{\pi^2 }{4}\frac{\Delta^2  k^2 }{\sinh^2\left(\frac{\pi  \Delta k}{2}\right)} -1\right] \times\frac{ 2e^{-|(j-j')\mathcal{P}|k}-e^{-|\mathcal{T}-(j-j')\mathcal{P}|k}-e^{-|\mathcal{T}+(j-j')\mathcal{P}|k}}{k^3}
\end{align}
Substituting $q =2\pi \Delta k$, $t = \frac{\mathcal{T}}{2\pi \Delta}$, $p = \frac{\mathcal{P}}{2\pi \Delta}$ we obtain:
\begin{align}
\sigma_{d, s}^{1, \mathcal{N}}& = \frac{\mu_{0}M_s^2}{\pi  \mathcal{N}\mathcal{P} } \sum_{j=0}^{\mathcal{N}-1} \sum_{j'=0}^{\mathcal{N}-1}  \int_{0}^{\infty}\d q \frac{1}{2\pi \Delta}\left[\frac{1 }{16}\frac{q^2 }{\sinh^2\left(\frac{q}{4}\right)} -1\right] \times\frac{ 2e^{-|(j-j')p|q}-e^{-|t-(j-j')p|q}-e^{-|t+(j-j')p|q}}{q^3}8\pi^3\Delta^3
\end{align}
or
\begin{align}
\sigma_{d, s}^{1, \mathcal{N}}& = \frac{4\pi \mu_{0}M_s^2\Delta^2}{ \mathcal{N}\mathcal{P} } \sum_{j=0}^{\mathcal{N}-1} \sum_{j'=0}^{\mathcal{N}-1}  \int_{0}^{\infty}\d q \left[\frac{1 }{16}\frac{q^2 }{\sinh^2\left(\frac{q}{4}\right)} -1\right] \times\frac{ 2e^{-|(j-j')p|q}-e^{-|t-(j-j')p|q}-e^{-|t+(j-j')p|q}}{q^3}
\end{align}
Finally, the surface-surface term can be expressed in more compact way as:
\begin{align}
\sigma_{d, s}^{1, \mathcal{N}}& = -\frac{\pi \mu_{0}M_s^2\Delta^2}{\mathcal{N}\mathcal{P} } \sum_{j=0}^{\mathcal{N}-1} \sum_{j'=0}^{\mathcal{N}-1}  \left[G_s\left(\frac{\left|(j-{j'})\mathcal{P}+\mathcal{T}\right|}{2\pi \Delta}\right)+ G_s\left(\frac{\left|(j-{j'})\mathcal{P}-\mathcal{T}\right|}{2\pi \Delta}\right)  -2G_s\left(\frac{\left|(j-{j'})\mathcal{P}\right|}{2\pi \Delta}\right)\right],
\end{align}
where the integral
\begin{align}
G_s(x) & =4\int_{0}^{\infty}\d q\frac{ e^{-q x}-1}{q^3}\left(\frac{q^2}{16\sinh ^2(\frac{q}{4})}-1\right)\notag\\
&=\Psi ^{(-2)}(2 x)+x^2 (2  \log (x)+\log (4)-1)-x(1+2  \ln[\Gamma (2 x)])
\end{align}
is solved analytically as shown below in the Section~\ref{sec:Gs}. 

		\subsubsection{Surface-Volume stray field energy}
Unlike in single layer film, in which mutual surface-volume stray field interactions are cancelled out due to the symmetry~\cite{Lemesh2017} of the system, multilayers should be treated differently. If the domain wall angle $\psi_i$ changes from layer to layer in the multilayer film with the multidomain state, then the state becomes asymmetric, so the surface $\rho_s$ and the volume  charges $\rho_v$ will start to interact.
For an isolated domain wall, the integral of interest is:
\begin{align}
\sigma_{d, sv}^{N_{dw}=1, N_l=\mathcal{N}}= 2\times \frac{\mu_{0}}{8\pi  L  \mathcal{N}\mathcal{P}} \iint \d ^3 \mathbf{r} \d ^3 \mathbf{r'} \rho_s(\mathbf{r} ) \frac{1}{|\mathbf{r}-\mathbf{r'}|} \rho_v(\mathbf{r'} ), \label{eq:surfvolsingle}
\end{align}
where the factor of 2 accounts for the fact that there is no double-counting, when we deal with interactions between the charges of different kind. Plugging the expressions for the respective charges from Eqs.~\eqref{eq:surfacechargesingle},~\eqref{eq:volumechargesingle}, we thus have
\begin{multline}
\sigma_{d, sv}^{1, \mathcal{N}}= 2\frac{\mu_{0}M_s^2}{8\pi  L \mathcal{N}\mathcal{P}} \sum_{i=0}^{\mathcal{N}-1} \sum_{j=0}^{\mathcal{N}-1}\sin(\psi_i)\iint \d ^3 \mathbf{r} \d ^3 \mathbf{r'} \left[\pm \tanh\left(\frac{x'}{\Delta_j}\right)\right] \frac{\tanh(x/\Delta_i)}{\Delta_i\cosh(x/\Delta_i)}\frac{1}{|\mathbf{r}-\mathbf{r'}|}\times \\\times \left[\delta(z'-\mathcal{P}j-\mathcal{T})- \delta(z'-\mathcal{P}j)\right]\theta(\mathcal{T}/2-|z-\mathcal{P}i-\mathcal{T}/2|)
\end{multline}
after the substitution $z'-\mathcal{P}j\rightarrow z'$, $z-\mathcal{P}i\rightarrow z$ we get
\begin{multline}
\sigma_{d, sv}^{1, \mathcal{N}}= \pm\frac{\mu_{0}M_s^2}{4\pi  L \mathcal{N}\mathcal{P} } \sum_{i=0}^{\mathcal{N}-1} \sum_{j=0}^{\mathcal{N}-1}\sin(\psi_i) \iint \d ^3 \mathbf{r} \d ^3 \mathbf{r'} \frac{\tanh\left(\frac{x'}{\Delta_j}\right) \frac{\tanh(x/\Delta_i)}{\Delta_i\cosh(x/\Delta_i)}}{\sqrt{(x-x')^2+(y-y')^2+(z-z'+(i-j)\mathcal{P})^2}}\times \\\times \left[\delta(z'-\mathcal{T})- \delta(z')\right]\theta(\mathcal{T}/2-|z-\mathcal{T}/2|)
\end{multline}
Now, our system is infinite in $y$ direction, so after evaluating the integral
\begin{align}
h_{sv}(x, z,\mathcal{T}, (i-j)\mathcal{P}) =  &\lim\limits_{L\rightarrow \infty} \frac{1}{L}\int_{0}^{L}dy\int_{0}^{L}dy'\int_{0}^{\mathcal{T}}dz' \frac{\delta(z'-\mathcal{T})- \delta(z')}{\sqrt{x^2+(y-y')^2+(z-z'+(i-j)\mathcal{P})^2}} =\notag\\&=h(x, z-\mathcal{T}+(i-j)\mathcal{P})-h(x, z+(i-j)\mathcal{P}) \notag\\
& = -\ln\left(\frac{x^2+(z-\mathcal{T}+(i-j)\mathcal{P})^2}{x^2+\left(z+(i-j)\mathcal{P}\right)^2}\right),
\end{align}
in which we used $h$ defined as \cite{buttner2015}:
\begin{equation}
h(x, z)=\lim\limits_{L\rightarrow \infty} \frac{1}{L}\int_{0}^L \d y\int_{0}^L\d y'\frac{1}{\sqrt{(y-y')^2+x^2+z^2}}=-\ln(x^2+z^2)-2+2\ln(2L)+\mathcal{O}(L^{-1})\label{eq:hone}
\end{equation}
We then obtain
\begin{align}
\sigma_{d, sv}^{1, \mathcal{N}}= \pm\frac{\mu_{0}M_s^2}{4\pi  \mathcal{N}\mathcal{P}} \sum_{i=0}^{\mathcal{N}-1} \sum_{j=0}^{\mathcal{N}-1}\sin(\psi_i)\int_{0}^{\mathcal{T}}\d z \iint_{-\infty}^{\infty} \d x \d x'  \tanh\left(\frac{x'}{\Delta_j}\right) \frac{\tanh(x/\Delta_i)}{\Delta_i\cosh(x/\Delta_i)} h_{sv}\left(x-x', z, \mathcal{T}, (i-j)\mathcal{P}\right)
\end{align}
Since both $\Lambda_v=\frac{\tanh(x/\Delta_i)}{\Delta_i\cosh(x/\Delta_i)}$ and $\Lambda_s=\tanh\left(\frac{x'}{\Delta_j}\right)$ are the functions that have analytical form of the Fourier transform, we can use the property of convolution to reduce the double integral to a single integral in $k$ space:
\begin{align}
\sigma_{d, sv}^{1, \mathcal{N}}=\pm \frac{\mu_{0}M_s^2\sqrt{2\pi}}{4\pi  \mathcal{N}\mathcal{P} } \sum_{i=0}^{\mathcal{N}-1} \sum_{j=0}^{\mathcal{N}-1}\sin(\psi_i)\int_{0}^{\mathcal{T}}\d z \int \d k \  \hat{\Lambda}_{s,k} \hat{\Lambda}_{v, k}^*\hat{h}_{sv}\left(k, z, \mathcal{T}, (i-j)\mathcal{P}\right)
\end{align}
where the Fourier coefficients for $\Lambda_s$, $\Lambda_v$ and the Fourier transform of $h_{sv}$ are 
\begin{align}
\hat{\Lambda}_{s,k}&= i \Delta_j \sqrt{\frac{\pi}{2}}\frac{1}{\sinh\left(\frac{\pi \Delta_j k}{2}\right)},\\
\hat{\Lambda}_{v, k}&= i k \Delta_i\sqrt{\frac{\pi}{2}}\frac{1}{\cosh \left(\frac{\pi \Delta_i k}{2}\right)},\\
\hat{\Lambda}_{s,k}\hat{\Lambda}_{v, k}^*&=   \frac{\pi k \Delta_i \Delta_j}{2\sinh \left(\frac{\pi \Delta_j k}{2}\right)\cosh \left(\frac{\pi \Delta_i k}{2}\right)},\\
\hat{h}_{sv}\left(k, z, \mathcal{T}, (i-j)\mathcal{P}\right) &= \frac{\sqrt{2\pi}}{|k|}\left(e^{-|z-\mathcal{T}+(i-j)\mathcal{P}||k|}-e^{-|z+(i-j)\mathcal{P}||k|}\right)
\end{align}
Thus, using the fact that the integrand is an even function, we can find:
\begin{align}
\sigma_{d, sv}^{1, \mathcal{N}}=\pm \frac{\pi\mu_{0}M_s^2}{2\mathcal{N}\mathcal{P}} \sum_{i=0}^{\mathcal{N}-1} \sum_{j=0}^{\mathcal{N}-1}\Delta_i\Delta_j\sin(\psi_j)\int_{0}^{\infty} \d k\frac{  1}{\sinh\left(\frac{\pi \Delta_j k}{2}\right)\cosh\left(\frac{\pi \Delta_i k}{2}\right)}\int_{0}^{\mathcal{T}}\d z \left(e^{-k|z-\mathcal{T}+(i-j)\mathcal{P}|}-e^{-k|z+(i-j)\mathcal{P}|}\right)
\end{align}
The integral over $z'$ can be carried out easily, since the multilayer period is always larger than the single magnetic layer thickness:
\begin{align}
\int_{0}^{\mathcal{T}}\d z \left(e^{-k|z-\mathcal{T}+(i-j)\mathcal{P}|}-e^{-k|z+(i-j)\mathcal{P}|}\right) = \begin{cases}0, & i=j \\-\frac{4\sinh^2(\frac{k\mathcal{T}}{2})e^{-k\mathcal{P}|i-j|}}{k},&i < j \\ \frac{4\sinh^2(\frac{k\mathcal{T}}{2})e^{-k\mathcal{P}|i-j|}}{k}, & i>j \end{cases}
\end{align}
Note that the system possesses no surface-volume interactions between charges of one specific layer. Thus, we can shorten the expression as follows:
\begin{align}
\sigma_{d, sv}^{1, \mathcal{N}}& = \pm\frac{2\pi \mu_{0}M_s^2}{\mathcal{N}\mathcal{P}} \sum_{i=0}^{\mathcal{N}-1}\sum_{j=0}^{\mathcal{N}-1} \Delta_i\Delta_j\sin(\psi_i)\text{sgn}(i-j)\int_{0}^{\infty}dk\frac{1}{k}\frac{\sinh^2(\frac{k\mathcal{T}}{2}) e^{-k\mathcal{P}|i-j|} }{\sinh\left(\frac{\pi \Delta_j k}{2}\right)\cosh\left(\frac{\pi \Delta_i k}{2}\right)}\notag\\
& = \pm\frac{2\pi \mu_{0}M_s^2}{\mathcal{N}\mathcal{P}} \sum_{j=0}^{\mathcal{N}-1}\sum_{i=0}^{\mathcal{N}-1} \Delta_i \Delta_j \sin(\psi_i)\text{sgn}(i-j)\int_{0}^{\infty}dk\frac{1}{k}\frac{  e^{-k|(i-j)P+\mathcal{T}|}+e^{-k|(i-j)P-\mathcal{T}|}-2 e^{-k\mathcal{P}|i-j|} }{4\sinh\left(\frac{\pi \Delta_j k}{2}\right)\cosh\left(\frac{\pi \Delta_i k}{2}\right)}\notag \\
\end{align}

Assuming $\Delta$ is constant, the integral can be solved analytically, resulting in:
\begin{align}
\sigma_{d, sv}^{1, \mathcal{N}} & = \pm\frac{4\pi \mu_{0}M_s^2}{\mathcal{N}\mathcal{P}} \sum_{j=0}^{\mathcal{N}-1}\sum_{i=0}^{\mathcal{N}-1} \Delta^2 \sin(\psi_i)\text{sgn}(i-j)\int_{0}^{\infty}dk\frac{1}{k}\frac{\sinh^2(\frac{k\mathcal{T}}{2}) e^{-k\mathcal{P}|i-j|} }{\sinh\left(\pi \Delta k\right)}\notag \\
& = \pm\frac{\pi \mu_{0}M_s^2}{\mathcal{N}\mathcal{P}} \sum_{j=0}^{\mathcal{N}-1}\sum_{i=0}^{\mathcal{N}-1} \Delta^2 \sin(\psi_i)\text{sgn}(i-j)\int_{0}^{\infty}dk\frac{1}{k}\frac{  e^{k(\mathcal{T}-P|i-j|)}+e^{k(-\mathcal{T}-P|i-j|)}-2 e^{-k\mathcal{P}|i-j|} }{\sinh\left(\pi \Delta k\right)}\notag \\
\end{align}
Finally, solving the integral analytically as shown below, we obtain:
\begin{align}
\sigma_{d, sv}^{1, \mathcal{N}} &=\pm\frac{\pi \mu_{0}M_s^2\Delta^2}{\mathcal{N}\mathcal{P}} \sum_{i=0}^{\mathcal{N}-1}\sum_{j=0}^{\mathcal{N}-1}  \sin(\psi_i)\text{sgn}(i-j)\left[G_{sv}\left(\frac{\left|(i-j)\mathcal{P}+\mathcal{T}\right|}{2\pi \Delta}\right)+ G_{sv}\left(\frac{\left|(i-j)\mathcal{P}-\mathcal{T}\right|}{2\pi \Delta}\right)  -2G_{sv}\left(\frac{\left|(i-j)\mathcal{P}\right|}{2\pi \Delta}\right)\right]
\end{align}
with 
\begin{align}
G_{sv}(x) = 2\ln\left[\Gamma\left(x+\frac{1}{2}\right)\right]
\end{align}

		\subsubsection{Integral $\mathbf{G_{sv}}$}\label{sec:Gsv}
The following integral reduce to:
\begin{align}
\int_{0}^{\infty}dk \frac{1}{k}\frac{\sinh^2(a k)(e^{-bk}+bk-1)}{\sinh(ck)} & = \int_{0}^{b} \d b'\int_{0}^{\infty}\d k\frac{\sinh^2(a k)(1-e^{-bk})}{\sinh(ck)} \notag\\&=\int_{0}^{b} \d b' \frac{\psi ^{(0)}\left(\frac{a+b'}{2 a}\right)+\gamma +\log (4)}{a}\notag \\ &=\frac{x (\gamma +\log (4))}{a}+2 \log \left(\Gamma \left(\frac{a+b}{2 a}\right)\right)-\log (\pi )
\end{align}
where we used the fact that
\begin{align}
\int_{0}^{\infty}dk \frac{1}{k}\frac{\sinh^2(a k)e^{-bk}}{\sinh(ck)} &= \frac{1}{2}\ln\left[\frac{\Gamma\left(\frac{1}{2}+\frac{b-2a}{2c}\right)\Gamma\left(\frac{1}{2}+\frac{b+2a}{2c}\right)}{\Gamma^2\left(\frac{1}{2}+\frac{b}{2c}\right)}\right]
\end{align}

		\subsubsection{Integral $\mathbf{G_s}$}	 \label{sec:Gs}
The following integration can be reduced to analytical form using a similar approach as in Appendix D of Ref.~\cite{buttner2015}:		 
\begin{align}
\int_{0}^{\infty}\d q\frac{ e^{-q x}-1}{q^3}\left(\frac{q^2}{\sinh ^2(\frac{q}{4})}-1\right)\notag\\
&=\frac{1}{4}\left[\Psi ^{(-2)}(2 x)+x^2 (2  \log (x)+\log (4)-1)-x(1+2  \ln[\Gamma (2 x)])\right]
\end{align}

\section{Theory of magnetic domains with twisted domain walls in multilayers }
	\subsection{Magnetostatic energy of multidomain state with twisted domain walls}
Now, consider a multilayer film with the periodic stripe domain pattern of periodicity $\lambda$ and a width $W$ of one of the domains. Assuming that we have an infinite film (in x and y direction) with large number $\mathcal{M}$ of domains, corresponding to $2\mathcal{M}$  domain walls, we are then interested in calculating the total magnetostatic energy normalized per single domain wall per single multilayer repeat:
\begin{align}
\sigma_{d}^{N_{dw}=\infty, N_l=\mathcal{N}}= \sum_{\alpha \beta = vv, ss, sv, vs}\lim\limits_{M\rightarrow \infty}\lim\limits_{L\rightarrow \infty}\frac{\mu_{0}}{16\pi  L  \mathcal{N}\mathcal{P}\mathcal{M}} \iint \d ^3 \mathbf{r} \d ^3 \mathbf{r'} \rho_\alpha(\mathbf{r} ) \frac{1}{|\mathbf{r}-\mathbf{r'}|} \rho_\beta(\mathbf{r'} ), \label{eq:magdomains}
\end{align}		
As we will see below, the total magnetostatic energy of such magnetized multidomain multilayers can be expressed as 
\begin{align}
\sigma_{d}^{\infty, \mathcal{N}}&=  \frac{\lambda}{4}\mu_0M_s^2 \left(\frac{2W}{\lambda}-1\right)^2\frac{\mathcal{T}}{\mathcal{P}}+ \sum_{i=0}^{\mathcal{N}-1}\sum_{j=0}^{\mathcal{N}-1} \tilde{F}_{s, ij} (\mathcal{T},\mathcal{P}, \Delta, \lambda, W)\notag\\&+ \sum_{i=0}^{\mathcal{N}-1}\sum_{j=0}^{\mathcal{N}-1} \left\{ \sin(\psi_i)\sin(\psi_j)\tilde{F}_{v, ij} (\mathcal{T},\mathcal{P}, \Delta, \lambda, W)+ \sin(\psi_i)\text{sgn}(i-j)\tilde{F}_{sv, ij} (\mathcal{T},\mathcal{P}, \Delta, \lambda, W)\right\}
\end{align}
with a generic  function $\tilde{F}_{\alpha, ij} $ and its dependencies  defined as follows
\begin{align}
\tilde{F}_{\alpha, ij} & = \frac{\pi \mu_{0}M_s^2\Delta^2}{\mathcal{N}\mathcal{P} }\sum_{n=1}^{\infty}\frac{  \sin^2\left(\frac{\pi n W}{\lambda}\right)}{n}\tilde{G}_{\alpha, ijn}(\mathcal{T},\mathcal{P}, \Delta, \lambda),\\
\tilde{G}_{v, ijn} & = \frac{2\sinh^2(\frac{\pi n \mathcal{T}}{\lambda})e^{-\frac{2\pi n \mathcal{P}|i-j|}{\lambda}}(1-\delta_{ij})+(e^{-\frac{2\pi n \mathcal{T}}{\lambda}}+\frac{2\pi n \mathcal{T}}{\lambda}-1)\delta_{ij}}{\cosh^2\left(\frac{\pi^2 n \Delta }{\lambda}\right)}\\
\tilde{G}_{s, ijn} & = \frac{2e^{-\frac{2\pi |(i-j)\mathcal{P}|n}{\lambda}}-e^{-\frac{2\pi|\mathcal{T}-(i-j)\mathcal{P}|n}{\lambda}}-e^{-\frac{2\pi |\mathcal{T}+(i-j)\mathcal{P}|n}{\lambda}}}{ 2\sinh^2\left(\frac{\pi^2 n \Delta}{\lambda}\right)}\\
\tilde{G}_{sv, ijn} & = \frac{8\sinh^2(\frac{\pi n \mathcal{T}}{\lambda})e^{-\frac{2\pi n\mathcal{P}|i-j|}{\lambda}} }{\sinh\left(\frac{2\pi^2 n \Delta}{\lambda}\right)}
\end{align}
The derived here expressions are valid as long as $W>7\Delta$~\cite{Lemesh2017}. Assuming the magnetic field applied in z direction (in the absence of currents), the total volumetric energy per single domain wall per layer therefore can be expressed as:
\begin{align}\mathcal{E}_{tot}^{ \infty, \mathcal{N}}(\lambda, W, \Delta, \psi_i)=\frac{2}{\lambda}\left[     \frac{2A}{\Delta}f+2K_{u}\Delta f -M_s\left(1-\frac{2W}{\lambda}\right)B_z\frac{f\lambda}{2}+\sigma_{d}^{\infty, \mathcal{N}}(\lambda, W, \Delta, \psi_i)-\frac{\pi D  f}{\mathcal{N}}  \sum_{i=0}^{\mathcal{N}-1}\sin(\psi_i) \right]
\end{align}

	\subsection{Statics of magnetic domains}
By performing the energy minimization, we will have the system of four equations:
\begin{align}
\sum_{i=0}^{\mathcal{N}-1}\sum_{j=0}^{\mathcal{N}-1} \left\{\left[\tilde{F}_{s, ij}-\lambda\frac{\partial \tilde{F}_{s, ij}}{\partial \lambda} \right]+\sin(\psi_{i})  \sin(\psi_{j}) \left[\tilde{F}_{v, ij}-\lambda\frac{\partial \tilde{F}_{v, ij}}{\partial \lambda} \right]  + \sin(\psi_{i})  \text{sgn}(i-j)\left[\tilde{F}_{sv, ij} -\lambda\frac{\partial \tilde{F}_{sv, ij}}{\partial \lambda} \right]\right\}\notag\\
+\left[\frac{2A}{\Delta}f+2K_{u}\Delta f- \frac{\pi D  f}{\mathcal{N}}  \sum_{i=0}^{\mathcal{N}-1}\sin(\psi_i)+WM_sB_zf+\mu_0M_s^2Wf\left(\frac{2W}{\lambda}-1\right) \right]=0\label{eq:domlambda}
\end{align}     
\begin{align}
M_s f \left[B_z+\mu_0M_s\left(\frac{2W}{\lambda}-1\right)\right]+\sum_{i=0}^{\mathcal{N}-1}\sum_{j=0}^{\mathcal{N}-1} \left\{\frac{\partial \tilde{F}_{s, ij}}{\partial W} +\sin(\psi_{i})  \sin(\psi_{j})  \frac{\partial \tilde{F}_{v, ij}}{\partial W}   + \sin(\psi_{i})  \text{sgn}(i-j)\frac{\partial \tilde{F}_{sv, ij}}{\partial W} \right\}=0 \label{eq:domw}
\end{align}  
\begin{align}
-\frac{2A}{\Delta^2}f+2K_{u} f+ \sum_{i=0}^{\mathcal{N}-1}\sum_{j=0}^{\mathcal{N}-1} \frac{\partial \tilde{F}_{s, ij}}{\partial \Delta} +\sum_{i=0}^{\mathcal{N}-1}\sum_{j=0}^{\mathcal{N}-1}\sin(\psi_{i})  \sin(\psi_{j})  \frac{\partial \tilde{F}_{v, ij}}{\partial \Delta}   + \sum_{i=0}^{\mathcal{N}-1}\sum_{j=0}^{\mathcal{N}-1}  \sin(\psi_{i})  \text{sgn}(i-j)\frac{\partial \tilde{F}_{sv, ij}}{\partial \Delta} =0\label{eq:domdelta}
\end{align}     
\begin{align}
-\frac{1}{\mathcal{N}}  \pi D  f  + \sum_{j=0}^{\mathcal{N}-1}\left(1+\delta_{ij}\right)   \tilde{F}_{v, ij} (\Delta, W)\sin(\psi_{j})+\sum_{j=0}^{\mathcal{N}-1} \tilde{F}_{sv, ij}(\Delta, W)\text{sgn}(i-j) =0\label{eq:dompsi0}
\end{align}
The number of independent equations can be reduced from four to three, since similarly to the isolated wall case (Eq.~\eqref{eq:psi}), the variable $\psi_i$ from Eq.~\eqref{eq:dompsi0} can be disentangled using similar matrix operations:
\begin{align}
\sin(\psi_{i}) =   \tilde{f}\left(\sum_{j=0}^{\mathcal{N}-1} [\left(1+\delta_{ij}\right)   \tilde{F}_{v, ij}(\Delta, W)]^{-1}\left[   \frac{\pi D  f}{\mathcal{N}}   1_j -\sum_{k=0}^{\mathcal{N}-1} \tilde{F}_{sv, jk}(\Delta, W)\text{sgn}(j-k) \right]\right)\label{eq:psi2}
\end{align}

	\subsection{Derivation of the stray field energy of magnetic domains with twisted walls}\label{sec:derivations}
	   \subsubsection{Surface-Surface stray field energy}
	   The surface stray field is generated by the following surface charge distribution:
	   \begin{align}
	   \rho_s(\mathbf{r}) = & \sum_{i=0}^{\mathcal{N}-1} \Lambda_s(x, \lambda, W)\left[\delta(z-\mathcal{P}i-\mathcal{T})- \delta(z-\mathcal{P}i)\right]\theta(L/2-|y-L/2|)\label{eq:surfacecharge}\\
	   &\Lambda_s(x, \lambda, W) =  M_s\sum_{m =- \mathcal{M}+1}^{\mathcal{M}-1}\left[1+\tanh\left(\frac{x-m'\lambda}{\Delta}\right)-\tanh\left(\frac{x-m'\lambda+W}{\Delta}\right)\right]
	   \end{align}
	   \begin{figure}[h!]
	   	\includegraphics[scale = 0.67]{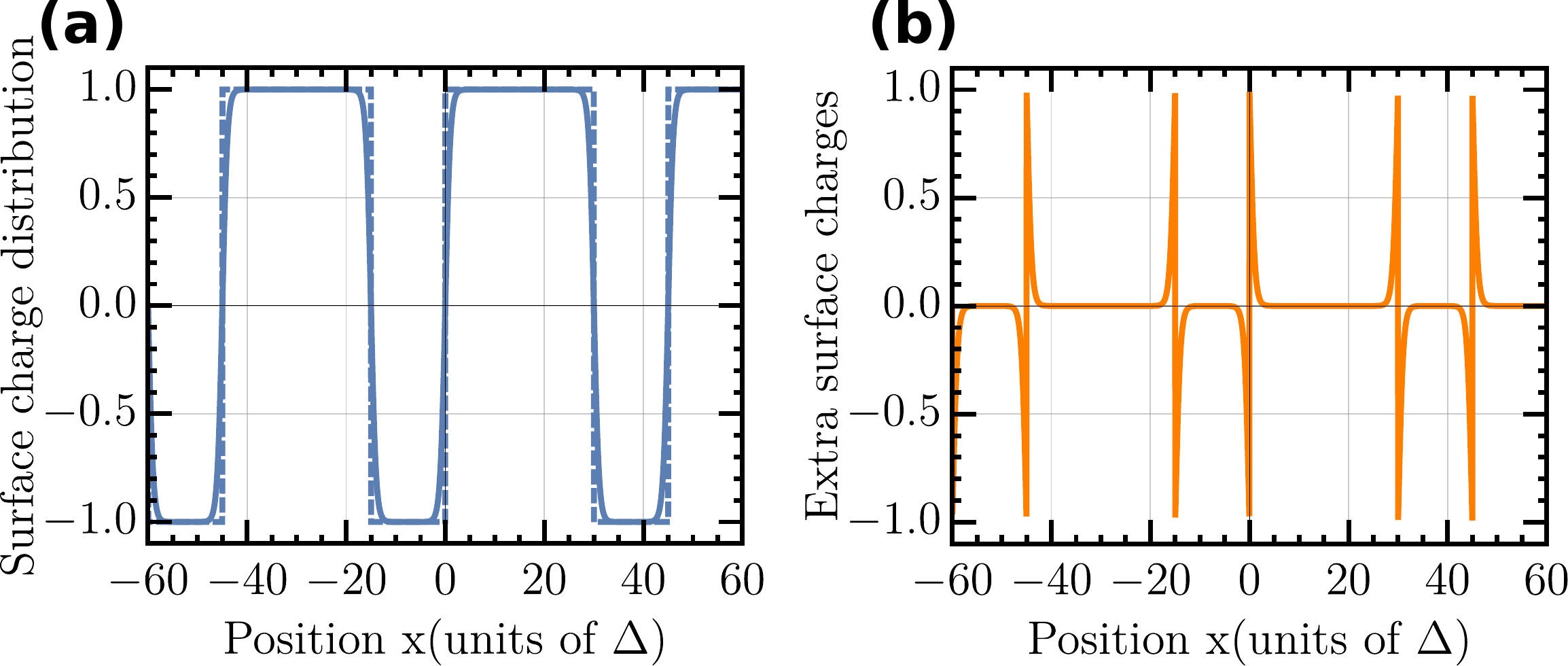}
	   	\caption{{\bf Distribution of surface charges in a magnetic layer with multidomain stripe state} with periodicity $\lambda=45\Delta$ and minority domain size of $W_{\text{\rm{min}}}=15\Delta$ possessing domain walls. {\bf (a)} Absolute values given by our theory (continuous lines) and by the binary stripe state (dashed lines) and {\bf (b)} their relative difference. }\label{fig:charges}
	   \end{figure}
	   Note that the function $\Lambda_{s}$ can be expressed as a convolved binary square wave $\Pi$:
	   \begin{align}
	   \Lambda_s(x, \lambda, W) & =(\rho_1*\Pi) (x) \\
	   & \Pi(x, \lambda, W) = \sum_{m = -\mathcal{M}+1}^{\mathcal{M}-1} \left[2\theta(x-m \lambda) - \theta(x-m\lambda+W)-\theta(x-m\lambda-\lambda+W)\right]\\
	   & \rho_1(x, \Delta) = \frac{M_{s}}{2\Delta}\frac{1}{\cosh^2(x/\Delta)}
	   \end{align}
Surface-surface component of Eq.~\eqref{eq:magdomains} for large number of domain walls then becomes:
	   \begin{multline}
	   \sigma_{d, s}^{2\mathcal{M}, \mathcal{N}}= \frac{\mu_{0}}{16\pi  L \mathcal{N}\mathcal{P} \mathcal{M}} \sum_{i=0}^{\mathcal{N}-1} \sum_{i'=0}^{\mathcal{N}-1}\iint \d ^3 \mathbf{r} \d ^3 \mathbf{r'} \Lambda_s(x, \lambda, W)\Lambda_s(x', \lambda, W)\frac{1}{|\mathbf{r}-\mathbf{r'}|}\times \\\times \left[\delta(z-\mathcal{P}i-\mathcal{T})- \delta(z-\mathcal{P}i)\right]\left[\delta(z'-\mathcal{P}i'-\mathcal{T})- \delta(z'-\mathcal{P}i')\right]
	   \end{multline}
	   after the substitution $z-\mathcal{P}j\rightarrow z$, $z'-\mathcal{P}j'\rightarrow z'$ we get
	   \begin{multline}
	   \sigma_{d, s}^{2\mathcal{M}, \mathcal{N}}= \frac{\mu_{0}}{16\pi  L \mathcal{N}\mathcal{P} \mathcal{M}} \sum_{i=0}^{\mathcal{N}-1} \sum_{i'=0}^{\mathcal{N}-1} \iint \d ^3 \mathbf{r} \d ^3 \mathbf{r'} \frac{\Lambda_s(x, \lambda, W)\Lambda_s(x', \lambda, W)}{\sqrt{(x-x')^2+(y-y')^2+(z-z'+(i-i')\mathcal{P})^2}}\times \\\times \left[\delta(z-\mathcal{T})- \delta(z)\right]\left[\delta(z'-\mathcal{T})- \delta(z')\right]
	   \end{multline}
	   With the tools provided in Ref.~\cite{buttner2015}, the integration along $y$ and $z$ can be performed analytically. In the limit $L\rightarrow\infty$, the integration kernel reads
	   \begin{align}
	   h_s\left(x,\mathcal{T}, (i-i')\mathcal{P}\right)&=\lim_{L\rightarrow\infty}\frac{1}{2L}\int_{0}^{L}\int_{0}^{L}\d y \d y^\prime\int_{0}^\mathcal{T}\int_{0}^\mathcal{T}\d z \d z^\prime \frac{(\delta(z-\mathcal{T})-\delta(z))(\delta(z^\prime-\mathcal{T})-\delta(z^\prime))}{\sqrt{x^2+(y-y^\prime)^2+(z-z'+(i-i')\mathcal{P})^2}}\notag\\
	   &=\frac{1}{2}\left[2h\left(x, (i-i')\mathcal{P}\right)-h\left(x, \mathcal{T}-(i-i')\mathcal{P}\right)-h\left(x, \mathcal{T}+(i-i')\mathcal{P}\right)\right],\label{eq:hs} 
	   \end{align}
	   where $h(x, z)=-\ln(x^2+z^2)$. We thus obtain
	   \begin{align}
	   \sigma_{d, s}^{2\mathcal{M}, \mathcal{N}}= \frac{\mu_{0}}{8\pi  \mathcal{N}\mathcal{P} \mathcal{M}} \sum_{i=0}^{\mathcal{N}-1} \sum_{i'=0}^{\mathcal{N}-1}  \iint_{-\infty}^{\infty}\d x \d x'  \Lambda_s(x, \lambda, W)\Lambda_s(x', \lambda, W) h_s\left(x-x', \mathcal{T}\right)
	   \end{align}
	   Recognizing that $\Lambda_s$  is a periodic function in the limit $\mathcal{M}\rightarrow \infty$, we can use Eq. \eqref{eq:fourier} to reduce the double integral to the sum in $k$ space:
	   \begin{align}
	   \sigma_{d, s}^{\infty, \mathcal{N}}= \frac{\mu_{0}\sqrt{2\pi}\mathcal{M}\lambda }{8\pi  \mathcal{N}\mathcal{P} \mathcal{M}} \sum_{i=0}^{\mathcal{N}-1} \sum_{i'=0}^{\mathcal{N}-1} \sum_{k}\hat{\Lambda}_{s,k} \hat{\Lambda}_{s, k}^* \hat{h}_s\left(k, \mathcal{T}, (i-i')\mathcal{P}\right)\label{eq:surfacesurfacetriple}
	   \end{align}
	   The Fourier coefficients for $\Lambda_s$ can be found using the derived property of convolution (Eq. \eqref{eq:fourier_convolution}),  
	   \begin{align}
	   \hat{\Lambda}_{s, k}[\hat{\Lambda}_{s, k}]^{*} & = \begin{cases}\frac{4\pi^2 M_s^2 \Delta_i \Delta_{i'} \sin^2\left(\frac{kW}{2} \right)}{\lambda \sinh\left(\frac{\pi \Delta_i k}{2}\right)\sinh\left(\frac{\pi \Delta_{i'} k}{2}\right)}, & k \neq 0 \\ M_s^2\left(1-\frac{2W}{\lambda}\right)^2, & k=0 \\\end{cases}\label{eq:lambs}
	   \end{align}
	   and the Fourier space function of $h_s$ is:
	   \begin{align}
	   \hat{h}_s\left(k,\mathcal{T}, j\mathcal{P}\right)&= \frac{\sqrt{2\pi}}{2|k|}\left(2e^{-|j\mathcal{P}||k|}-e^{-|\mathcal{T}-j\mathcal{P}||k|}-e^{-|\mathcal{T}+j\mathcal{P}||k|}\right)
	   \end{align}
	   We can use $\Lambda_{s}$ from Eq. \eqref{eq:lambs} and the fact that the function under the sum remains the same after the substitution $k\rightarrow -k$, resulting in:
	   \begin{align}
	   \sigma_{d, s}^{\infty, \mathcal{N}}&= \frac{\lambda}{4}\mu_0M_s^2 \left(\frac{2W}{\lambda}-1\right)^2\frac{\mathcal{T}}{\mathcal{P}}+\notag\\+& \frac{2\mu_{0}\sqrt{2\pi}\mathcal{M}\lambda }{8\pi  \mathcal{N}\mathcal{P} \mathcal{M}} \sum_{i=0}^{\mathcal{N}-1} \sum_{i'=0}^{\mathcal{N}-1} \sum_{k=2\pi/\lambda}^{\infty}\frac{4\pi^2 M_s^2 \Delta_i \Delta_{i'} \sin^2\left(\frac{kW}{2} \right)}{\lambda^2 \sinh\left(\frac{\pi \Delta_i k}{2}\right)\sinh\left(\frac{\pi \Delta_{i'} k}{2}\right)}
	   \frac{\sqrt{2\pi}}{2k}\left(2e^{-|(i-i')\mathcal{P}|k}-e^{-|\mathcal{T}-(i-i')\mathcal{P}|k}-e^{-|\mathcal{T}+(i-i')\mathcal{P}|k}\right)
	   \end{align}
	   and further,	   
	   \begin{align}
	   \sigma_{d, s}^{\infty, \mathcal{N}}&= \frac{\lambda}{4}\mu_0M_s^2 \left(\frac{2W}{\lambda}-1\right)^2\frac{\mathcal{T}}{\mathcal{P}}+\notag\\+& \frac{\pi^2\mu_{0}M_s^2 }{\mathcal{N}\mathcal{P}\lambda}\sum_{k=2\pi/\lambda}^{\infty} \sum_{i=0}^{\mathcal{N}-1} \sum_{i'=0}^{\mathcal{N}-1} \frac{ \Delta_i \Delta_{i'} \sin^2\left(\frac{kW}{2} \right)}{k \sinh\left(\frac{\pi \Delta_i k}{2}\right)\sinh\left(\frac{\pi \Delta_{i'} k}{2}\right)}
	   \left(2e^{-|(i-i')\mathcal{P}|k}-e^{-|\mathcal{T}-(i-i')\mathcal{P}|k}-e^{-|\mathcal{T}+(i-i')\mathcal{P}|k}\right)
	   \end{align}
	   The  frequencies $k$ in Fourier space can be expressed in terms of integer numbers as $k=2\pi n/\lambda$. Thus,  
	   \begin{align}
	   \sigma_{d, s}^{\infty, \mathcal{N}}&= \frac{\lambda}{4}\mu_0M_s^2 \left(\frac{2W}{\lambda}-1\right)^2\frac{\mathcal{T}}{\mathcal{P}}+\notag\\+& \frac{\pi\mu_{0}M_s^2 }{2\mathcal{N}\mathcal{P}}\sum_{n=1}^{\infty} \sum_{i=0}^{\mathcal{N}-1} \sum_{i'=0}^{\mathcal{N}-1} \frac{ \Delta_i \Delta_{i'} \sin^2(\frac{\pi n W}{\lambda}) }{n \sinh\left(\frac{\pi^2 n \Delta_i }{\lambda}\right)\sinh\left(\frac{\pi^2 n \Delta_{i'}}{\lambda}\right)}
	   \left(2e^{-\frac{2\pi |(i-i')\mathcal{P}|n}{\lambda}}-e^{-\frac{2\pi|\mathcal{T}-(i-i')\mathcal{P}|n}{\lambda}}-e^{-\frac{2\pi |\mathcal{T}+(i-i')\mathcal{P}|n}{\lambda}}\right)
	   \end{align}
	   Or, in a shorter form:
	   \begin{align}
	   \sigma_{d, s}^{\infty, \mathcal{N}}&= \frac{\lambda}{4}\mu_0M_s^2 \left(\frac{2W}{\lambda}-1\right)^2\frac{\mathcal{T}}{\mathcal{P}}+\notag\\+& \frac{\pi\mu_{0}M_s^2 }{2\mathcal{N}\mathcal{P}}\sum_{n=1}^{\infty} \frac{\sin^2(\frac{\pi n W}{\lambda})}{n} \sum_{i=0}^{\mathcal{N}-1} \sum_{i'=0}^{\mathcal{N}-1} \frac{ \Delta_i \Delta_{i'}  }{\sinh\left(\frac{\pi^2 n \Delta_i }{\lambda}\right)\sinh\left(\frac{\pi^2 n \Delta_{i'}}{\lambda}\right)}f_n(i-i')
	   \end{align}
	   with 
	   \begin{align}
	   f_n(\alpha)=2e^{-\frac{2\pi |\alpha\mathcal{P}|n}{\lambda}}-e^{-\frac{2\pi|\mathcal{T}-\alpha\mathcal{P}|n}{\lambda}}-e^{-\frac{2\pi |\mathcal{T}+\alpha\mathcal{P}|n}{\lambda}}
	   \end{align}

	   \subsubsection{Volume-Volume stray field energy}
 Now consider the volume-volume component of magnetostatic energy. Analogously to the previous section, the volume charges integral of interest can be expressed as
	   with the following volume charge distribution:
	   \begin{align}
	   \rho_v(\mathbf{r}) = &\sum_{j=0}^{\mathcal{N}-1}\sin(\psi_j)\Lambda_v(x, \lambda, W, \Delta)\theta(\mathcal{T}/2-|z-j\mathcal{P}-\mathcal{T}/2|)\theta(L/2-|y-L/2|) \label{eq:volumecharge}\\
	   &\Lambda_v(x, \lambda, W, \Delta)=(\rho_2*\Sh)(x)\\
	   &\Sh(x, \lambda, W)=\sum_{m = -\mathcal{M}+1}^{\mathcal{M}-1}\left[\delta(x-m\lambda) - \delta(x-m\lambda+W)\right]\\
	   &\rho_{2}(x, \Delta) = -\nabla \cdot \mathbf{M} = \frac{M_{s}}{\Delta}\frac{\tanh(x/\Delta)}{\cosh(x/\Delta)},
	   \end{align}
	   Where $\Sh$ is a general notation for the Dirac comb function.
	   \begin{figure}[h!]
	   	\includegraphics[scale = 0.67]{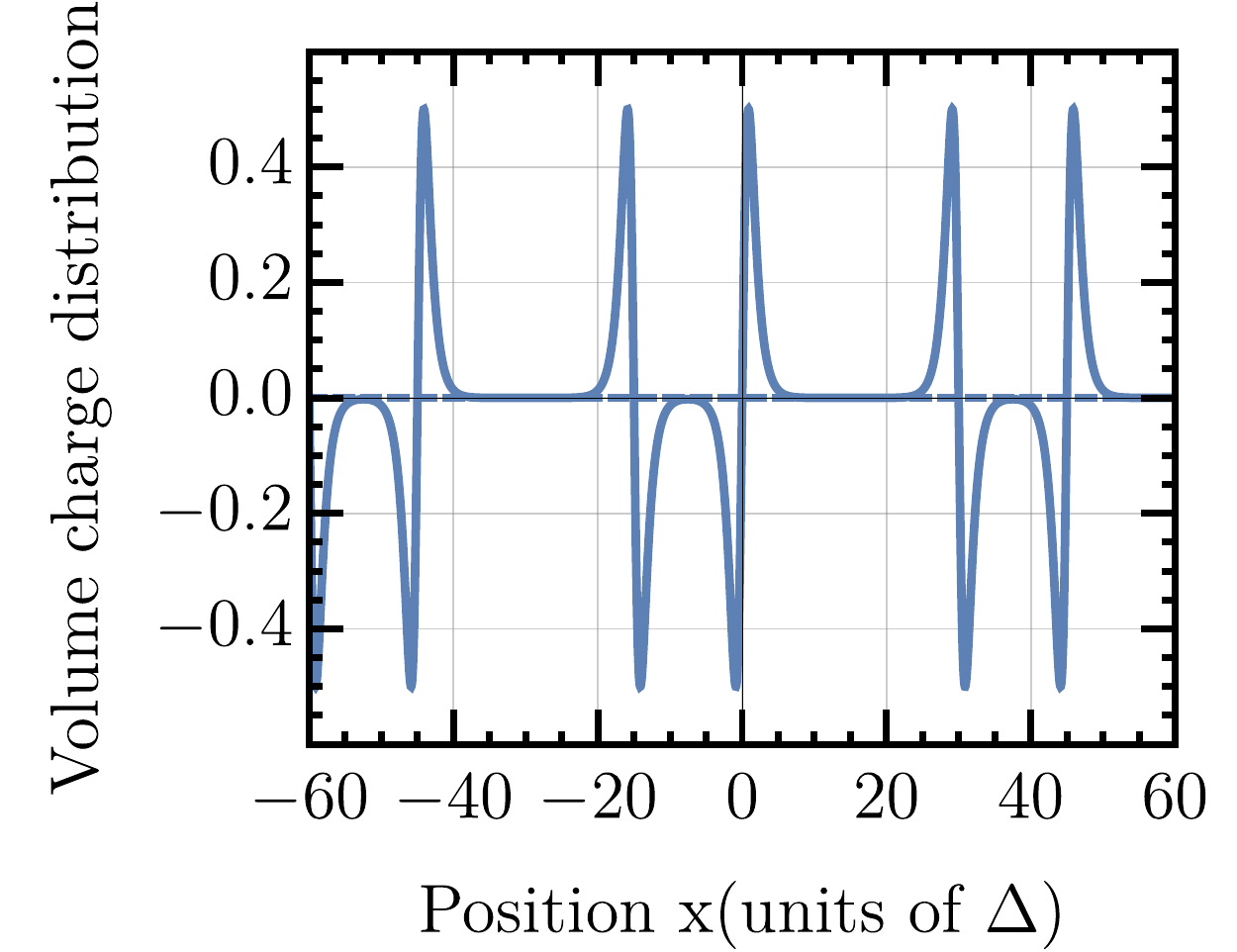}
	   	\caption{{\bf Distribution of volume charges in a magnetic layer with multidomain stripe state} with periodicity $\lambda=45\Delta$ and minority domain size of $W_{\rm{min}}=15\Delta$, possessing Neel domain walls of fixed chirality.}\label{fig:charges2}
	   \end{figure}
	   Volume-volume component of Eq.~\eqref{eq:magdomains} then becomes:
	   \begin{multline}
	   \sigma_{d, v}^{2\mathcal{M}, \mathcal{N}}= \frac{\mu_{0}}{16\pi  L \mathcal{N}\mathcal{P} \mathcal{M}} \sum_{j=0}^{\mathcal{N}-1} \sum_{j'=0}^{\mathcal{N}-1}\sin(\psi_j)\sin(\psi_{j'})\iint \d ^3 \mathbf{r} \d ^3 \mathbf{r'} \Lambda_v(x, \lambda, W)\Lambda_v(x', \lambda, W)\frac{1}{|\mathbf{r}-\mathbf{r'}|}\times \\\times \theta(\mathcal{T}/2-|z-j\mathcal{P}-\mathcal{T}/2|)\theta(\mathcal{T}/2-|z'-j'\mathcal{P}-\mathcal{T}/2|)
	   \end{multline}
	   after the substitution $z-\mathcal{P}j\rightarrow z$, $z'-\mathcal{P}j'\rightarrow z'$ we get
	   \begin{multline}
	   \sigma_{d, v}^{2\mathcal{M}, \mathcal{N}}= \frac{\mu_{0}}{16\pi  L \mathcal{N}\mathcal{P} \mathcal{M}} \sum_{j=0}^{\mathcal{N}-1} \sum_{j'=0}^{\mathcal{N}-1}\sin(\psi_j)\sin(\psi_{j'})  \iint \d ^3 \mathbf{r} \d ^3 \mathbf{r'} \frac{\Lambda_v(x, \lambda, W)\Lambda_v(x', \lambda, W)}{\sqrt{(x-x')^2+(y-y')^2+(z-z'+(j-j')\mathcal{P})^2}}\times \\\times \left[\delta(z'-\mathcal{T})- \delta(z')\right]\theta(\mathcal{T}/2-|z-\mathcal{T}/2|)
	   \end{multline}
	   Now, our system is infinite in $y$ direction, so results of B\"uttner \cite{buttner2015} et al.,
	   \begin{equation}
	   h(x, z)=\lim\limits_{L\rightarrow \infty} \frac{1}{L}\int_{0}^L \d y\int_{0}^L\d y'\frac{1}{\sqrt{(y-y')^2+x^2+z^2}}=-\ln(x^2+z^2)-2+2\ln(2L)+\mathcal{O}(L^{-1})\label{eq:hone}
	   \end{equation}
	   For which by using the antisymmetry of the system  we can remove the terms that would vanishing after the integration over x, leaving
	   \begin{equation}
	   h_v(x, z)=-\ln(x^2+z^2)
	   \end{equation}
	   we thus obtain
	   \begin{align}
	   \sigma_{d, v}^{2\mathcal{M}, \mathcal{N}}= \frac{\mu_{0}}{16\pi  \mathcal{N}\mathcal{P} \mathcal{M}} \sum_{j=0}^{\mathcal{N}-1} \sum_{j'=0}^{\mathcal{N}-1}\sin(\psi_j)\sin(\psi_{j'}) \iint_{0}^{\mathcal{T}} \d z \d z' \iint_{-\infty}^{\infty}\d x \d x'  \Lambda_v(x, \lambda, W)\Lambda_v(x', \lambda, W) h_v\left(x-x', z-z'+(j-j')\mathcal{P}\right)
	   \end{align}
	   Recognizing that $\Lambda_v$  is a periodic function in the limit $\mathcal{M}\rightarrow \infty$, we can use Eq. \eqref{eq:fourier} to reduce the double integral to the sum in $k$ space:
	   \begin{align}
	   \sigma_{d, v}^{\infty, \mathcal{N}}= \frac{\mu_{0}\sqrt{2\pi}\mathcal{M}\lambda }{16\pi  \mathcal{N}\mathcal{P} \mathcal{M}} \sum_{j=0}^{\mathcal{N}-1} \sum_{j'=0}^{\mathcal{N}-1}\sin(\psi_j)\sin(\psi_{j'}) \sum_{k}\hat{\Lambda}_{v,k} \hat{\Lambda}_{v, k}^* \iint_{0}^{\mathcal{T}} \d z \d z' \hat{h}_v\left(k, z-z'+(j-j')\mathcal{P}\right)
	   \end{align}
	   The fourier coefficients for $\Lambda_v$ can be found using the derived property of convolution (Eq. \eqref{eq:fourier_convolution}), resulting in: 
	   \begin{align}
	   \hat{\Lambda}_{v, k}[\hat{\Lambda}_{v, k}]^{*}=\frac{4\pi^2 M_s^2 k^2 \Delta_j \Delta_{j'} \sin^2\left(\frac{kW}{2} \right)}{\lambda^2 \cosh\left(\frac{\pi \Delta_j k}{2}\right)\cosh\left(\frac{\pi \Delta_{j'} k}{2}\right)},
	   \end{align}
	   The Fourier transform of $\hat{h}_{v}(k, z)$ reads:
	   \begin{align}
	   \hat{h}_v(k, z)=\sqrt{2\pi}\frac{1}{|k|}e^{-|z||k|}
	   \end{align}
	   Also, the double  integral over $dz$, $dz'$ has already been found by B\"uttner et al.\cite{buttner2015}:
	   \begin{align}
	   \iint_{0}^{\mathcal{T}}\d z \d z' \hat{h}_v\left(k, z-z'+j\mathcal{P}\right) & =\notag\\= \frac{\sqrt{2\pi}}{|k|^3}(e^{-|j\mathcal{P}-\mathcal{T}||k|}+e^{-|j\mathcal{P}+\mathcal{T}||k|}-2e^{-|j\mathcal{P}||k|}+|j\mathcal{P}-\mathcal{T}||k|+|j\mathcal{P}+\mathcal{T}||k|-2|j\mathcal{P}||k|)
	   \end{align}
	   Thus, we have
	   \begin{align}
	   \sigma_{d, v}^{\infty, \mathcal{N}}&= \frac{\pi^2 \mu_{0} M_s^2}{2\mathcal{P}\mathcal{N}\lambda} \sum_{j=0}^{\mathcal{N}-1} \sum_{j'=0}^{\mathcal{N}-1}\Delta_{j}\Delta_{j'}\sin(\psi_j)\sin(\psi_{j'}) \sum_{k=-\infty}^{\infty}\frac{ \sin^2\left(\frac{kW}{2} \right)}{ |k|\cosh\left(\frac{\pi \Delta_{j} k}{2}\right)\cosh\left(\frac{\pi \Delta_{j'} k}{2}\right)} \notag\\
	   &\times (e^{-|(j-j')\mathcal{P}-\mathcal{T}||k|}+e^{-|(j-j')\mathcal{P}+\mathcal{T}||k|}-2e^{-|(j-j')\mathcal{P}||k|}+|(j-j')\mathcal{P}-\mathcal{T}||k|+|(j-j')\mathcal{P}+\mathcal{T}||k|-2|(j-j')\mathcal{P}||k|)
	   \end{align}   
	   The  frequencies $k$ in Fourier space can be expressed in terms of integer numbers as $k=2\pi n/\lambda$.  Also, the function under the integral remains the same after the substitution $k\rightarrow -k$, and also becomes zero at $k=0$. Thus     
	   \begin{multline}
	   \sigma_{d, v}^{\infty, \mathcal{N}}=\frac{\pi \mu_{0} M_s^2}{2\mathcal{P}\mathcal{N}}  \sum_{j=0}^{\mathcal{N}-1} \sum_{j'=0}^{\mathcal{N}-1}\Delta_{j}\Delta_{j'}\sin(\psi_j)\sin(\psi_{j'}) \sum_{n=1}^{\infty}\sin^2\left(\frac{\pi nW }{\lambda}\right) \\
	   \times  \left[ \frac{\exp[-\frac{2\pi n |(j-j')\mathcal{P}+\mathcal{T}|}{\lambda}]+\exp[-\frac{2\pi n |(j-j')\mathcal{P}-\mathcal{T}|}{\lambda}]-2\exp[-\frac{2\pi n |(j-j')\mathcal{P}|}{\lambda}]+\frac{2\pi n|(j-j')\mathcal{P}+\mathcal{T}|}{\lambda}+\frac{2\pi n|(j-j')\mathcal{P}-\mathcal{T}|}{\lambda}-2\frac{2\pi n|(j-j')\mathcal{P}|}{\lambda}}{n\cosh\left(\frac{\pi^2 n \Delta_{j} }{\lambda}\right)\cosh\left(\frac{\pi^2 n \Delta_{j'} }{\lambda}\right)}\right]\\
	   \end{multline}
	   where we also recognized that $n=0$ term vanishes. Simplifying it, we get:
	   \begin{align}
	   \sigma_{d, v}^{\infty, \mathcal{N}}&=\frac{\pi \mu_{0} M_s^2}{\mathcal{P}\mathcal{N}}  \sum_{n=1}^{\infty}   \frac{ \sin^2\left(\frac{\pi nW }{\lambda}\right)}{n} \sum_{j=0}^{\mathcal{N}-1}\sum_{j'=0}^{\mathcal{N}-1}\frac{\Delta_{j}\Delta_{j'} \sin(\psi_j)\sin(\psi_{j'})}{\cosh\left(\frac{\pi^2 n \Delta_{j} }{\lambda}\right)\cosh\left(\frac{\pi^2 n \Delta_{j'} }{\lambda}\right)}g_n(j-j')\notag\\
	   & g_{n}(\alpha)=\begin{cases}2\sinh^2(\frac{\pi n \mathcal{T}}{\lambda})e^{-\frac{2\pi n \mathcal{P}|\alpha|}{\lambda}}, & \alpha\neq 0\\ e^{-\frac{2\pi n \mathcal{T}}{\lambda}}+\frac{2\pi n \mathcal{T}}{\lambda}-1, & \alpha=0 \end{cases}
	   \end{align}

	   \subsubsection{Surface-Volume stray field energy}
	   If the domain wall angle $\psi$ changes from layer to layer in the multilayer film with the multidomain state, then the state becomes asymmetric, so the surface and volume charges will start to interact. Assuming the system possesses an infinite number of domains, the integral  of interest is
	   with the surface charge distribution from Eq. \eqref{eq:surfacecharge} and the  volume charge distribution from Eq. \eqref{eq:volumecharge}. Therefore, the surface-volume component of magnetostatic energy (Eq.~\eqref{eq:magdomains}) can be expressed as:
	   \begin{multline}
	   \sigma_{d, sv}^{2\mathcal{M}, \mathcal{N}}= 2\frac{\mu_{0}}{16\pi  L \mathcal{N}\mathcal{P} \mathcal{M}} \sum_{i=0}^{\mathcal{N}-1} \sum_{j=0}^{\mathcal{N}-1}\sin(\psi_j)\iint \d ^3 \mathbf{r} \d ^3 \mathbf{r'} \Lambda_s(x, \lambda, W)\Lambda_v(x', \lambda, W)\frac{1}{|\mathbf{r}-\mathbf{r'}|}\times \\\times \left[\delta(z-\mathcal{P}i-\mathcal{T})- \delta(z-\mathcal{P}i)\right]\theta(\mathcal{T}/2-|z'-j\mathcal{P}-\mathcal{T}/2|)
	   \end{multline}
	   after the substitution $z-\mathcal{P}i\rightarrow z$, $z'-\mathcal{P}j\rightarrow z'$ we get
	   \begin{multline}
	   \sigma_{d, sv}^{2\mathcal{M}, \mathcal{N}}= \frac{\mu_{0}}{8\pi  L \mathcal{N}\mathcal{P} \mathcal{M}} \sum_{i=0}^{\mathcal{N}-1} \sum_{j=0}^{\mathcal{N}-1}\sin(\psi_j) \iint \d ^3 \mathbf{r} \d ^3 \mathbf{r'} \frac{\Lambda_s(x, \lambda, W)\Lambda_v(x', \lambda, W)}{\sqrt{(x-x')^2+(y-y')^2+(z-z'+(i-j)\mathcal{P})^2}}\times \\\times \left[\delta(z-\mathcal{T})- \delta(z)\right]\theta(\mathcal{T}/2-|z'-\mathcal{T}/2|)
	   \end{multline}
	   Now, our system is infinite in $y$ direction, so after evaluating the integral
	   \begin{align}
	   h_{sv}(x, z',\mathcal{T}, (i-j)\mathcal{P}) =  &\lim\limits_{L\rightarrow \infty} \frac{1}{L}\int_{0}^{L}dy\int_{0}^{L}dy'\int_{0}^{\mathcal{T}}dz \frac{\delta(z-\mathcal{T})- \delta(z)}{\sqrt{x^2+(y-y')^2+(z-z'+(i-j)\mathcal{P})^2}} =\notag\\&=h(x, \mathcal{T}-z'+(i-j)\mathcal{P})-h(x, -z'+(i-j)\mathcal{P}) \notag\\
	   & = -\ln\left(\frac{x^2+(z'-\mathcal{T}+(j-i)\mathcal{P})^2}{x^2+\left(z'+(j-i)\mathcal{P}\right)^2}\right)
	   \end{align}
	   in which we used $h$ from Eq. \eqref{eq:hone}, we thus obtain
	   \begin{align}
	   \sigma_{d, sv}^{\infty, \mathcal{N}}= \frac{\mu_{0}}{8\pi  \mathcal{N}\mathcal{P} \mathcal{M}} \sum_{i=0}^{\mathcal{N}-1} \sum_{j=0}^{\mathcal{N}-1}\sin(\psi_j)\int_{0}^{\mathcal{T}}\d z' \iint_{-\infty}^{\infty} \d x \d x'  \Lambda_s(x, \lambda, W)\Lambda_v(x', \lambda, W) h_{sv}\left(x-x', z', \mathcal{T}, (i-j)\mathcal{P}\right)
	   \end{align}
	   Since both $\Lambda_v$ and $\Lambda_s$ are periodic functions, we can use Eq. \eqref{eq:fourier} to reduce the double integral to a single integral in $k$ space:
	   \begin{align}
	   \sigma_{d, sv}^{\infty, \mathcal{N}}= \frac{\mu_{0}\sqrt{2\pi}\mathcal{M}\lambda}{8\pi  \mathcal{N}\mathcal{P} \mathcal{M}} \sum_{i=0}^{\mathcal{N}-1} \sum_{j=0}^{\mathcal{N}-1}\sin(\psi_j)\int_{0}^{\mathcal{T}}\d z' \sum_{k}  \hat{\Lambda}_{s,k} \hat{\Lambda}_{v, k}^*\hat{h}_{sv}\left(k, z', \mathcal{T}, (i-j)\mathcal{P}\right)
	   \end{align}
	   where the Fourier coefficients for $\Lambda_s$, $\Lambda_v$ and the Fourier transform of $h_{sv}$ are 
	   \begin{align}
	   \hat{\Lambda}_{s,k} \hat{\Lambda}_{v, k}^*&= \frac{4\pi^2 M_s^2 k \Delta_i \Delta_j \sin^2\left(\frac{kW}{2} \right)}{\lambda^2 \sinh\left(\frac{\pi \Delta_i k}{2}\right)\cosh\left(\frac{\pi \Delta_j k}{2}\right)},\\
	   \hat{h}_{sv}\left(k, z', \mathcal{T}, (i-j)\mathcal{P}\right) &= \frac{\sqrt{2\pi}}{|k|}\left(e^{-|z'-\mathcal{T}+(j-i)\mathcal{P}||k|}-e^{-|z'+(j-i)\mathcal{P}||k|}\right)
	   \end{align}
	   Now, the sum in $k$ space can be simplified as:
	   \begin{align}
	   \sum_{k=-\infty}^{\infty} \hat{\Lambda}_{s,k} \hat{\Lambda}_{v, k}^*\hat{h}_{sv}\left(k, z', \mathcal{T}, (i-j)\mathcal{P}\right)  
	   =\frac{8\pi^2\sqrt{2\pi} M_s^2 \Delta_i\Delta_j}{\lambda^2 }\sum_{k=2\pi/\lambda}^{\infty}\frac{  \sin^2\left(\frac{kW}{2} \right)}{\sinh\left(\frac{\pi \Delta_i k}{2}\right)\cosh\left(\frac{\pi \Delta_j k}{2}\right)}\left(e^{-k|z'-\mathcal{T}+(j-i)\mathcal{P}|}-e^{-k|z'+(j-i)\mathcal{P}|}\right)
	   \end{align}
	   Thus,
	   \begin{align}
	   \sigma_{d, sv}^{\infty, \mathcal{N}}= \frac{2\pi^2\mu_{0}M_s^2}{\mathcal{N}\mathcal{P} \lambda} \sum_{i=0}^{\mathcal{N}-1} \sum_{j=0}^{\mathcal{N}-1}\Delta_i\Delta_j\sin(\psi_j)\sum_{k=2\pi/\lambda}^{\infty}\frac{  \sin^2\left(\frac{kW}{2} \right)}{\sinh\left(\frac{\pi \Delta_i k}{2}\right)\cosh\left(\frac{\pi \Delta_j k}{2}\right)}\int_{0}^{\mathcal{T}}\d z' \left(e^{-k|z'-\mathcal{T}+(j-i)\mathcal{P}|}-e^{-k|z'+(j-i)\mathcal{P}|}\right)
	   \end{align}
	   The integral over $z'$ can be carried out easily, since the multilayer period is always larger than the single magnetic layer thickness:
	   \begin{align}
	   \int_{0}^{\mathcal{T}}\d z' \left(e^{-k|z'-\mathcal{T}+(j-i)\mathcal{P}|}-e^{-k|z'+(j-i)\mathcal{P}|}\right) = \begin{cases}0, & i=j \\\frac{4\sinh^2(\frac{k\mathcal{T}}{2})e^{-k\mathcal{P}|i-j|}}{k},&i < j \\ -\frac{4\sinh^2(\frac{k\mathcal{T}}{2})e^{-k\mathcal{P}|i-j|}}{k}, & i>j \end{cases}
	   \end{align}
	   Note that the system possesses no surface-volume interactions between charges of one specific layer.
	   \begin{align}
	   \sigma_{d, sv}^{\infty, \mathcal{N}}= \frac{8\pi^2\mu_{0}M_s^2}{\mathcal{N}\mathcal{P} \lambda} \sum_{k=2\pi/\lambda}^{\infty}\frac{  \sin^2\left(\frac{kW}{2} \right)\sinh^2(\frac{k\mathcal{T}}{2})}{k}\sum_{j=0}^{\mathcal{N}-1}\sum_{i=0}^{\mathcal{N}-1}\frac{\Delta_i\Delta_j\sin(\psi_j) e^{-k\mathcal{P}|i-j|} \text{sgn}(j-i)}{\sinh\left(\frac{\pi \Delta_i k}{2}\right)\cosh\left(\frac{\pi \Delta_j k}{2}\right)}
	   \end{align}
	   Plugging in $k=2\pi n/\lambda$, we finally obtain:
	   \begin{align}
	   \sigma_{d, sv}^{\infty, \mathcal{N}}= \frac{4\pi \mu_{0}M_s^2}{\mathcal{N}\mathcal{P} } \sum_{n=1}^{\infty}\frac{  \sin^2\left(\frac{\pi n W}{\lambda}\right)\sinh^2(\frac{\pi n \mathcal{T}}{\lambda})}{n}\sum_{i=0}^{\mathcal{N}-1}\sum_{j=0}^{\mathcal{N}-1} \frac{\Delta_i\Delta_j\sin(\psi_j)e^{-\frac{2\pi n\mathcal{P}|i-j|}{\lambda}} \text{sgn}(j-i)}{\sinh\left(\frac{\pi^2 n \Delta_i}{\lambda}\right)\cosh\left(\frac{\pi^2 n \Delta_j}{\lambda}\right)}
	   \end{align}

	   \subsubsection{Discrete Fourier space identities}\label{sec:discrete_Fourier_identities}
	   Analogously to Ref~\citenum{Lemesh2017}, consider periodic real valued functions $f(x)$ and $g(x)$. They have the following discrete representation in Fourier space:
	   \begin{align}
	   f(x) = \sum_{k} \hat{f}_k e^{ikx} = \sum_{k}\hat{f}_k^\ast e^{-i kx}\\
	   g(x) = \sum_{k^\prime} \hat{g}_{k^\prime} e^{ik^\prime x} = \sum_{k^\prime}\hat{g}_{k^\prime}^\ast e^{-i k^\prime x},
	   \end{align}
	   where $\hat{f}_k$, $\hat{g}_{k^\prime}$ are the Fourier coefficients of $g$ and $f$. Suppose we want to calculate the following integral:
	   \begin{align}
	   I = \iint \d x \d x^\prime f(x)g(x^\prime) h(x-x^\prime),
	   \end{align}
	   which also has an additional function $h$. The integral can be expressed as
	   \begin{align}
	   I &= \iint \d x \d x^\prime \sum_{k, k^\prime} \hat{f}_k\hat{g}_{k^\prime}^\ast e^{ikx}e^{-ik^\prime x^\prime} h(x-x^\prime) \notag\\
	   &= \iint \d x \d x^\prime \sum_{k, k^\prime} \hat{f}_k\hat{g}_{k^\prime}^\ast e^{ik(x-x')}e^{i(k-k^\prime) x^\prime} h(x-x^\prime)
	   \end{align}
	   Introducing the substitution $(x\text{, } x^\prime) \rightarrow (y\text{, } x')$, where $y=x-x^{\prime}$ we get:
	   \begin{align}
	   I &= \int \d y  \sum_{k, k^{\prime}} \hat{f}_k\hat{g}_{k^\prime}^\ast e^{iky}h(y)\int \d x^\prime e^{i(k-k^\prime) x^\prime} 
	   \end{align}
	   We can express $\int \d x^\prime e^{i(k-k^\prime) x^\prime}$ as $N\lambda\delta_{k, k^{\prime}}$, where $\delta_{k, k^{\prime}}$ is the Kronecker delta, $N$ is the number of periods along the $x$ dimension (which we will set to infinity at a later stage) and $\lambda$ is the period length, so $N\lambda$ is the total length of the sample.  Therefore,
	   \begin{align}
	   I &= N\lambda\int \d y  \sum_{k, k^{\prime}} \hat{f}_k\hat{g}_{k^\prime}^\ast e^{iky}h(y)\delta_{k,k^{\prime}} \notag\\
	   &= N\lambda \sum_{k} \hat{f}_k\hat{g}_{k}^\ast \int \d y h(y) e^{iky}\notag\\
	   &= \sqrt{2\pi}N\lambda  \sum_{k} \hat{f}_k\hat{g}_{k}^\ast \hat{h}(k) \label{eq:fourier}
	   \end{align}
	   \subsubsection{Fourier coefficients of convolved functions}
	   First, consider the function:
	   \begin{equation}
	   F(x) = (f*g)(x),
	   \end{equation}
	   where $g(x)$ is a periodic function with a period $\lambda$, and $f(x)$ is a regular real valued function. Then their convolution $F(x)$ is also a periodic function with a period $\lambda$:
	   \begin{align}
	   \hat{F}_k&=\frac{1}{\lambda}\int\d x(f*g)(x)e^{-ikx}\notag\\
	   &=\frac{1}{\lambda}\int\d x \int \d x'g(x-x')f(x')e^{-ikx} \notag\\
	   &=\frac{1}{\lambda}\int\d x \int \d x'g(x-x')f(x')e^{-ik(x-x')}e^{-ikx'}
	   \end{align}
	   Introducing the substitution $(x\text{, } x^\prime) \rightarrow (y\text{, } x')$, where $y=x-x^{\prime}$ we obtain:
	   \begin{align}
	   \hat{F}_k&=\frac{1}{\lambda}\int\d y g(y)e^{-iky}\int \d x'f(x') e^{-ikx'}\notag\\
	   &=\sqrt{2\pi}\hat{g}_{k} [\hat{f}(k)]^* \label{eq:fourier_convolution}
	   \end{align}
	   
	   \subsubsection{Fourier coefficients of $\Lambda_s$}
	   The fourier coefficients for $\Lambda_s$ can be found using the derived property of convolution (Eq. \eqref{eq:fourier_convolution}) as
	   \begin{align}
	   \hat{\Lambda}_{s, k} & = \sqrt{2\pi }[\hat{\Pi}_{k}]^* \hat{\rho}_1( k),
	   \end{align}   
	   where
	   \begin{align}
	   \hat{\Pi}_{k} & = \frac{1}{\lambda} \int_{-W}^{0} -1 \d x e^{-ikx}+\frac{1}{\lambda} \int_{0}^{\lambda-W} 1 \d x e^{-ikx} \notag\\
	   &=\frac{(2-e^{ikW}(1+e^{-ik\Lambda}))}{ik \lambda}  \\
	   \hat{\rho}_1(k) & = \frac{1}{\sqrt{2\pi}}\int_{-\infty}^{\infty}\frac{M_{s}}{2\Delta}\frac{1}{\cosh^2(x/\Delta)}e^{ikx}dx \notag\\
	   &= \frac{M_s k \Delta}{2}\sqrt{\frac{\pi}{2}}\frac{1}{\sinh\left(\frac{\pi  \Delta  k}{2}\right)}
	   \end{align}
	   Note that $k=2\pi n/\lambda$, with $n\in\mathbb{N}$, therefore $e^{-ik\lambda}\equiv 0$ unless $k=0$. We then finally obtain $\hat{\Lambda}_{s, k}$
	   \begin{align}
	   \hat{\Lambda}_{s, k} = & \begin{cases}\frac{M_s \pi\Delta(1-e^{-ikW})i}{\lambda \sinh\left(\frac{\pi \Delta k}{2}\right)}, & k \neq 0 \\ M_s\left(1-\frac{2W}{\lambda}\right), & k=0 \end{cases} \\
	   \hat{\Lambda}_{s, k}[\hat{\Lambda}_{s, k}]^{*} & = \begin{cases}\left(\frac{2\pi M_s \Delta  \sin\left(\frac{kW}{2} \right)}{\lambda \sinh\left(\frac{\pi \Delta k}{2}\right)}\right)^2, & k \neq 0 \\ M_s^2\left(1-\frac{2W}{\lambda}\right)^2, & k=0 \\\end{cases},
	   \end{align}
	   
	   \subsubsection{Fourier coefficients of $\Lambda_v$}
	   Analogously, the fourier coefficients for $\Lambda_v$ can be found as
	   \begin{align}
	   \hat{\Lambda}_{v, k} & = \sqrt{2\pi }[\hat{\Sh}_{k}]^* \hat{\rho}_2( k),
	   \end{align}
	   where
	   \begin{align}
	   \hat{\Sh}_{k} & = \frac{1}{\lambda} \int_{-\lambda/2}^{\lambda/2}\d x[\delta(x)-\delta(x+W)]e^{-ikx} \notag\\
	   & =\frac{1}{\lambda} \int_{-\lambda/2}^{\lambda/2}\d x \int \d k' e^{ik'x}(1-e^{ik'W})e^{-ikx}\notag\\
	   & =\frac{1}{\lambda} \int \d k' (1-e^{ik'W})\int_{-\lambda/2}^{\lambda/2}\d x e^{i(k'-k)x}\notag\\
	   & =\frac{1}{\lambda} \int \d k' (1-e^{ik'W})\delta(k'-k)\notag\\
	   &=\frac{1-e^{ikW}}{\lambda}  \\
	   \hat{\rho}_2(k) & =\frac{1}{\sqrt{2\pi}} \int_{-\infty}^{\infty}\frac{M_{s}}{\Delta}\frac{\tanh(x/\Delta)}{\cosh(x/\Delta)}e^{ikx}dx \notag\\
	   &=M_s i k \Delta \sqrt{\frac{\pi}{2}}\frac{1}{\cosh\left(\frac{\pi \Delta k}{2}\right)}
	   \end{align}
	   Therefore,
	   \begin{align}
	   \hat{\Lambda}_{v, k} & = \frac{\pi M_s k \Delta  \left(1-e^{-ikW}\right)i}{\lambda \cosh\left(\frac{\pi \Delta k}{2}\right)},\\
	   \hat{\Lambda}_{v, k}[\hat{\Lambda}_{v, k}]^{*}&=\frac{2\pi^2 M_s^2 k^2 \Delta^2  \left[1-\cos\left(kW \right)\right]}{\lambda^2 \cosh^2\left(\frac{\pi^2 \Delta k}{\lambda}\right)}=\left(\frac{2\pi M_s k \Delta  \sin\left(\frac{kW}{2} \right)}{\lambda \cosh\left(\frac{\pi \Delta k}{2}\right)}\right)^2,
	   \end{align}

\section{Theory of multilayer skyrmions}
The statics and current-driven dynamics of skyrmions in multilayers can be derived by extending the recent single layer (large R) skyrmion theory~\cite{Buttner2017}. Here, we assume that the equilibrium domain wall parameters $\Delta$, $\psi_i$ are known parameters (see Section~\ref{sec:dw} for details). Assuming that skyrmions are circular, magnetostatically coupled, and possess no defects, we can re-express the Eq.~S37 from Ref~\cite{Buttner2017} to account for the multilayer formalism~\footnote{The typical skyrmion collapse radii in ferromagnetic multilayers ~\cite{Buttner2017} are typically above the limitations of this theory $R> \mathcal{O}(\Delta)$}:
\begin{align}
E_{\rm{tot}}^{\text{sk}, \mathcal{N}}  (R, \Delta, \psi_i, B_z) &  =2\pi d R \sigma_{tot}^{ 1, \mathcal{N}}(\Delta, \psi_i)\notag  \\& +aR-bR\ln\left(R/d\right)+cB_z R^2\label{eq:skyrmionenergy}
\end{align}
with constants defined as
\begin{align}
a & = -\mu_0 M_s^2 (\mathcal{P}\mathcal{N})^2[6\ln(2)-1]\\
b& = 2\mu_0 M_s^2(\mathcal{P}\mathcal{N})^2\\
c& = -2\pi \mathcal{P}\mathcal{N} M_s\\
d& = \mathcal{P}\mathcal{N}
\end{align}
The equilibrium radius can be found by minimizing 
$E_{\rm{tot}}^{\text{sk}, \mathcal{N}} $ with respect to $R$ (assuming $\Delta$, $\psi_i$, $B_z$ are given).

 Now, for skyrmion dynamics, we need to start from the Thiele equation~\cite{Thiele1973} extended to multilayers via the First Newton's law:
\begin{align}
\mathcal{N}(\vec{\tilde{G}}\times \vec{v}-\alpha \tilde{D} 
\vec{v})+\sum_{i=0}^{\mathcal{N}-1}\vec{F}_i=\vec{0},\label{eq:thielle}
\end{align}
with the following parameters defined in Ref.~\citenum{Buttner2017}:
\begin{align}
\vec{\tilde{G}} &= (0, 0, -4\pi N)^{T}\\
\tilde{D} &= \pi I_{A}(R/\Delta)\\
\vec{F}_i & = -\frac{\hbar \gamma \theta_{\text{SH}}}{2eM_s\mathcal{T}}N\Delta I_{D}(R/\Delta)\hat{R}(\psi_i-\pi/2)\vec{j}\label{eq:skpar},
\end{align} 
where the factor $\pi/2$ stems from the differences in our definition of $\psi$ and in Ref.~\citenum{Buttner2017}, and
\begin{align}
I_{\rm{A}} (\rho) &= 2\rho+\frac{2}{\rho}+1.93(\rho-0.65)\exp[-1.48(\rho-0.65)]\\
I_{\rm{D}} (\rho) &= \pi \rho+\frac{1}{2}\exp(-\rho)
\end{align}
Steady-state skyrmion velocity and hall angle (i.e. the angle between the velocity and the current direction) can be found in a straightforward manner by assuming that current flows in x direction, and that skyrmions remains circular, without topological defects, magnetostatically coupled and preserving its static configuration~\footnote{In contrast to straight domain walls, the injected current has a minor influence on the domain wall angle of skyrmions, since  $\psi$ and $q$ are not conjugated variables for a skyrmion. However, as was found from our explicit multilayer simulations, this property starts to break down at very large currents}. We then can apply a similar logic as in Ref~\cite{Buttner2017}, but substituting $\sin(\psi)\rightarrow\sum_{i=0}^{\mathcal{N}-1}\sin(\psi_i)/\mathcal{N}$, and $\cos(\psi)\rightarrow\sum_{i=0}^{\mathcal{N}-1}\cos(\psi_i)/\mathcal{N}$, which eventually results in 
\begin{align}
|v| & = j \frac{\pi \hbar \gamma \Delta  \theta_{\text{SH}} I_{D}(\rho)}{2 e M_s \mathcal{T}\sqrt{\tilde{G}^2+\tilde{D}^2\alpha^2}} \tilde{f}\\
\xi' & = \rm{atan}2(\tilde{G},\tilde{D}\alpha) -(\tilde{\psi}-\pi/2)+\pi\Theta(\theta_{\rm{SH}}N),
\end{align}
With the following constants that  are captured only in our 3D model
\begin{align}
\tilde{f} & = \frac{\sqrt{\left(\sum_{i=0}^{\mathcal{N}-1} \cos(\psi_i)\right)^2+\left(\sum_{i=0}^{\mathcal{N}-1} \sin(\psi_i)\right)^2}}{\mathcal{N}} \\
\tilde{ \psi} & = \rm{atan2} \left(\sum_{i=0}^{\mathcal{N}-1} \sin(\psi_i), \sum_{i=0}^{\mathcal{N}-1} \cos(\psi_i)\right)
\end{align}


%